\documentclass[%
reprint,
superscriptaddress,
 amsmath,amssymb,
 aps,
 pra,
]{revtex4-2}

\usepackage{graphicx}
\usepackage{dcolumn}
\usepackage{bm}
\usepackage{xcolor}
\usepackage{float}
\usepackage{comment}
\usepackage{hyperref}

\usepackage[T1]{fontenc}

\usepackage[caption=false]{subfig}

\usepackage{amsmath}
\usepackage{amssymb}
\usepackage{mathtools}
\allowdisplaybreaks

\usepackage[normalem]{ulem}
\usepackage{cancel}

\usepackage{amsthm}
\newtheorem{theorem}{Theorem}[section]

\newtheorem{conjecture}[theorem]{Conjecture}

\makeatletter
\newsavebox{\@brx}
\newcommand{\llangle}[1][]{\savebox{\@brx}{\(\m@th{#1\langle}\)}%
  \mathopen{\copy\@brx\kern-0.5\wd\@brx\usebox{\@brx}}}
\newcommand{\rrangle}[1][]{\savebox{\@brx}{\(\m@th{#1\rangle}\)}%
  \mathclose{\copy\@brx\kern-0.5\wd\@brx\usebox{\@brx}}}
\makeatother

\begin{document}

\title{Enhancing Noisy Quantum Sensing by GHZ State Partitioning}

\author{Allen Zang}
\affiliation{Pritzker School of Molecular Engineering, The University of Chicago, Chicago, IL, USA}

\author{Tian-Xing Zheng}
\affiliation{Pritzker School of Molecular Engineering, University of Chicago, Chicago, IL, USA}
\affiliation{Department of Physics, The University of Chicago, Chicago, IL, USA}

\author{Peter C. Maurer}
\affiliation{Pritzker School of Molecular Engineering, University of Chicago, Chicago, IL, USA}

\author{Frederic T. Chong}
\affiliation{Department of Computer Science, The University of Chicago, Chicago, IL, USA}
\affiliation{Infleqtion Inc., Chicago, IL, USA}

\author{Martin Suchara}
\affiliation{Microsoft Azure Quantum, Microsoft Corporation, Redmond, WA, USA}

\author{Tian Zhong}
\affiliation{Pritzker School of Molecular Engineering, The University of Chicago, Chicago, IL, USA}

\date{\today}

\begin{abstract}
    Presence of harmful noise is inevitable in entanglement-enhanced sensing systems, requiring careful allocation of resources to optimize sensing performance in practical scenarios. We advocate a simple but effective strategy to improve sensing performance in the presence of noise. Given a fixed number of quantum sensors, we partition the preparation of GHZ states by preparing smaller, independent sub-ensembles of GHZ states instead of a GHZ state across all sensors. We perform extensive analytical studies of the phase estimation performance when using partitioned GHZ states under realistic noise -- including state preparation error, particle loss during parameter encoding, and sensor dephasing during parameter encoding. We derive simple, closed-form expressions that quantify the optimal number of sub-ensembles for partitioned GHZ states. We also examine the explicit noisy quantum sensing dynamics under dephasing and loss, where we demonstrate the advantage from partitioning for maximal QFI, short-time QFI increase, and the sensing performance in the sequential scheme. The results offer quantitative insights into the sensing performance impact of different noise sources and reinforce the importance of resource allocation optimization in realistic quantum applications. 
\end{abstract}

\maketitle

\section{Introduction}\label{sec:intro}

Quantum sensors provide precise signal detection capabilities by harnessing coherence, a unique property of quantum bits (qubits). The sensor qubit in a superposition state accumulates a phase between its "0" and "1" state that is proportional to the external field coupling to the qubit and the evolution of time. By reading out the phase, one can obtain information about the environment of the quantum sensor. 

The quantum sensing concept~\cite{degen2017quantum} has been applied to fundamental scientific studies~\cite{safronova2018search,ye2024essay} such as measurement of time~\cite{ludlow2015optical}, gravity~\cite{muller2010precision,abbott2016observation} and the search of dark matter~\cite{bass2024quantum}. It also has the potential to realize practical quantum advantage including nanoscale magnetic ~\cite{taylor2008high, maze2008nanoscale, barry2020sensitivity} and electric ~\cite{dolde2011electric,bian2021nanoscale} field imaging for material ~\cite{casola2018probing, rovny2024nanoscale} and biological~\cite{schirhagl2014nitrogen,aslam2023quantum} systems. However, most of the sensing platforms so far have been treating each sensor qubit in the ensemble separately~\cite{pham2011magnetic,glenn2018high,barry2020sensitivity,aeppli2024clock} and haven not fully utilized the correlation among the sensors. When using a separable state as a quantum sensor, the standard deviation goes down as $1/\sqrt{n}$, the standard-quantum limit (SQL). When using an entangled state, the precision could go beyond the SQL and reach the theoretical upper bound of quantum sensing $1/n$, the Heisenberg Limit (HL)~\cite{giovannetti2004quantum, pezze2009entanglement}. A typical example uses Greenberger-Horne-Zeilinger (GHZ) states $\frac{1}{\sqrt{2}}|00\dots 0\rangle + \frac{1}{\sqrt{2}}e^{in\theta}|11\dots 1\rangle$ as quantum sensors~\cite{greenberger1989going}, where the signal accumulation is $n$ times faster than an uncorrelated $n$-spin ensemble. This so-called entanglement-enhanced sensing has been subject of theoretical studies for a long time~\cite{wineland1992spin,kitagawa1993squeezed,cappellaro2005entanglement,giovannetti2006quantum,cappellaro2009quantum,goldstein2011environment,pezze2018quantum}, but rarely realized experimentally~\cite{pedrozo2020entanglement,yu2020quantum,xie2021beating,wu2025spin}, and with exceptions~\cite{cooper2019environment} did not show advantages over the classical case.

\begin{figure}[t]
    \centering
    \includegraphics[width=\linewidth]{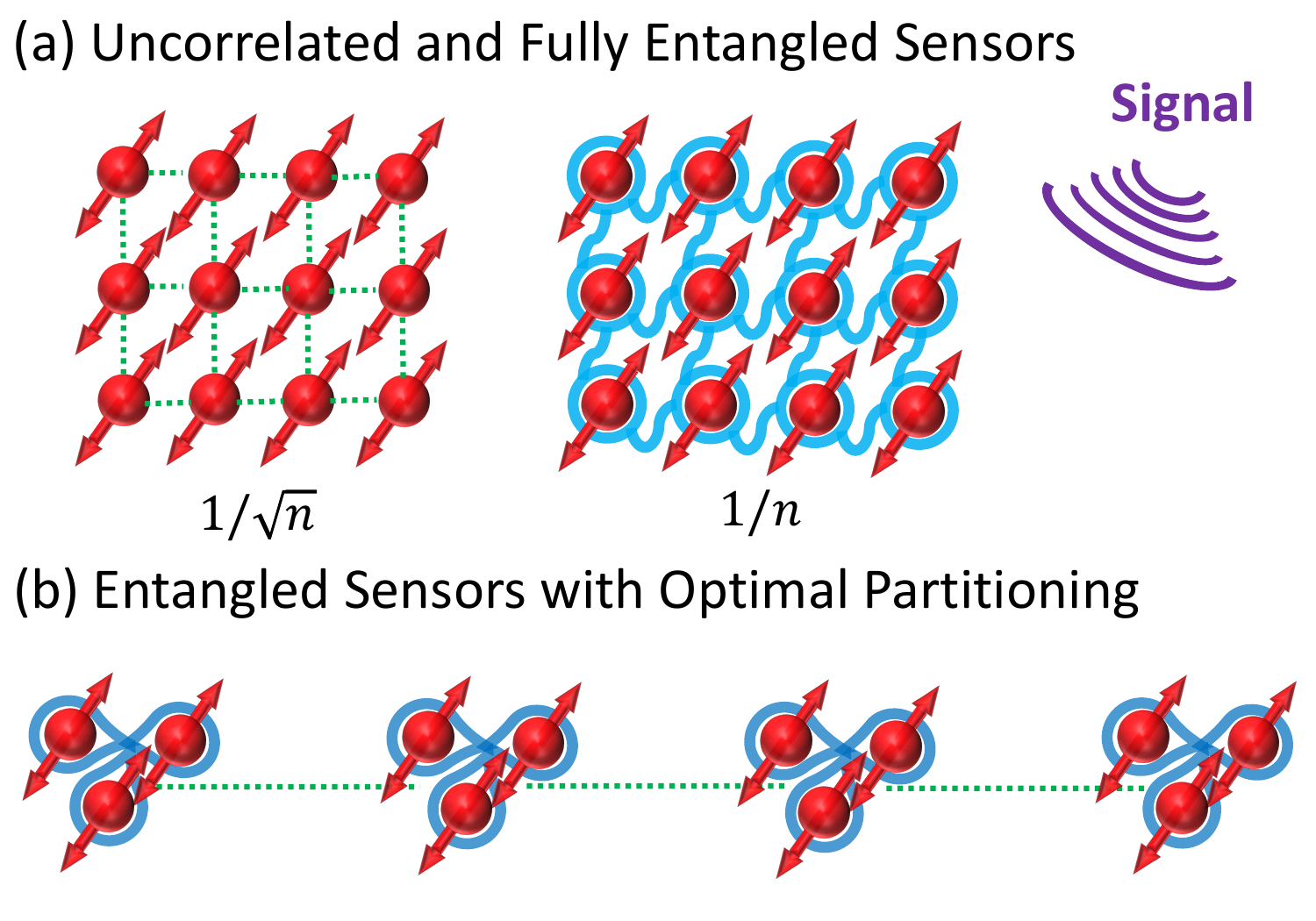}
    \caption{Schematic of using the same number of quantum sensors as (a) one ensemble of uncorrelated sensors (SQL) or a large GHZ state (HL). (b) Partitioned quantum sensor composed of several smaller GHZ states. Blue curves represent entanglement across individual quantum sensors.
    } 
    \label{fig:schematic}
\end{figure}

The major bottleneck for realizing the entanglement-enhanced sensing is noise that includes the imperfections during state preparation, signal accumulation and readout~\cite{huelga1997improvement,escher2011general,demkowicz2012elusive,cooper2019environment}. Moreover, certain types of imperfections will make the entangled states' sensitivity worse compared to even the non-entangled case. For example, a deeper quantum circuit is needed for preparing a larger GHZ state that reaches higher quantum sensing performance. The unavoidable infidelity in such a circuit (e.g. caused by two-qubit gate errors) adds up with the circuit depth and eventually damages the sensitivity. 

To overcome the limitations introduced by the imperfections, we propose a partitioning strategy which breaks down the all-to-all entangled GHZ states into smaller GHZ-state ensembles. From a computational perspective, it makes sense to use shallower circuits to create a number of different GHZ states, but it is unclear what the performance will be for the GHZ-state sensing case. 
Partitioning quantum sensors into multiple sub-ensembles has been suggested for atomic clocks~\cite{borregaard2013efficient,rosenband2013exponential,kessler2014heisenberg,schulte2020prospects,li2022improved}. Our results emphasizes that even simpler partitioning strategies can be advantageous in a general quantum sensing primitive of phase estimation under realistic imperfections, as long as we can obtain information about the imperfections through quantum characterization and benchmarking~\cite{martinis2015qubit,harper2020efficient,eisert2020quantum,kliesch2021theory}.
Moreover, recent experimental advancements have demonstrated the capability of controlling multiple qubit ensembles~\cite{schioppo2017ultrastable,hines2023spin,cao2024multi,finkelstein2024universal,shaw2024multi,robinson2024direct,cooper2024graph,zheng2024reducing,hassan2024ratchet}, which justifies the realism of the partitioning strategy.
Specifically, we build the analytical model to analyze the sensing performance of the partitioned sensor, and find that under certain imperfections, the partitioned sensor outperforms the traditional sensor which entangles all available qubits. Our analytical solution indicates that optimal resource allocation depends on the type and level of imperfections, which corroborates with recent theoretical results~\cite{yin2024small} and previous numerical results~\cite{jarzyna2013matrix,chabuda2020tensor}, in that the size of a sub-ensemble under optimal partitioning scales with inverse error rate. Our analysis of the noisy quantum sensing dynamics further elucidates the benefits that partitioning can bring in practical scenarios. Our results can serve as the practical guidance for realizing entanglement-enhanced sensing in experimental systems with resource limitations and control imperfections. 

The paper is organized as follows. In Section~\ref{sec:model}, we formulate the problem and describe the error models under consideration. We derive the optimized partitioning strategy for three types of imperfections. In Section~\ref{sec:state_prep} we start with the case where state preparation is the only error source. In Section~\ref{sec:loss} we take into account sensor qubit losses in addition to the state preparation errors. We consider dephasing in Section \ref{sec:dephasing}. The noisy quantum sensing dynamics with losses and dephasing is investigated in Section~\ref{sec:dynamics}. In Section~\ref{sec:conclusion} we summarize our results.

\section{Problem formulation and models}\label{sec:model}
We study the problem of phase estimation in quantum sensing. We consider an ensemble of $n$ qubit sensors, each coupled to a local parameter $\theta$ through the canonical $z$-direction phase accumulation generated by the Hermitian operator $G = \frac{\theta}{2}\sum_{i=1}^{n}Z^{(i)}$, where $Z^{(i)}$ denotes the spin $z$-component/Pauli $Z$ for the $i$-th qubit. The ideal parameter encoding channel is the unitary $U(\theta) = \exp\left[-i\frac{\theta}{2}\sum_{i=1}^{n}Z^{(i)}\right]$.
In the following, we first elaborate on the modeling of the three error sources in noisy quantum phase estimation: state preparation error, particle loss, and sensor dephasing. Then we describe the figures of merit we use to evaluate sensing performance.

\subsection{GHZ state preparation}
For concreteness and ease of analysis, we assume that the result of preparing an $n$-qubit GHZ state is a depolarized GHZ state
\begin{align}\label{eqn:noisy_ghz}
    \rho = V(n)|\mathrm{GHZ}_n\rangle\langle\mathrm{GHZ}_n| + [1-V(n)]I_n/2^n,
\end{align}
where $V(n)$ denotes the visibility under white noise, $|\mathrm{GHZ}_n\rangle=(|00\dots 0\rangle+|11\dots 1\rangle)/\sqrt{2}$ is the standard pure $n$-qubit GHZ state, and $I_n$ is the identity operator applied on $n$ qubits. We assume that its fidelity decays with the number of qubits in the GHZ state~\cite{zang2024quantum}
\begin{align}
    F(n) = \langle\mathrm{GHZ}_n|\rho|\mathrm{GHZ}_n\rangle = k^{n-1}F,
\end{align}
where $k,F\in(0,1]$. The factor $F$ can be understood as the fidelity of initialization, e.g. the preparation of a first qubit in the $|+\rangle$ state before successive application of entangling gates in the standard GHZ state generation circuit. The parameter $k$ corresponds to the quality of local entanglement generation and it can be interpreted as the fidelity of noisy CNOT gates used to generate the GHZ states. Larger values of $k$ are better. We emphasize that the above model is not unique for modeling realistic systems, but it is operationally meaningful and simple so that no further details are needed. The model also captures the phenomenological aspect of noisy state preparation that the state fidelity decays as the size of the state increases, which applies to but is not limited to gate-based state preparation~\cite{monz201114,song2019generation,omran2019generation,ho2019ultrafast,pogorelov2021compact,mooney2021generation,zhao2021creation,comparin2022multipartite,zhang2024fast,cao2024multi,yin2024fast}. Furthermore, the exponential decay in the GHZ state fidelity is a good approximation under the assumption that the circuit-level noise is not too biased; for a more detailed discussion see~\cite{zang2024quantum}.

\subsection{Sensor qubit loss}
In some physical platforms, such as ultracold atoms, the physical particles serving as quantum sensors can be lost during system evolution, especially during the parameter encoding. These systems, however, usually demonstrate long coherence times. For such systems, we focus on the regime where the parameter encoding duration is shorter than the coherence time, and therefore the parameter encoding is effectively unitary under the assumption that no sensor is lost. We treat particle loss as a binary process, i.e., one sensor is either lost or not lost during the parameter encoding. This error can be described by a probability $p$ for each sensor indicating survival of the encoding process.

\subsection{Sensor qubit dephasing}
In solid state systems, sensors coherence lifetime is usually the major limitation of its signal accumulation time. This means that decoherence should be considered for practically meaningful duration of parameter encoding. In this work, we specifically focus on the effect of individual qubit dephasing, which corresponds to the Pauli $Z$ error. The dephasing error is an important example of Markovian noise. Moreover, for our problem setup the $Z^{(i)}$ Lindbladian is exactly the same as the Hamiltonian which generates the signal, and the improvement from using advanced error suppression techniques such as quantum error correction is greatly limited~\cite{sekatski2017quantum,demkowicz2017adaptive,zhou2018achieving} because the encoding dynamics with dephasing does not satisfy Hamiltonian-not-in-Lindblad span (HNLS) condition~\cite{zhou2018achieving}. This fact further justifies the consideration of simpler approaches such as partitioning.

\subsection{Figures of merit}
In this work we use the Quantum Fisher information (QFI) as the performance evaluation metric. The QFI $\mathcal{F}$ quantifies the ultimate accuracy limit for parameter estimation through the quantum Cram\'er-Rao bound (QCRB)~\cite{helstrom1969quantum,paris2009quantum,petz2011introduction,toth2014quantum,liu2020quantum} $\mathrm{Var}(\theta)\geq 1/\mathcal{F}$ (ignoring the number of individual experiments). When using $n$-qubit GHZ-diagonal states as the initial probe state $\rho_0=\sum_a\lambda_{a}|\psi_{a}\rangle\langle\psi_{a}|$ where $|\psi_{a}\rangle$ is the GHZ basis state with index $a$, for the $z$-direction phase accumulation unitary $U(\theta) = \exp\left[-i\frac{\theta}{2}\sum_{i=1}^{n}Z^{(i)}\right]$, the QFI is~\cite{zang2024quantum}
\begin{align}\label{eqn:QFI}
    \mathcal{F} = n^2\sum_{(a,b)\in\mathcal{S}}\frac{(\lambda_{a}-\lambda_{b})^2}{\lambda_{a}+\lambda_{b}},
\end{align}
where $\mathcal{S}$ is the set of index pairs $(a,b)$ such that $|\psi_{a}\rangle$ and $|\psi_{b}\rangle$ are GHZ states which can be expressed as superposition of the same pair of computational basis states but with the opposite relative phase, e.g. $(|000\rangle\pm|111\rangle)/\sqrt{2}$, without double counting $(b,a)$.

\section{State preparation errors}\label{sec:state_prep}
We begin with the simplest case where only state preparation errors are considered. Through this minimal example, we demonstrate the advantage that can be obtained from partitioning when errors are present, and also showcase some common properties in phase estimation using GHZ states with imperfections.

\subsection{Quantum Fisher information}
According to our error model, using Eqn.~\ref{eqn:QFI} we have the closed-form expression of its QFI
\begin{align}
    \mathcal{F}(F,k,n) = \frac{[F(2k)^n-k]^2n^2}{k(2^n-1)[k + F(2^n-2)k^n]}.
    \label{eqn:qfi_state_prep_monolithic}
\end{align}
We may refer to the case where all $n$ sensors are used to create an $n$-qubit GHZ state as the monolithic strategy. In partitioning, suppose instead of preparing the $n$-qubit GHZ state, we prepare $m$ GHZ states each with equal number\footnote{The number of qubits must be an integer, thus we actually refer to $\lfloor n/m\rfloor$ or $\lceil n/m\rceil$. These two differ by at most one, so they will not make a big difference in practical regimes with large $n$.} of qubits $n/m$. The QFI for the partitioned case can then be straightforwardly obtained from Eqn.~\ref{eqn:qfi_state_prep_monolithic} through
\begin{align}
    \mathcal{F}(F,k,n,m) = m\mathcal{F}(F,k,n/m),
\end{align}
according to the additivity of QFI under tensor product~\cite{meyer2021fisher}. Note that even though we are using identical symbols, the QFI for the monolithic strategy and the QFI for the partitioned strategy have different numbers of arguments, and to avoid confusion we will write out the arguments explicitly when referring to a specific function. We have also assumed that sub-ensembles can be independently prepared with uniform errors so that $k$ is unchanged and each sub-ensemble is identical.

\begin{figure}[t]
    \centering
    \includegraphics[width=\linewidth]{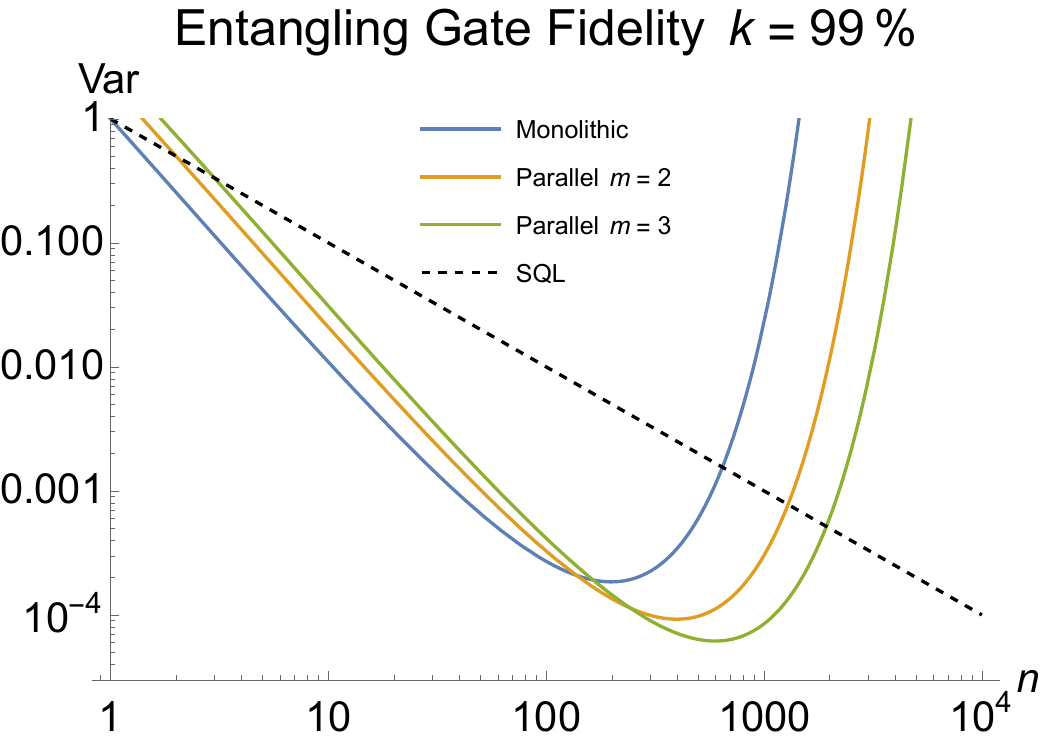}
    \caption{Visualization of the parameter estimation variance bound (QCRB) as the inverse of QFI for monolithic and partitioned strategies under state preparation errors, plotted as a function of the total number of sensors $n$ on a log-log scale. The blue curve represents the monolithic strategy, the yellow curve represents the partitioned strategy with $m=2$, and the green curve represents the partitioned strategy with $m=3$. The dashed line shows the standard quantum limit (SQL) $1/n$ that can be achieved by a fully separable probe state as a baseline. We assumed $k=99\%$ entangling gate fidelity for preparing the GHZ states. This imperfection causes the trade-off between the amount of resources, i.e. the number of quantum sensors, and the quality of the entanglement, which leads to the observed advantage of partitioning.}
    \label{fig:state_prep_err}
\end{figure}

We demonstrate the parameter estimation performance of monolithic and partitioned strategies in Fig.~\ref{fig:state_prep_err} with an example for $F=1$ and $k=0.99$. We observe that for small total number of quantum sensors $n$, the monolithic strategy wins over partitioned strategy by achieving a lower estimation variance. In contrast, partitioning allows the system to achieve lower estimation variance than the monolithic strategy, and thus making the utilization of larger amount of sensors operationally meaningful. This clearly demonstrates the advantage of using partitioning for quantum sensing under realistic imperfections.

\subsection{Optimal total number of sensors under fixed partitioning}
It is a universal feature that the QCRB will first decrease and then increase as the total number of sensors in the system increases. This demonstrates the tradeoff between the amount of information gain by using a larger probe state and the amount of errors introduced during the preparation of the larger probe state. Therefore there is an optimal total number of sensors $n^*(k,m,F)$ for any choice of number of sub-ensembles $m$, local entanglement generation quality $k$, and initialization fidelity $F$. It is meaningless to scale the sensing system above $n^*(k,m,F)$.

To obtain the optimal total number of sensors $n^*(k,m,F)$, we can take the partial derivative of the QFI with respect to $n$ and set it to zero, $\partial\mathcal{F}(F,k,n,m)/\partial n = 0$. Although the explicit expression is complicated, we can perform analytical approximations under the assumptions that $F$ and $k$ are both close to 1 for the sensing scenario to be practically meaningful. The approximation leads to a simple expression for the optimal total number of sensors
\begin{align}\label{eqn:opt_tot_num}
    n^*\approx -\frac{2m}{\ln k} \approx \frac{2m}{1-k}.
\end{align}

\subsection{Optimal partitioning under fixed total number of sensors}
Similarly, when we are given a specific amount of quantum sensors $n$, it is important to ask what is the \textit{optimal partitioning}, i.e. the optimal number of sub-ensembles $m^*$ to maximize the achievable QFI. This optimal partitioning number $m^*$ can be obtained again through explicit evaluation of the partial derivative $\partial\mathcal{F}(F,k,n,m)/\partial m = 0$. Through analytical approximations we obtain a simple closed-form expression for the optimal partitioning
\begin{align}\label{eqn:opt_part_num_st_prep}
    m^*\approx - n\ln k \approx n(1-k).
\end{align}
It is an interesting observation that under the optimal partitioning, each sub-ensemble is expected to contain $n/m^*\approx -1/\ln k$ sensors, which is almost exactly half the optimal total number of sensors that one GHZ state should contain under the initial state preparation errors. This is the result of diminishing returns in the QFI gain when the number of sensors in one GHZ state approaches $n^*$ from below. 

\begin{figure}[t]
    \centering
    \includegraphics[width=\linewidth]{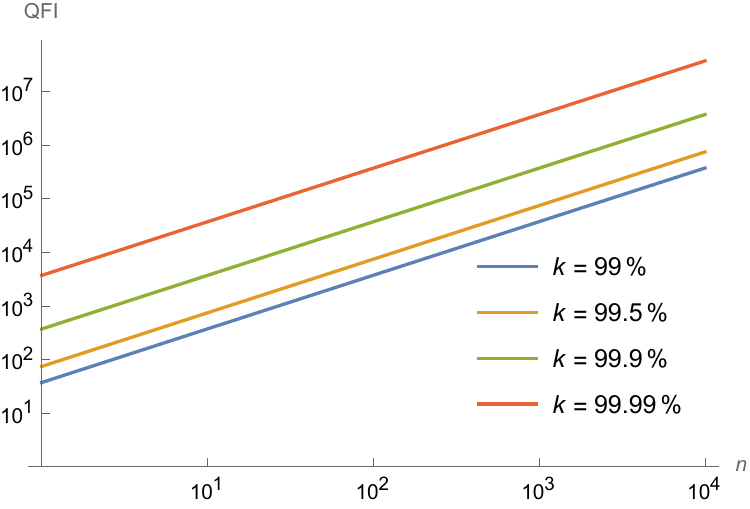}
    \caption{Visualization of the highest achievable QFI under optimal partitioning $\mathcal{F}(F,k,n,-n\ln k)$ as a function of the total number of sensors $n$ for different entangling gate fidelities $k$. The initialization fidelity is fixed to $F=1$. We consider the following gate fidelity values: $k=99\%, 99.5\%,99.9\%,99.99\%$.}
    \label{fig:opt_qfi_vs_n_diff_k}
\end{figure}

An intuitive way to understand this result is that the depth of the state preparation circuit ($n/m$) cannot go beyond the level which the gate fidelity allows ($- 1/\ln k \approx 1/(1-k)$). We further demonstrate the impact of gate fidelity $k$ with the highest achievable QFI of an optimally partitioned sensing system, estimated by $\mathcal{F}(F,k,n,m^*)\approx\mathcal{F}(F,k,n,-n\ln k)$. The results are visualized in Fig.~\ref{fig:opt_qfi_vs_n_diff_k}. For this specific figure we have fixed $F=1$ without loss of generality, as it can be verified that near-unit $F$ does not affect the results by much. For the choices of entangling gate fidelities, we consider a conservative value $k=99\%$, a state-of-the-art value $k=99.5\%$~\cite{wang2020high,google2023suppressing,bluvstein2024logical,bartling2024universal}, a near-future value $k=99.9\%$, and an optimistic value $k=99.99\%$.

It is noteworthy that the same amount of increase in gate fidelity does not result in the same portion of increase in the highest achievable QFI for a fixed number of sensors under the optimal partitioning. For instance, increasing from $k=99.5\%$ to $k=99.9\%$ results in higher improvement than increasing from $k=99\%$ to $k=99.5\%$. This clearly demonstrates the importance and benefit of achieving higher gate fidelity. More explicitly, this can be demonstrated through
\begin{align}
    \frac{\partial}{\partial k}\mathcal{F}(F,k,n,-n\ln k) \approx \frac{1}{e}\frac{1+\ln k}{k^2\ln^2k}Fn,
\end{align}
where we have considered $k\lesssim 1$ to obtain the simple approximation on the right hand side. The $k$-dependent part increases non-linearly as $k$ increases.

\subsection{Advantage of partitioning}
The advantage of partitioning can be demonstrated as follows. First, imagine that one is able to pack the optimal number of sensors in the partitioned sensing system. Then the optimal sensing performance for each partitioning number $m$ of the system is characterized by the highest achievable QFI. Then according to Eqn.~\ref{eqn:opt_tot_num} and the additivity of QFI under tensor product, we have
\begin{align}
    \mathcal{F}(F,k,n^*,m) \approx& \mathcal{F}(F,k,-\frac{2m}{\ln k},m)\nonumber\\
    =& m\mathcal{F}(F,k,-\frac{2}{\ln k}).
\end{align}
Therefore, partitioning in $m$ sub-ensembles results in an $m$-fold enhancement in the sensing performance in terms of the highest achievable QFI. 

Second, suppose that the total number of sensors $n$ is fixed. Then we can compare the achievable QFI with the optimal partitioning number $m^*$ and the monolithic strategy by calculating the ratio between the former and the latter. Using similar approximations as before, we have
\begin{align}\label{eqn:qfi_ratio_opt_par_vs_mono}
    \frac{\mathcal{F}(F,k,n,m^*)}{\mathcal{F}(F,k,n)} \approx -\frac{1}{e}\frac{k^{-n}}{n\ln k}.
\end{align}
Therefore, partitioning offers an exponential advantage over na\"ively creating a monolithic GHZ state. 

Alternatively, one may decide to only use $\tilde{n}=\min\{n,n^*\}$ sensors when partitioning cannot be used. Then the comparison will be between $\mathcal{F}(F,k,\tilde{n})$ and $\mathcal{F}(F,k,n,m^*)$. In the regime with a larger number of sensors, we have $\tilde{n}=n^*$ and the comparison between monolithic and partitioned strategies can be approximated as 
\begin{align}\label{eqn:qfi_ratio_opt_par_vs_mono_opt_num}
    \frac{\mathcal{F}(F,k,n,m^*)}{\mathcal{F}(F,k,n^*)} \approx -\frac{e\ln k}{4} n.
\end{align}
This demonstrates that partitioning with an efficient allocation of quantum sensor resource offers an advantage over the monolithic strategy. This is  because increasing the number of quantum sensors can continue enhancing the sensing performance instead of accumulating too much noise. 

\subsection{Additional comments on real-world implementations}
\begin{figure}[t]
    \centering
    \includegraphics[width=\linewidth]{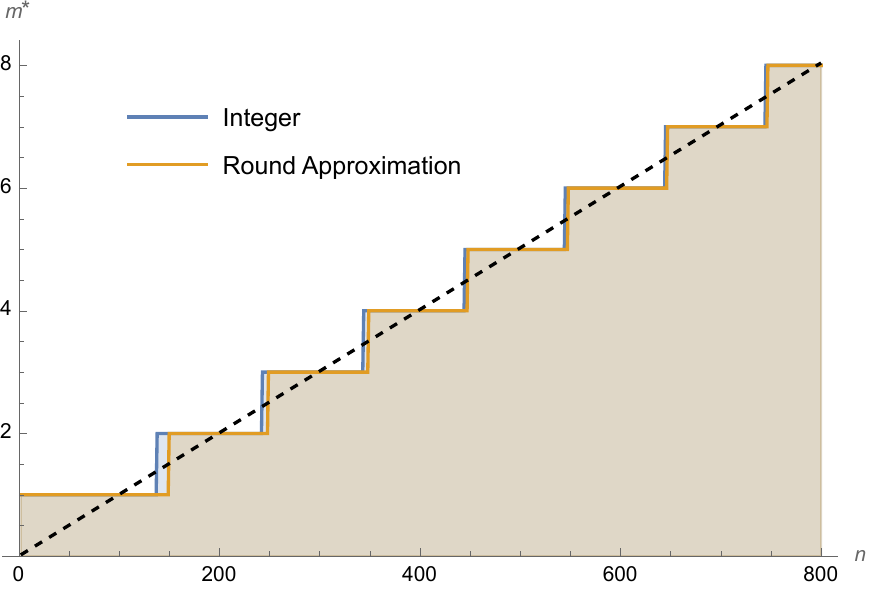}
    \caption{Visualization of the optimal partitioning number $m^*$ for different total number of sensors $n$ with $k=0.99$ and $F=1$. The Blue curve represents the realistic integer $m^*$, while the yellow curve corresponds to rounding the analytical approximation to its nearest positive integer, and the black dashed line depicts the original analytical approximation.
    }
    \label{fig:int_opt_partition}
\end{figure}
It is important to emphasize that the total number of sensors $n$, the partitioning number (number of sub-ensembles) $m$, and the number of sensors in each sub-ensemble should all be integers. The previous formulas do not necessarily yield integers for common system parameters. For instance, given $k=0.999$ and $n=1500$, we may calculate $m^*\approx 1.5$. In such a case we need to check explicitly $\mathcal{F}(1,0.999,1500,1)\approx 5.02\times 10^5$ while $\mathcal{F}(1,0.999,1500,2)\approx 5.32\times 10^5$ to determine that the partitioning with $m=2$ is more desirable. We visualize the optimal partitioning integer number in Fig.~\ref{fig:int_opt_partition}, where the integer result is the solution of $\arg\max_{m\geq1,m\in\mathbb{Z}}\mathcal{F}(F,k,n,m)$ for $k=0.99$ and $F=1$. The analytical approximation also demonstrates a very good fit. Additionally, we also considered the simpler way to obtain a reasonable integer partition number by rounding the analytical approximation to its nearest positive integer. This simpler approach demonstrates a good fit to the results from integer programming, with the only subtlety around the regimes where the analytical approximation is close to half integer, and it is observable that the fit becomes better for larger $n$ system sizes.

So far we have assumed equal partitioning. In fact, such equal partitioning can be justified in practice where each sub-ensemble contains a similar number of sensors. We evaluate the concavity of the quantum Fisher information in Eqn.~\ref{eqn:qfi_state_prep_monolithic}. We can show that $\mathcal{F}(F,k,n)$ is concave for $n\in[-(2-\sqrt{2})/\ln k, -(2+\sqrt{2})/\ln k]=[n_\mathrm{low},n_\mathrm{up}]$ to a very good approximation. Recall that the optimal total number of sensors with the monolithic strategy is $n^*\approx -2/\ln k\in[n_\mathrm{low},n_\mathrm{up}]$, and with the optimal (equal) partitioning the number of sensors per sub-ensemble is approximately $n/m^*\approx-1/\ln k\in[n_\mathrm{low},n_\mathrm{up}]$. For $m$ sub-ensembles each with $n_i\in[n_\mathrm{low},n_\mathrm{up}]$ sensors with $i=1,2,\dots,m$ and a total number of sensor $n = \sum_{i=1}^m n_i$we have the following for the total QFI
\begin{align}
    \mathcal{F}_\mathrm{partition} =& \sum_{i=1}^m\mathcal{F}(n_i)\nonumber\\
    =& \left(\sum_{i=1}^m 1\right)\frac{\sum_{i=1}^m 1\times \mathcal{F}(n_i)}{\sum_{i=1}^m 1}\nonumber\\
    \leq& \left(\sum_{i=1}^m 1\right)\mathcal{F}\left(\frac{\sum_{i=1}^m 1\times n_i}{\sum_{i=1}^m 1}\right)\nonumber\\
    =& \left(\sum_{i=1}^m 1\right)\mathcal{F}\left(\frac{n}{m}\right) = m\mathcal{F}\left(\frac{n}{m}\right),
\end{align}
where we have used Jensen's inequality for a function $f(x)$ that is concave on a domain $\mathcal{I}$: $f\left(\frac{\sum_ia_ix_i}{\sum_ia_i}\right) \geq \frac{\sum_ia_if(x_i)}{\sum_ia_i}$ for $x_i\in\mathcal{I}$ and for positive weights $a_i$. The equality holds if and only if $f(x)$ is linear or $x_i$ are all identical, so the upper bound of the partitioning QFI on the right hand side is achieved when we create identical sub-ensembles, i.e. with equal partitioning.

On the other hand, the optimal integer partitioning number found by $\arg\max_{m\geq1,m\in\mathbb{Z}}\mathcal{F}(F,k,n,m)$ is not necessarily a factor of $n$. However, even though equal partitioning is desirable, it does not mean that we should insist on choosing $n$'s factor as the partitioning number. In practice, we may allow the number of sensors contained in each of the $m^*$ sub-ensembles to be slightly different, but still close to the mean value $n/m^*$. As a concrete example, we consider a total number of sensors $n=350$ and entanglement generation quality $k=0.99$. One can solve the integer program to infer that the positive integer $m^*=4$ optimizes the QFI as $\mathcal{F}(1,0.99,350,4)\approx 12838.7$. However, as $350=2\times 5^2\times 7$, 4 is not its factor. Nevertheless, we can consider 4 sub-ensembles each having $87,87,88,88$ sensors, respectively, and a straightforward calculation gives $\mathcal{F}_{87\times2,88\times 2}\approx 12838.5$, which is very close to the optimal value achieved by a non-integer number of sensors in each sub-ensemble, with only $\approx 0.0012\%$ decrease. In contrast, if we insist on using exactly equal number of sensors in every sub-ensemble, we may significantly lose performance, for instance $\mathcal{F}_{175\times 2}\approx 10656.9$, $\mathcal{F}_{70\times 5}\approx 12246.0$, $\mathcal{F}_{50\times 7}\approx 10694.6$, with at least $\approx 4.6\%$ decrease.

\begin{figure}[t]
    \centering
    \includegraphics[width=\linewidth]{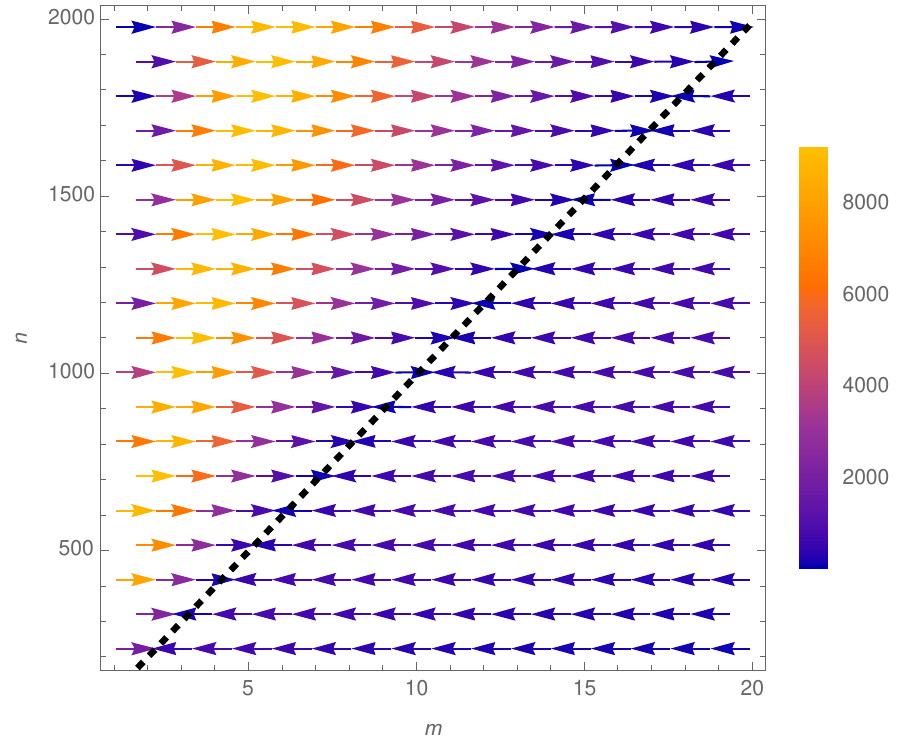}
    \caption{Visualization of $(\partial\mathcal{F}/\partial m,\partial\mathcal{F}/\partial n)$, the gradient of $\mathcal{F}(F,k,n,m)$, as a vector field (the ``optimization flow'') on the $(m,n)$ space for $F=1$ and $k=0.99$. Different colors denote different magnitudes of the vector field at different points. The black dashed line represents $m = -n\ln 0.99$.
    }
    \label{fig:total_sys_opt}
\end{figure}

In the above we have focused on either a fixed total number of sensors $n$, or a fixed partitioning number $m$. To improve the total system's sensing performance in practice, one may want to jointly tune $n$ and $m$. Here we show the gradient of $\mathcal{F}(F,k,n,m)$, i.e. $(\partial\mathcal{F}/\partial m,\partial\mathcal{F}/\partial n)$, as a vector field on the $(m,n)$ space in Fig.~\ref{fig:total_sys_opt} for $F=1$ and $k=0.99$. It is then obvious that to obtain an optimized performance of the total system, it is most desirable to increase $n$ and $m$ simultaneously according to $m = -n\ln k$ so that the increase in one is compatible with the other. This is exactly the optimal partitioning number we derived before. Moreover, the visualization of the gradient field on the $(m,n)$ space further validates the previous analytical approximations, showing that they represent the unique optimality conditions since there are no other maximum points in the parameter space.

We note that the integer partitioning number, equal partitioning, fine tuning of the number of sensors in each sub-ensemble, and the total system optimization all naturally apply to the following discussions which involve additional error sources. The analysis from this subsection will not be repeated in the following case studies beacuse the qualitative behavior will be similar.

\section{Sensor qubit losses}\label{sec:loss}
So far we have considered state preparation errors only. In this section, we take into account an additional realistic error source, particle losses, which are common in atomic systems.

\subsection{Effect of particle losses}
Given particle losses, we consider two scenarios: 
\begin{enumerate}
    \item When we do not have the capability of detecting particle losses we measure all resulting states regardless of whether any loss has ocurred.
    \item We can detect particle losses in non-demolition ways, so we only perform the final measurement on those GHZ states with no particle loss. 
\end{enumerate}

In the first case, the effective encoded state is a statistical mixture of the resulting states of all possible particle loss patterns, weighted by the probability of each pattern. Here by particle loss pattern we mean what sensor qubits have been lost during the encoding process, since particle loss is a binary process. 

We can show that the resulting state of any particle loss pattern is a GHZ-diagonal state where the eigenvalue corresponding to the GHZ basis state $|\psi_a\rangle$ is equal to that of $|\psi_a\rangle$ when $|\psi_{a}\rangle$ and $|\psi_{b}\rangle$ can be both expressed as superpositions of the same pair of computational basis states but with the opposite relative phase, i.e. $\lambda_a=\lambda_b, \forall (a,b)\in\mathcal{S}$ (see the notation in Sec.~\ref{sec:model}). Such a state is also diagonal in the computational basis, and again with identical eigenvalue for $|i\rangle$ and $|\overline{i}\rangle$. Any such resulting state is independent of the parameter to estimate, while it can still be interpreted as having undergone the unitary encoding channel which leaves the state invariant. Consequently, the effect of particle losses can be combined into state preparation errors. Moreover, as we focus on the QFI, we only need to consider the eigenvalues for $(|00\dots 0\rangle\pm|11\dots 1\rangle)/\sqrt{2}$ modified by particle losses, according to our assumption that state preparation errors result in depolarized GHZ states (recall that all other $\lambda_a=\lambda_b$ for $(a,b)\in\mathcal{S}$, except for $(|00\dots 0\rangle\pm|11\dots 1\rangle)/\sqrt{2}$). We emphasize that here we have assumed that the lost qubits are replaced by maximally mixed state.

Now consider a single ensemble of $n$ qubits. The effective QFI which takes into account particle loss effects has the follow form:
\begin{align}
    &\mathcal{F}_{\mathrm{loss},1}(F,k,p,n)\nonumber\\
    =& \frac{p^{2n}\left(k^{n-1}F - \frac{1-k^{n-1}F}{2^n-1}\right)^2n^2}{p^n\left(k^{n-1}F - \frac{1-k^{n-1}F}{2^n-1}\right) + \frac{(1+p)^n - (2p)^n}{2^n}},
    \label{eqn:qfi_loss_monolithic}
\end{align}
where the subscript 1 emphasizes that the above expression of partitioned QFI corresponds to the first case where we measure after every shot to accumulate data. When we consider partitioning, the QFI of the system with $m$ sub-ensembles can be obtained from Eqn.~\ref{eqn:qfi_loss_monolithic}
\begin{align}
    \mathcal{F}_{\mathrm{loss},1}(F,k,p,n,m) = m\mathcal{F}_{\mathrm{loss},1}(F,k,p,n/m).
\end{align}

In the second case we perform post-selection when we detect GHZ state particle losses (when using GHZ states as a probe, any particle loss will completely destroy the usefulness for sensing). Conditioned on no particle loss, the QFI corresponds to the derivations in the previous section with only state preparation errors. It is important to emphasize that in practice the sensing cycle needs to run for a sufficient number of cycles to accumulate enough measurements for classical post-processing to estimate the parameter values. The QCRB is also dependent on the number of available samples in the following sense: $\mathrm{Var}\geq 1/(\mu\mathcal{F})$, where $\mu$ is the number of samples, or equivalently the number of successful sensing cycles. Meanwhile, post-selection based on loss detection results in a trade-off between $\mu$ and $\mathcal{F}$: while the QFI conditioned on no loss is higher than without loss detection, fewer sensing cycles lead to accepted measurement results which reduces the amount of available data after using the same amount of sensors and time than without loss detection.

For this second case, we consider that a single ensemble of $n$ qubits will be prepared and used for parameter encoding and readout $\mu$ times. The assumption of loss detection means that only $p^n$ portion of the $\mu$ sensing cycles will accumulate data. The QCRB thus gives a lower bound for the parameter estimation variance
\begin{align}
    \mathrm{Var} \geq \frac{1}{(p^n\mu)\mathcal{F}(F,k,n)} = \frac{1}{\mu[p^n\mathcal{F}(F,k,n)]},
\end{align}
where $\mathcal{F}(F,k,n)$ is the QFI for the case with only state preparation errors in Eqn.~\ref{eqn:qfi_state_prep_monolithic}. We can then define the effective QFI for the second case as
\begin{align}
    \mathcal{F}_{\mathrm{loss},2}(F,k,p,n) = p^n\mathcal{F}(F,k,n),
\end{align}
so that $\mathrm{Var}\geq 1/(\mu\mathcal{F}_{\mathrm{loss},2})$ for a fair comparison with the first case, where the subscript 2 denotes the second case assuming a loss detection capability. If we equally partition the $n$ qubits into $m$ sub-ensembles for every single run of the sensing cycle, and the cycles are again repeated $\mu$ times, the scenario can be interpreted that $m\mu$ sensing cycles are run and each single cycle only involves $n/m$ qubits. There is no correlation between the partitioned sub-ensembles. If we assume the capability of detecting losses in each sub-ensemble, then the probability for each sub-ensemble to survive is $p^{n/m}$, and the QCRB gives
\begin{align}
    \mathrm{Var} \geq& \frac{1}{(m\mu p^{n/m})\mathcal{F}(F,k,n/m)}\nonumber\\
    =& \frac{1}{\mu[mp^{n/m}\mathcal{F}(F,k,n/m)]},
\end{align}
where $\mathcal{F}(F,k,n,m)$ is again for state preparation errors only. Similar to the monolithic strategy, we can thus define the effective QFI for the $m$-partition case with loss detection as
\begin{align}
    \mathcal{F}_{\mathrm{loss},2}(F,k,p,n,m) =& mp^{n/m}\mathcal{F}(F,k,n/m)\nonumber\\
    =& p^{n/m}\mathcal{F}(F,k,n,m).
\end{align}

\subsection{Optimal system configuration}

\begin{figure}[t]
    \centering
    \includegraphics[width=\linewidth]{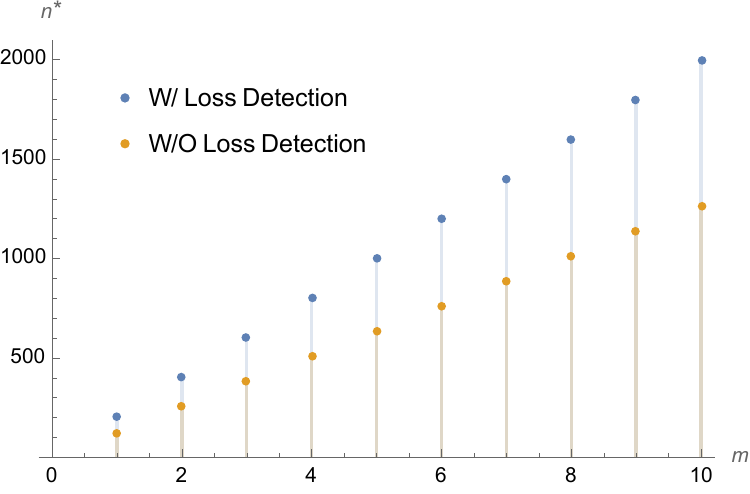}
    \caption{Visualization of $n_{\mathrm{loss},1}^*$ and $n_{\mathrm{loss},2}^*$ as functions of the partition number $m$. Other system parameters are fixed to $F=1,k=99.5\%,p=0.995$. Blue dots represent $n_{\mathrm{loss},2}^*$ while yellow dots denote $n_{\mathrm{loss},1}^*$.}
    \label{fig:opt_num_1vs2}
\end{figure}

Similar to the previous section in which we focused on state preparation errors only, we still encounter the same qualitative features when we include particle losses during the encoding process. Here we present the optimal configurations of the sensing system under partitioning. We again consider the optimal number of sensors $n^*$ that the total system of $m$ partitions contains to achieve the highest possible sensing performance, and the optimal number of partitions $m^*$ one should apply to an ensemble of $n$ sensors.
The optimal total number of sensors $n_{\mathrm{loss},1(2)}^*$ is the value of $n$ for the equation $\partial\mathcal{F}_{\mathrm{loss},1(2)}(F,k,p,n,m)/\partial n = 0$ to hold, and the optimal partition number $m_{\mathrm{loss},1(2)}^*$ is the value of $m$ for the equation $\partial\mathcal{F}_{\mathrm{loss},1(2)}(F,k,p,n,m)/\partial m = 0$ to hold.

First we consider the optimal total number of sensors $n^*$. As in the case with loss detection capability, the effective QFI is essentially the QFI for state preparation error scaled by a multiplicative factor $p^{n/m}$. Using the approximation $F,k,p\lesssim 1$ and the fact that $n,n/m$ are not small, we derive the following closed-form approximation of the optimal total number of sensors
\begin{align}\label{eqn:opt_n_loss_2}
    n_{\mathrm{loss},2}^*\approx -\frac{2m}{\ln k + \ln p} \approx \frac{2m}{2 - k - p}.
\end{align}
One can see that the above expression resembles Eqn.~\ref{eqn:opt_tot_num}, and the only difference is in the denominator. When considering particle losses during the encoding process, the denominator is no longer $\ln k$ but $\ln k + \ln p = \ln(kp)$. This shows that $k$ and $p$ contribute to $n_{\mathrm{loss},2}^*$ equally even though they do not affect each other since they characterize errors in different stages of the sensing process. 

The case where we do not have the ability to detect losses results in more complicated analytical expressions. Repeating similar approximations we find that the root for the following equation of $n$ is a good approximation of $n_{\mathrm{loss},1}^*$
\begin{align}\label{eqn:opt_n_loss1}
    0 =& F(2kp)^\frac{n}{m}[2m + n\ln(kp)] - k(2p)^\frac{n}{m}[2m + n\ln(k^2p)]\nonumber\\
    &+ k(1+p)^\frac{n}{m}\left[2m + n\ln\frac{2k^2p^2}{1+p}\right],
\end{align}
from which we can see that $-\frac{2m}{\ln k + \ln p}$ is not the root. Attempts to further simplify the equation to obtain a more concise closed-form expression will significantly decrease the approximation accuracy. We demonstrate the comparison between $n_{\mathrm{loss},1}^*$ and $n_{\mathrm{loss},2}^*$ in Fig.~\ref{fig:opt_num_1vs2}. Without loss of generality, we have chosen $F=1,k=99.5\%,p=0.995$ for this figure. In the figure we see that $n_{\mathrm{loss},1}^*$ is almost linear with $m$ as intuitively expected. Moreover, we observe that in general $n_{\mathrm{loss},1}^* < n_{\mathrm{loss},2}^*$, which demonstrates that loss detection allows the sensing system to hold more quantum sensors before the sensing performance decreases. 

\begin{figure}[t]
    \centering
    \includegraphics[width=\linewidth]{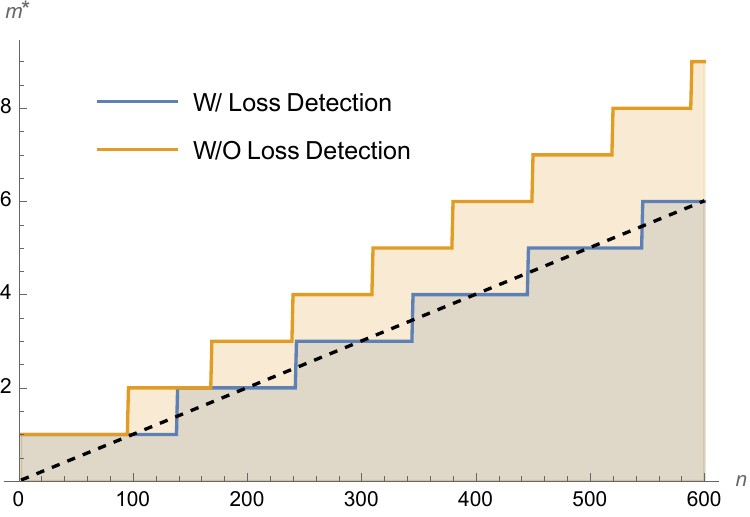}
    \caption{Visualization of $m_{\mathrm{loss},1}^*$ and $m_{\mathrm{loss},2}^*$ as functions of the total number of sensors $n$. Other system parameters are fixed to $F=1,k=99.5\%,p=0.995$. The blue curve represents integer $m_{\mathrm{loss},2}^*$ while the yellow curve denotes integer $m_{\mathrm{loss},1}^*$. The black dashed line shows the closed-form approximation of $m_{\mathrm{loss},2}^*$.}
    \label{fig:opt_part_1vs2}
\end{figure}

Next we consider the optimal partition number $m^*$ given $n$ total sensors. We start with the case where we can detect particle losses, and through explicit derivations and approximations obtain the optimal partition number
\begin{align}
    m_{\mathrm{loss},2}^*\approx - n(\ln k + \ln p) \approx n(2 - k - p).
\end{align}
Again this expression is very similar to the results for state preparation error only, and similar to $n_{\mathrm{loss},2}^*$, $k$ and $p$ contribute equally. For the case without the ability to detect particle losses, similar to $n_{\mathrm{loss},1}^*$, we find that the root for the following equation of $n$ is a good approximation of $m_{\mathrm{loss},1}^*$
\begin{align}
    0 =& F(2kp)^\frac{n}{m}[m + n\ln(kp)] - k(2p)^\frac{n}{m}[m + n\ln(k^2p)]\nonumber\\
    &+ k(1+p)^\frac{n}{m}\left[m + n\ln\frac{2k^2p^2}{1+p}\right],
\end{align}
from which we can see that $- n(\ln k + \ln p)$ is not the root. Notably, the above equation resembles Eqn.~\ref{eqn:opt_n_loss1} in that we only changes the $2m$ into $m$ in the brackets. However, due to the unchanged exponential factors before the brackets, for this case we do not have $n/m_{\mathrm{loss},1}^* \approx n_{\mathrm{loss},1}^*/2$. We compare $m_{\mathrm{loss},1}^*$ and $m_{\mathrm{loss},2}^*$ in Fig.~\ref{fig:opt_part_1vs2}. Note that we obtain the optimal integer partition numbers from integer programming $\arg\max_{m\geq1,m\in\mathbb{Z}}\mathcal{F}_{\mathrm{loss},1(2)}(F,k,p,n,m)$, and we have used the same parameters as in Fig.~\ref{fig:opt_num_1vs2}. The figure not only demonstrates that the analytical approximation for the second case is a good fit, but also that in general we have $m_{\mathrm{loss},1}^* > n_{\mathrm{loss},1}^*$, which is compatible with the observation that $n_{\mathrm{loss},1}^* < n_{\mathrm{loss},2}^*$.

\subsection{Advantage of loss detection}
Here we focus on the sensing performance difference of the two scenarios -- with and wiothout the ability to detect particle losses. Since both $\mathcal{F}_1$ and $\mathcal{F}_2$ have the multiplicative factor $m$ for partitioning, we only need to compare $\mathcal{F}_{\mathrm{loss},1}(F,k,p,n/m)$ and $p^{n/m}\mathcal{F}(F,k,n/m)$.

In fact, according to the expressions of the effective QFI's for both cases, we are able to show that
\begin{align}
    \mathcal{F}_{\mathrm{loss},1}(F,k,p,n,m) \leq \mathcal{F}_{\mathrm{loss},2}(F,k,p,n,m),
\end{align}
where the equality only holds when $p=1$, i.e. there is no particle loss. This demonstrates that the assumed capability of loss detection yileds better sensing performance in comparison to the case without loss detection.

How much advantage can we obtain from loss detection? To quantify this advantage, we consider the ratio between the two QFI's, and through explicit derivation obtain the following approximation for this ratio
\begin{align}\label{eqn:loss_qfi_ratio_1st}
    &\frac{\mathcal{F}_{\mathrm{loss},2}(F,k,p,n,m)}{\mathcal{F}_{\mathrm{loss},1}(F,k,p,n,m)}\nonumber\\
    \approx& 1 + \frac{k}{F}\left(\frac{1+p}{4kp}\right)^\frac{n}{m}\left[2^\frac{n}{m} - \left(\frac{4p}{1+p}\right)^\frac{n}{m}\right],
\end{align}
and for a relatively higher loss probability $1-p$ we can further simplify the above approximation by eliminating the second term in the bracket
\begin{align}
    \frac{\mathcal{F}_{\mathrm{loss},2}(F,k,p,n,m)}{\mathcal{F}_{\mathrm{loss},1}(F,k,p,n,m)} \approx 1 + \frac{k}{F}\left(\frac{1+p}{2kp}\right)^\frac{n}{m}.
\end{align}

\begin{figure}[t]
    \centering
    \includegraphics[width=\linewidth]{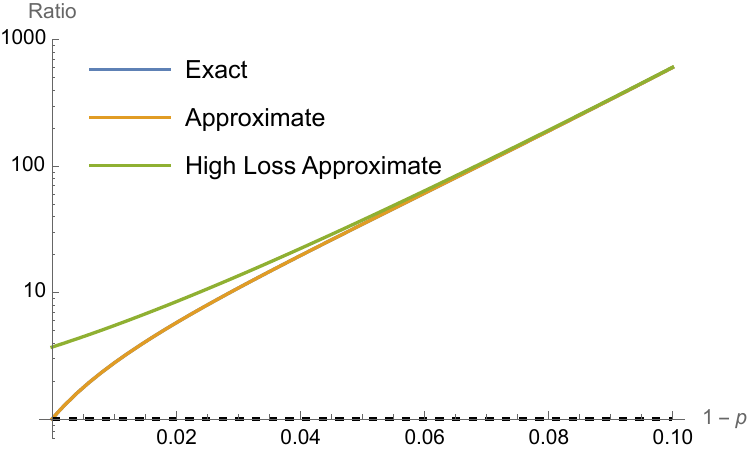}
    \caption{Ratio between the QFI for the case with loss detection and the QFI for the case without loss detection, i.e. $\mathcal{F}_2(F,k,p,n,m)/\mathcal{F}_1(F,k,p,n,m)$, as a function of loss probability $1-p$ in a log plot. While we vary $p$, the other system parameters are fixed to $F=1, k=0.99, n=1000, m=10$. The blue curve shows the exact ratio. The yellow curve corresponds to the first approximation of the ratio in Eqn.~\ref{eqn:loss_qfi_ratio_1st}. The green curve depicts the approximation assuming a relatively large loss probability. The horizontal black dashed line denotes the unit-value floor for the ratio.}
    \label{fig:loss_detect_adv_vs_p}
\end{figure}

We visualize the advantage of loss detection in Fig.~\ref{fig:loss_detect_adv_vs_p} by plotting the exact ratio between QFI's $\mathcal{F}_{\mathrm{loss},2}(F,k,p,n,m)/\mathcal{F}_{\mathrm{loss},1}(F,k,p,n,m)$ and its approximations, as a function of the loss probability during encoding $1-p$. Without loss of generality, for this figure we choose parameters $F=1, k=0.99, n=1000, m=10$. Notably the blue and yellow curves almost completely overlap so that only one curve can be effectively observed. This demonstrates that Eqn.~\ref{eqn:loss_qfi_ratio_1st} accurately approximates the exact value of the QFI ratio, while having a relatively simple form. On the other hand, the further approximation which is tailored towards relatively high loss significantly overestimates the QFI ratio in a low loss regime. Nevertheless, it does converge to the exact result as $1-p$ increases as expected. Meanwhile, the QFI ratio increases as $1-p$ increases, which validates the intuition that loss detection becomes more useful when loss is more significant. In another extreme of low loss, the QFI ratio converges to 1 which means that the cases with and without loss detection are almost identical in terms of sensing performance.

While the large value of the QFI ratio in the regime of a relatively large loss probability is attractive, this regime is not necessarily ideal for practical quantum sensing systems. Therefore, it is important to consider the low loss regime. Specifically, we can obtain the leading orders of the QFI ratio in the low loss regime in terms of $1-p$
\begin{align}
    &\frac{\mathcal{F}_{\mathrm{loss},2}(F,k,p,n,m)}{\mathcal{F}_{\mathrm{loss},1}(F,k,p,n,m)}\nonumber\\
    =& 1 + \frac{kn\left(2^\frac{n}{m}-1\right)}{k + F\left(2^\frac{n}{m}-2\right)k^\frac{n}{m}}\frac{1}{2m}(1-p)\nonumber\\
    &+ \frac{kn\left(2^\frac{n}{m}-1\right)}{k + F\left(2^\frac{n}{m}-2\right)k^\frac{n}{m}}\frac{3m+n}{8m^2}(1-p)^2 + O\left((1-p)^3\right).
\end{align}
From the above expansion one can see that smaller $k$ induces higher relative advantage from loss detection. 

We emphasize that in the above discussion of the advantage of loss detection we have assumed identical $n$ and $m$ for both cases, while we have also known from the previous subsection that optimal system configurations for both cases are different under identical system parameters. Recall that generally we have $m_{\mathrm{loss},1}^* > n_{\mathrm{loss},1}^*$ and $n_{\mathrm{loss},1}^* < n_{\mathrm{loss},2}^*$. Suppose we fix $n=n_{\mathrm{loss},1}^*$ for both cases under a fixed partition number $m$. Then the case with loss detection still demonstrates advantage, and this advantage can be amplified by including more sensors in the system. If instead $m=m_{\mathrm{loss},1}^*$ for both cases under a fixed total number of sensors $n$, the advantage of loss detection still holds, and we can reduce the number of partitions to further increase the advantage. This can also potentially relax the requirement on experimental control.

\subsection{Advantage of partitioning}
The advantage of partitioning over the monolithic strategy is qualitatively the same as when we only consider state preparation errors. Note that in practice we prefer to implement the sensing protocol with low loss systems, the QFI of which for both cases converge as shown in the previous subsection. Therefore, to demonstrate such advantage quantitatively, we focus on the case with loss detection capability without loss of generality. 

First of all, we consider the highest achievable QFI. By partitioning the sensing system into $m$ sub-ensembles, the highest achievable QFI has an $m$-fold enhancement over the single-ensemble setup, directly coming from the additivity of the QFI, which is the same as before.

Then we consider a fixed total number of sensors $n$, and compare the achievable QFI under optimal partitioning with the monolithic strategy. The ratio between the QFI is
\begin{align}
    \frac{\mathcal{F}_{\mathrm{loss},2}(F,k,p,n,m_{\mathrm{loss},2}^*)}{\mathcal{F}_{\mathrm{loss},2}(F,k,p,n)} \approx -\frac{1}{e}\frac{(kp)^{-n}}{n(\ln k + \ln p)},
\end{align}
which is similar to Eqn.~\ref{eqn:qfi_ratio_opt_par_vs_mono}, while demonstrating equivalent contribution from initial state preparation errors $k$ and sensor losses $p$.

Last but not least, we compare $\mathcal{F}_{\mathrm{loss},2}(F,k,p,\tilde{n}_{\mathrm{loss},2})$ and $\mathcal{F}_{\mathrm{loss},2}(F,k,p,n,m^*)$, where $\tilde{n}_{\mathrm{loss},2}=\min\{n,n_{\mathrm{loss},2}^*\}$. In the regime with a larger number of sensors we have $\tilde{n}_{\mathrm{loss},2}=n_{\mathrm{loss},2}^*$, then the ratio becomes
\begin{align}
    \frac{\mathcal{F}_{\mathrm{loss},2}(F,k,p,n,m_{\mathrm{loss},2}^*)}{\mathcal{F}_{\mathrm{loss},2}(F,k,p,n_{\mathrm{loss},2}^*)} \approx -\frac{e(\ln k + \ln p)}{4} n,
\end{align}
which is similar to Eqn.~\ref{eqn:qfi_ratio_opt_par_vs_mono_opt_num}. The difference is simply to replace $\ln k$ with $\ln k + \ln p$, again demonstrating the equivalent contribution from $k$ and $p$.

\section{Sensor qubit dephasing}\label{sec:dephasing}
In the previous section, we considered particle losses which are common in atomic sensing systems. In this section we focus on qubit dephasing motivated by solid-state sensing systems where the sensors are almost never lost, but instead their decoherence during sensing can be greater due to more complicated and noisy surrounding solid-state environment~\cite{barry2020sensitivity}. 

\subsection{Effect of dephasing}
We consider the independent encoding dynamics for each qubit sensor under dephasing described by the single-qubit Lindblad master equation
\begin{align}\label{eqn:encode_dephase}
    \frac{d}{dt}\rho = -i\frac{\omega}{2}[Z,\rho] + \frac{\gamma}{2}(Z\rho Z - \rho),
\end{align}
where $\omega$ is the precession frequency of each qubit sensor, and $\gamma$ is the single-qubit dephasing rate. One can verify that the quantum channel described by the above master equation is $\mathcal{E}_\mathrm{enc}(t)\circ\mathcal{E}_\mathrm{dp}(t) = \mathcal{E}_\mathrm{dp}(t)\circ\mathcal{E}_\mathrm{enc}(t)$, where $\mathcal{E}_\mathrm{enc}(t)[\rho] = e^{-iZ\omega t/2}\rho e^{iZ\omega t/2}$ is the unitary encoding channel on a single qubit, with the parameter to estimate being $\theta = \omega t$, and 
\begin{align}\label{eqn:dephasing}
    \mathcal{E}_\mathrm{dp}(t)[\rho] = \frac{1 + e^{-\gamma t}}{2}\rho + \frac{1 - e^{-\gamma t}}{2}Z\rho Z
\end{align}
is the pure dephasing/phase flip channel. 

We use $q$ to denote the probability that no phase flip happens during the encoding process. Then the $n$-qubit error channel which effectively acts on the initially prepared noisy GHZ state is $\mathcal{E}(q)^{\otimes n}$, where $\mathcal{E}(q)[\rho] = q\rho + (1-q)Z\rho Z$ with $q\in(1/2,1]$. The total noise channel is thus an $n$-qubit Pauli channel whose non-trivial Pauli strings only costist of the identity operator $I$ and the Pauli $Z$ operator, i.e. an element of $\{I,Z\}^{\otimes n}$. In this $n$-qubit Pauli channel the probability for each Pauli string to be applied to the initial state is determined by the number of Pauli $Z$ operators in it $\#Z$. More explicitly we have $p_{\#Z} =q^{n-\#Z}(1-q)^{\#Z}$.

For our ideal objective state, the standard $n$-qubit GHZ state $|\mathrm{GHZ}_n\rangle$, it is known that its stabilizer is generated by the following $n$ Pauli strings 
\begin{align}
    S =& \{X^{(1)}\otimes X^{(2)}\otimes \dots\otimes X^{(n)},Z^{(1)}\otimes Z^{(2)},\nonumber\\
    &~~~~Z^{(2)}\otimes Z^{(3)},\dots,Z^{(n-1)}\otimes Z^{(n)}\},
\end{align}
from which one can observe that any Pauli string that contains an even number of Pauli $Z$ operators will stabilize $|\mathrm{GHZ}_n\rangle$. On the other hand, the remaining Pauli strings in the $n$-qubit Pauli channel from dephasing still only have identity operators and the Pauli $Z$ operators, so they will only result in a phase flip in $|\mathrm{GHZ}_n\rangle$.

Consider the initially prepared depolarized GHZ state as in Eqn.~\ref{eqn:noisy_ghz}
\begin{align}
    \rho = V(n)|\mathrm{GHZ}_n\rangle\langle\mathrm{GHZ}_n| + [1-V(n)]I_n/2^n,
\end{align}
where $V(n)$ again denotes visibility, and according to our state preparation error model we have its GHZ fidelity $F(n) = V(n) + [1-V(n)]/2^n = Fk^{n-1}$. The above explicit decomposition into a pure GHZ projector and a maximally mixed state is convenient in that the maximally mixed state is invariant under the Pauli channel according to the unitality. For the GHZ projector, the probability for it to remain unchanged is thus the total probability of its stabilizing Pauli strings in the noise channel 
\begin{align}
    \tilde{q} =& \sum_{\mathrm{even}~\#Z}\binom{n}{\#Z}p_{\#Z} = \sum_{i=0}^{\lfloor n/2\rfloor}\binom{n}{2i}q^{n-2i}(1-q)^{2i}\nonumber\\
    =& \frac{1 + (2q-1)^n}{2},
\end{align}
for positive integer $n$. The resulting state is then
\begin{align}\label{eqn:eff_probe_dephasing}
    \rho' =& \tilde{q}V(n)\mathrm{Proj}_{\mathrm{GHZ}^+} + (1-\tilde{q})V(n)\mathrm{Proj}_{\mathrm{GHZ}^-}\nonumber\\
    &+ [1-V(n)]I_n/2^n,
\end{align}
where $\mathrm{Proj}_{\mathrm{GHZ}^+} = |\mathrm{GHZ}_n\rangle\langle\mathrm{GHZ}_n|$ and $\mathrm{Proj}_{\mathrm{GHZ}^-} = |\mathrm{GHZ}_n^-\rangle\langle\mathrm{GHZ}_n^-|$ with $|\mathrm{GHZ}_n^-\rangle=(|00\dots 0\rangle-|11\dots 1\rangle)/\sqrt{2}$ is the phase-flipped $|\mathrm{GHZ}_n\rangle$.

The effective QFI under dephasing during the encoding process is thus determined by the effective initial probe state $\rho'$. Recalling Eqn.~\ref{eqn:QFI}, the analytical formula for QFI, it is clear that we only need to focus on the eigenvalues of $|\mathrm{GHZ}_n\rangle$ and $|\mathrm{GHZ}_n^-\rangle$. Through explicit derivation we obtain the effective QFI for a single ensemble of $n$ qubits
\begin{align}
    \mathcal{F}_\mathrm{dp}(F,k,q,n) =& \frac{[F(2k)^n-k]^2(2q-1)^{2n}n^2}{k(2^n-1)[k + F(2^n-2)k^n]}\nonumber\\
    =& (2q-1)^{2n}\mathcal{F}(F,k,n),
\end{align}
which corresponds to Eqn.~\ref{eqn:qfi_loss_monolithic} scaled by $(2q-1)^{2n}$, qualitatively similar to the case with loss detection capability in the previous section. Then the effective QFI for the $m$-partition case under dephasing is
\begin{align}
    \mathcal{F}_\mathrm{dp}(F,k,q,n,m) =& m\mathcal{F}_\mathrm{dp}(F,k,q,n/m)\nonumber\\
    =& (2q-1)^{2n/m}\mathcal{F}(F,k,n,m).
\end{align}

\subsection{Optimal system configuration}
Having derived the effective QFI, we can use it to determine the optimal configuration of the sensing system as before. 

Under dephasing the optimal total number of sensors $n^*_\mathrm{dp}$ in a sensing system with $m$ partitions can be obtained by finding $n$ that makes $\partial\mathcal{F}_\mathrm{dp}(F,k,q,n,m)/\partial n = 0$. Through straightforward analytical derivations, under realistic conditions $F,k,p\lesssim 1$, and assuming $n,n/m$ are not small, we arrive at the following closed-form approximation of the optimal total number of sensors
\begin{align}
    n_\mathrm{dp}^*\approx -\frac{2m}{\ln k + 2\ln(2q-1)} \approx \frac{2m}{5-k-4q}.
\end{align}
The above expression is similar to $n_{\mathrm{loss},2}^*$ in Eqn.~\ref{eqn:opt_n_loss_2}, while under dephasing, the probability $q$ that each single qubit remains unaffected by a phase-flip has a different contribution to $n_\mathrm{dp}^*$ than the initial state preparation entangling gate fidelity $k$. More specifically, we have
\begin{align}
    \frac{\partial n_\mathrm{dp}^*/\partial k}{\partial n_\mathrm{dp}^*/\partial q} \approx \frac{2q-1}{4k} \approx \frac{1}{4},
\end{align}
which means that $n_\mathrm{dp}^*$ is approximately 4 times more sensitive to the change in the dephasing amplitude than the state preparation gate fidelity. 

Then, for a given total number of sensors $n$, the optimal partition number $m^*_\mathrm{dp}$ is the $m$ that achieves $\partial\mathcal{F}_\mathrm{dp}(F,k,q,n,m)/\partial m = 0$. Again, after some derivations and approximations, we obtain the simple approximation of the optimal partition number
\begin{align}
    m^*_\mathrm{dp} \approx -n\left[\ln k + 2\ln(2q-1)\right] \approx n(5-k-4q),
\end{align}
which clearly aligns with $n_\mathrm{dp}^*$.

\subsection{Advantage of partitioning}
As before, we examine the advantage of partitioning over the monolithic strategy. First of all, the $m$-fold enhancement in the highest achievable QFI remains the same.

We compare the achievable QFI under optimal partitioning with the monolithic strategy for a fixed total number of sensors $n$. The ratio between the QFI is
\begin{align}
    \frac{\mathcal{F}_\mathrm{dp}(F,k,q,n,m_\mathrm{dp}^*)}{\mathcal{F}_\mathrm{dp}(F,k,q,n)} \approx -\frac{1}{e}\frac{\left[k(2q-1)^2\right]^{-n}}{n\left[\ln k + 2\ln(2q-1)\right]}.
\end{align}
We also compare $\mathcal{F}_\mathrm{dp}(F,k,q,\tilde{n}_\mathrm{dp})$ and $\mathcal{F}_\mathrm{dp}(F,k,q,n,m_\mathrm{dp}^*)$, where $\tilde{n}_\mathrm{dp}=\min\{n,n_\mathrm{dp}^*\}$. In the regime with a larger number of sensors we have $\tilde{n}_\mathrm{dp}=n_\mathrm{dp}^*$, then the ratio becomes
\begin{align}
    \frac{\mathcal{F}_\mathrm{dp}(F,k,p,n,m_\mathrm{dp}^*)}{\mathcal{F}_\mathrm{dp}(F,k,p,n_\mathrm{dp}^*)} \approx -\frac{e\left[\ln k + 2\ln(2q-1)\right]}{4} n.
\end{align}

The above results are compatible with previous results for state preparation errors only, and when including sensor losses. The difference from the scenario with sensor losses is simply to replace $p$ with $(2q-1)^2$, demonstrating the unequal contribution from $k$ and $q$.

\section{Dynamics of the sensing process}\label{sec:dynamics}
Throughout the previous exploration, we have focused on the ``static'' picture, where we ignore the time dependence of the parameter to estimate and error parameters. In practice, it is common that the ultimate parameter of interest is not the accumulated phase $\theta$ in the unitary $U(\theta) = \exp\left[-i\frac{\theta}{2}\sum_{i=1}^{n}Z^{(i)}\right]$, but the rate of phase accumulation (frequency) $\omega$ s.t. $\theta = \omega t$ where $t$ is the duration of system evolution. Therefore, in this section we take into account the system evolution explicitly.

The QFI considered before bounded the estimation variance of $\theta = \omega t$. The estimation variance of $\omega$ satisfies $\mathrm{Var}(\omega) = \mathrm{Var}(\theta)/t^2$, where $t$ is considered as a constant independent of $\theta$. As the QFI is the upper bound of the inverse of the estimation variance, then intuitively the QFI for $\omega$ should satisfy $\mathcal{F}^{(\omega)}(t) = t^2\mathcal{F}^{(\theta)}(t)$, where the superscript denotes the parameter with respect to which the QFI is defined. This relation can be seen from the definition of the QFI as well, where the derivative with respect to the parameter to estimate happens twice. 

The explicit modeling of the sensing dynamics depend on the following parameterizations that include $t$ explicitly. We consider that the probability each sensor to survive undergoes an exponential decrease as the evolution time increases: $p = e^{-\eta t}$. For sensor dephasing, as we have derived in Eqn.~\ref{eqn:dephasing}, the probability that no phase flip happens is $q = \left(1 + e^{-\gamma t}\right)/2$. Note that the state preparation errors described by $k$ happens before the parameter-encoding evolution, so $k$ is independent of evolution time $t$. 

When considering time dynamics, we are able to obtain the explicit analytical time dependence of the QFI on the evolution time. From the time dependence we examine the optimal evolution time~\cite{saleem2023optimal} that maximize the achievable QFI. Then we also demonstrate the advantage of partitioning in terms of parameter estimation bandwidth which is characterized by inverse evolution time. 

\subsection{Sensor qubit losses}
Here we consider sensor qubit losses during the parameter encoding evolution, and the same as before we will consider two cases where we either have the capability to detect losses or not. 

\subsubsection{Without loss detection}
When we do not have the capability to detect losses that happen during the parameter encoding evolution, the QFI as a function of the evolution time $t$ is
\begin{align}
    \mathcal{F}_{\mathrm{loss},1}(t) =& \mathcal{F}_{\mathrm{loss},1}(F,k,e^{-\eta t},n,m)t^2\nonumber\\
    =& \frac{n^2 \left[k - F(2k)^\frac{n}{m}\right]^2 e^{-\frac{2n}{m}\eta t} t^2 / \left[m k^2 \left(2^\frac{n}{m} - 1\right)\right]}{\left(\frac{1 + e^{-\eta t}}{2}\right)^\frac{n}{m} + \frac{e^{-\frac{n}{m}\eta t} \left[k + Fk^\frac{n}{m} \left(2^\frac{n}{m} - 2\right)\right]}{k\left(2^\frac{n}{m} - 1\right)} - e^{-\frac{n}{m}\eta t}}\nonumber\\
    \approx& \frac{n^2 F^2\left(k e^{-\eta t}\right)^\frac{2n}{m} t^2 / (mk^2)}{\left(\frac{1 + e^{-\eta t}}{2}\right)^\frac{n}{m} + e^{-\frac{n}{m}\eta t}\left(\frac{F}{k}k^\frac{n}{m} - 1\right)},
\end{align}
where for the approximation we have discarded exponentially smaller terms, and we have omitted the superscript of the QFI without loss of clarity. 
\begin{figure}[t]
    \centering
    \includegraphics[width=\linewidth]{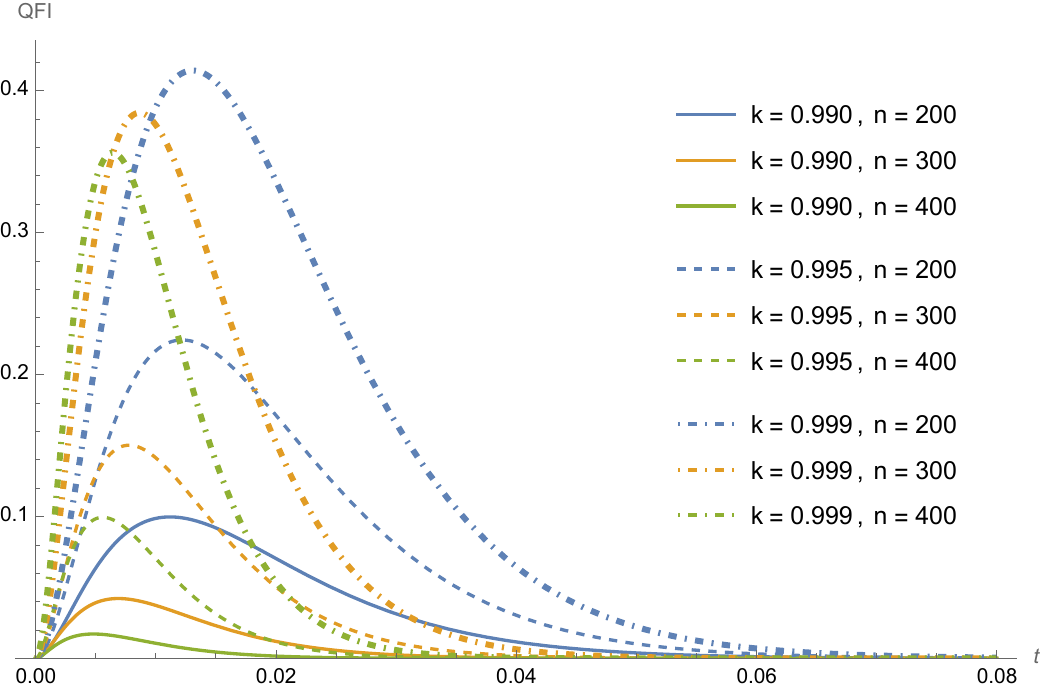}
    \caption{Time dynamics of the (rescaled) QFI, i.e. QFI multiplied by $\eta^2$, when there is no capability of detecting losses. We fix $F=1$ and $m=2$ for all the curves. We choose the following system parameters: entanglement generation quality (entangling gate fidelity) $k=99\%, 99.5\%, 99.9\%$, and total number of sensors $n=200, 300, 400$. Blue curves correspond to $n=200$, yellow curves correspond to $n=300$, and green curves correspond to $n=400$. Solid curves correspond to $k=99\%$, dashed curves correspond to $k=99.5\%$, and thick dot-dashed curves correspond to $k=99.9\%$.}
    \label{fig:dynamics_loss1}
\end{figure}

To make the time dependence explicit and intuitive, we visualize the QFI dynamics in Fig.~\ref{fig:dynamics_loss1}, where we have redefined the time as $t\rightarrow \eta t$ which is dimensionless. Correspondingly, to compensate the $t^2$ factor of the QFI, we rescale the QFI by multiplying a factor of $\eta^2$. In this figure, we have fixed $F=1$ and $m=2$ without loss of generality. First of all, it is observed that the QFI always first increases and then decreases as the duration of the parameter encoding evolution increases, clearly illustrating the competition between information gain and decoherence. Furthermore, through the comparison among different curves, we demonstrate the effect of different values of other system parameters, especially the quality of entanglement generation (entangling gate fidelity) $k$, and the total number of sensors $n$. The sensing performance improves as we have better initial state preparation, i.e. higher $k$, as expected. More specifically, when fixing other system parameters, the QFI for higher $k$ is always higher than the QFI for lower $k$ at the same duration of evolution $\eta t$, which is proved in the appendix. Having higher $k$ also allows the achievement of higher maximal QFI, while it is noteworthy that the time to achieve the maximal QFI, which we may call the \textit{peak time}, depends on $k$. We obtain the equation whose solution is a good approximation of the peak time (in unit of $1/\eta$) as follows
\begin{align}
    \frac{m - n t}{2m - n t} = \left(1 - \frac{F}{k}k^\frac{n}{m}\right)\frac{1 + e^t}{2 e^t}\left(\frac{2e^{-t}}{1 + e^{-t}}\right)^\frac{n}{m} - \frac{1}{2 e^t},
\end{align}
which can be solved analytically based on small $t$ approximation which is practically valid. We show a closed-form expression of the peak time in the appendix based on first order Taylor expansion of the terms which are non-linear in $t$.

On the other hand, it is worth emphasizing that for larger total number of sensors $n$ the peak time of evolution is longer while the achieved maximal QFI is also higher than that for larger $n$. This demonstrates the role of more sensors in the context of sensing with decoherence: Larger amount of sensors will allow one to accumulate information faster, while the maximal amount of obtainable information for parameter estimation in one evolution is actually lower than smaller amount of sensors, \textit{if there is no limit on the duration of evolution}.

\subsubsection{With loss detection}
In the case where we are able to detect losses, the QFI dynamics becomes
\begin{align}
    \mathcal{F}_{\mathrm{loss},2}(t) =& \mathcal{F}_{\mathrm{loss},2}(F,k,e^{-\eta t},n,m)t^2\nonumber\\
    =& \frac{n^2\left[F(2k)^\frac{n}{m} - k\right]^2e^{-\frac{n}{m}\eta t} t^2}{mk \left(2^\frac{n}{m} - 1\right)\left[F\left(2^\frac{n}{m} - 2\right) k^\frac{n}{m} + k\right]}.
\end{align}
Qualitatively, the behavior of the QFI dynamics is the same as the first case in that the QFI will first increase and then decrease, since the competition between information gain and decoherence is universal. And we can also confirm in the appendix that the QFI monotonically increases as the initial state preparation is better (when $k$ is higher). Different from the first case without loss detection capability, when we can detect loss the peak time only depends on the total number of sensors $n$ and the partition number $m$. This can be shown by explicitly taking the partial derivative of $\mathcal{F}_{\mathrm{loss},2}(t)$ with respect to $t$
\begin{align}
    \frac{\partial}{\partial t}\mathcal{F}_{\mathrm{loss},2}(t) = \frac{n^2\left[F(2k)^\frac{n}{m} - k\right]^2e^{-\frac{n}{m}\eta t} (2m - n\eta t) t}{m^2k \left(2^\frac{n}{m} - 1\right)\left[F\left(2^\frac{n}{m} - 2\right) k^\frac{n}{m} + k\right]},
\end{align}
which means that $\mathcal{F}_{\mathrm{loss},2}(t)$ attains its highest value at the peak time $\eta t_{\mathrm{loss},2}^* = 2m/n$.

\subsubsection{Advantage of loss detection}
\begin{figure}[t]
    \centering
    \includegraphics[width=\linewidth]{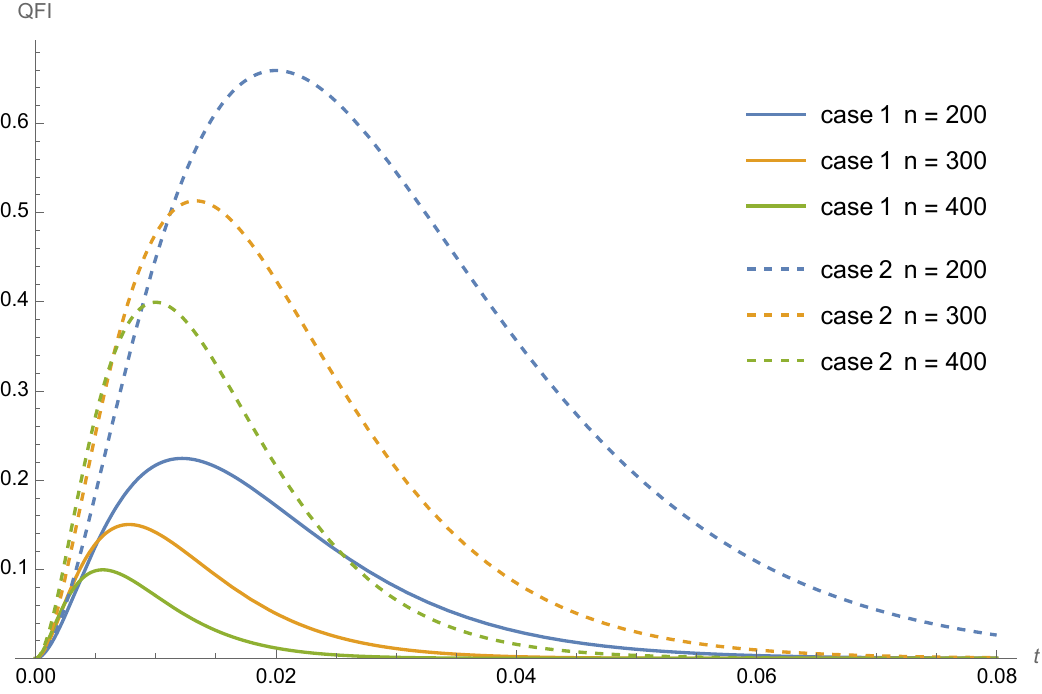}
    \caption{Time dynamics of the (rescaled) QFI (QFI multiplied by $\eta^2$), when there is or is not capability of detecting losses. Time $t$ is in unit of $1/\eta$. We fix $F=1$, $k=99.5\%$ and $m=2$ for all the curves. For both cases we consider total number of sensors $n=200,300,400$. Solid curves correspond to the first case without loss detection and dashed curves correspond to the second case with loss detection.}
    \label{fig:dynamics_loss1vs2}
\end{figure}
When there are particle losses during the probing evolution, whether having the capability of detecting the losses will make a difference in the performance. Here we compare the two cases considered above. A basic example is illustrated in Fig.~\ref{fig:dynamics_loss1vs2}, where we visualize $\mathcal{F}_{\mathrm{loss},1}(t)$ and $\mathcal{F}_{\mathrm{loss},2}(t)$ for $n=200,300,400$, while fixing $F=1$, $k=99.5\%$ and $m=2$. It can be seen that having loss detection capability leads to higher QFI at all evolution time $t$. Indeed, this can be proved through taking the ratio $\mathcal{F}_{\mathrm{loss},2}(t)/\mathcal{F}_{\mathrm{loss},1}(t)$
\begin{align}
    \frac{\mathcal{F}_{\mathrm{loss},2}(t)}{\mathcal{F}_{\mathrm{loss},1}(t)} - 1 = \frac{k\left(1 - 2^{-\frac{n}{m}}\right) \left[\left(1 + e^{\eta t}\right)^\frac{n}{m} - 2^\frac{n}{m}\right]}{F\left(2^{\frac{n}{m}} - 2\right)k^\frac{n}{m} + k} \geq 0.
\end{align}
In addition, one can see that the peak time for the first case is earlier than the second case. This is because longer evolution time leads to smaller probability for each sensor to survive the whole evolution, according to $p = e^{-\eta t}$. Recalling the discussion in Sec.~\ref{sec:loss}, as $p$ decreases the ratio between $\mathcal{F}_{\mathrm{loss},2}$ and $\mathcal{F}_{\mathrm{loss},1}$ increases exponentially. Therefore, in the first case the highest QFI must have been achieved before the second case achieves its highest QFI.

\subsection{Sensor qubit dephasing}
Here we consider sensor qubit dephasing during the parameter encoding evolution. 

When every qubit endures independent dephasing during the parameter encoding process, the total probe evolution is described by Eqn.~\ref{eqn:encode_dephase}. Then we are able to derive the QFI dynamics as
\begin{align}
    \mathcal{F}_\mathrm{dp}(t) = & \mathcal{F}_\mathrm{dp}\left(F,k,\frac{1 + e^{-\gamma t}}{2},n,m\right)t^2\nonumber\\
    =& \frac{n^2\left[F(2k)^\frac{n}{m} - k\right]^2e^{-\frac{2n}{m}\gamma t} t^2}{mk \left(2^\frac{n}{m} - 1\right)\left[F\left(2^\frac{n}{m} - 2\right) k^\frac{n}{m} + k\right]}.
\end{align}
Note that the expression resembles $\mathcal{F}_{\mathrm{loss},2}(t)$, and the only difference is in the exponential decay factor from dephasing v.s. loss, as expected according the discussion in Sec.~\ref{sec:dephasing}. Since the $k$-dependence of $\mathcal{F}_\mathrm{dp}(t)$ is the same as $\mathcal{F}_{\mathrm{loss},2}(t)$, we can immediately see the monotonic increase of QFI with increasing $k$. Then we can also evaluate the partial derivative with respect to $t$ to derive the peak time
\begin{align}
    \frac{\partial}{\partial t}\mathcal{F}_\mathrm{dp}(t) = \frac{n^2\left[F(2k)^\frac{n}{m} - k\right]^2e^{-\frac{2n}{m}\gamma t} (m - n\gamma t) t}{m^2k \left(2^\frac{n}{m} - 1\right)\left[F\left(2^\frac{n}{m} - 2\right) k^\frac{n}{m} + k\right]},
\end{align}
from which we can see that the peak time is $\eta t_\mathrm{dp}^* = m/n$. Similar to the case of detectable particle loss, the peak time only depends on $\gamma,m,n$, while the difference in the constant factor demonstrates the difference between dephasing and loss. 

In fact, as particle losses and dephasing commute with each other, it is analytically convenient to combine both error sources when assuming loss detectability. The QFI dynamics under the combined errors is simply
\begin{align}
    \mathcal{F}_\mathrm{loss,dp}(t) =& \mathcal{F}_\mathrm{dp}(t)e^{-\frac{n}{m}\eta t}.
\end{align}
which is simply a parameter substitution $\eta\rightarrow\eta+2\gamma$ for $\mathcal{F}_{\mathrm{loss},2}(t)$, or $\gamma\rightarrow\gamma+\eta/2$ for $\mathcal{F}_\mathrm{dp}(t)$. Then we can immediately identify the peak time for the combined dynamics as $(2\gamma+\eta)t^*_\mathrm{loss,dp} = 2m/n$.

\subsection{Sequential scheme and the advantage of partitioning}
\begin{figure}[t]
    \centering
    \includegraphics[width=\linewidth]{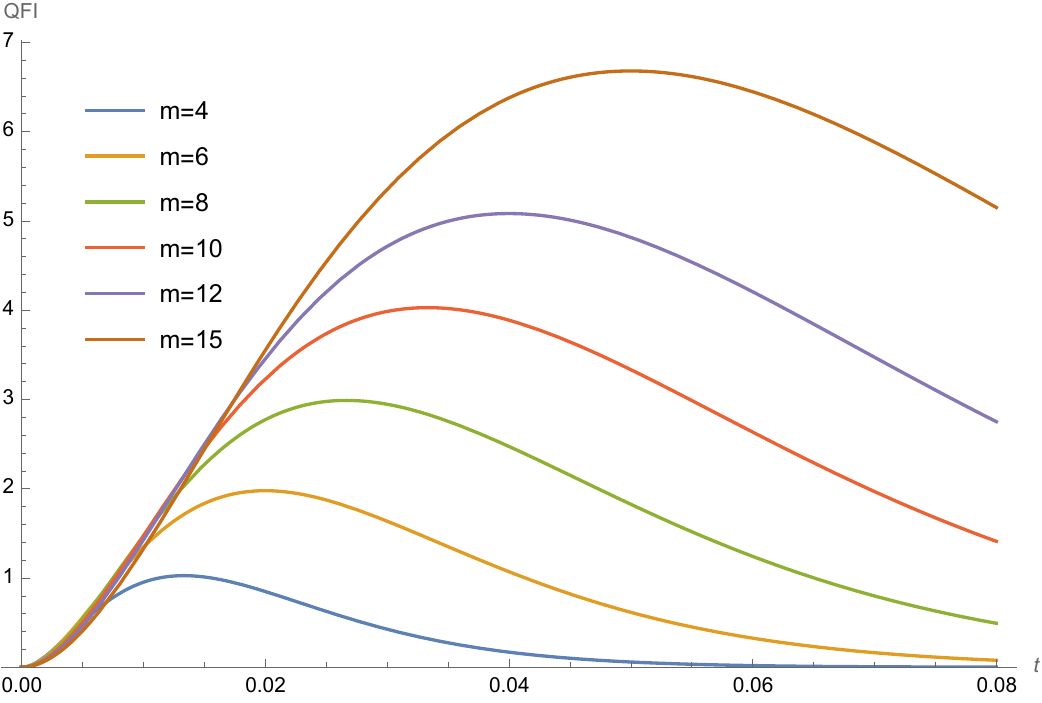}
    \caption{Time dynamics of the (rescaled) QFI (QFI multiplied by $(2\gamma+\eta)^2$) under sensor loss and dephasing with loss detectability. Time $t$ is in unit of $1/(2\gamma+\eta)$. We fix $F=1$, $k=99.5\%$ and $n=600$ for all the curves. The number of sub-ensembles is varied for $m=4,6,8,10,12,15$.}
    \label{fig:partition_adv_peak}
\end{figure}
Previously we have focused on individual parameter encoding processes under different error sources. From such a perspective, partitioning introduces conspicuous advantage in terms of the highest QFI achievable, shown in Fig.~\ref{fig:partition_adv_peak} taking the dephasing case as an example. Larger partition number $m$ leads to later peak time, while the peak QFI will also be higher. Quantitatively, we can substitute the peak time $(2\gamma+\eta)t^*_\mathrm{loss,dp} = 2m/n$ into the QFI dynamics to obtain
\begin{align}\label{eqn:peak_qfi}
    \mathcal{F}_\mathrm{loss,dp}(t^*_\mathrm{loss,dp}) \approx \frac{4F}{ek(2\gamma + \eta)^2}k^\frac{n}{m}m.
\end{align}
One can see an apparent linear scaling with $m$. In addition to that, the $k^\frac{n}{m}$ factor introduces an additional degree of advantage in the maximal QFI with increasing $m$: As $k<1$ and $n$ are generally fixed for a specific sensing system, larger $m$ will lead to higher $k^\frac{n}{m}$. On the other hand, we can evaluate the $t$ derivative of QFI at $t\rightarrow 0$, whose lowest order is 
\begin{align}
    \frac{\partial}{\partial t}\mathcal{F}_\mathrm{loss,dp}(t) \approx \frac{2F}{k}\frac{k^\frac{n}{m}n^2}{m}t + O(t^2).
\end{align}
One can see that the factor of the linear term monotonically increases as $m$ increases before $m\geq -n\log k = m^*$, which is the optimal partition number we derived in Eqn.~\ref{eqn:opt_part_num_st_prep}. This demonstrates that partitioning also offers advantage in the speed of QFI increase at short evolution times. Such short evolution time requirement can come from demand for measurement bandwidth and experimental limitations~\cite{ludlow2015optical,degen2017quantum}.

\begin{figure}[t]
    \centering
    \includegraphics[width=\linewidth]{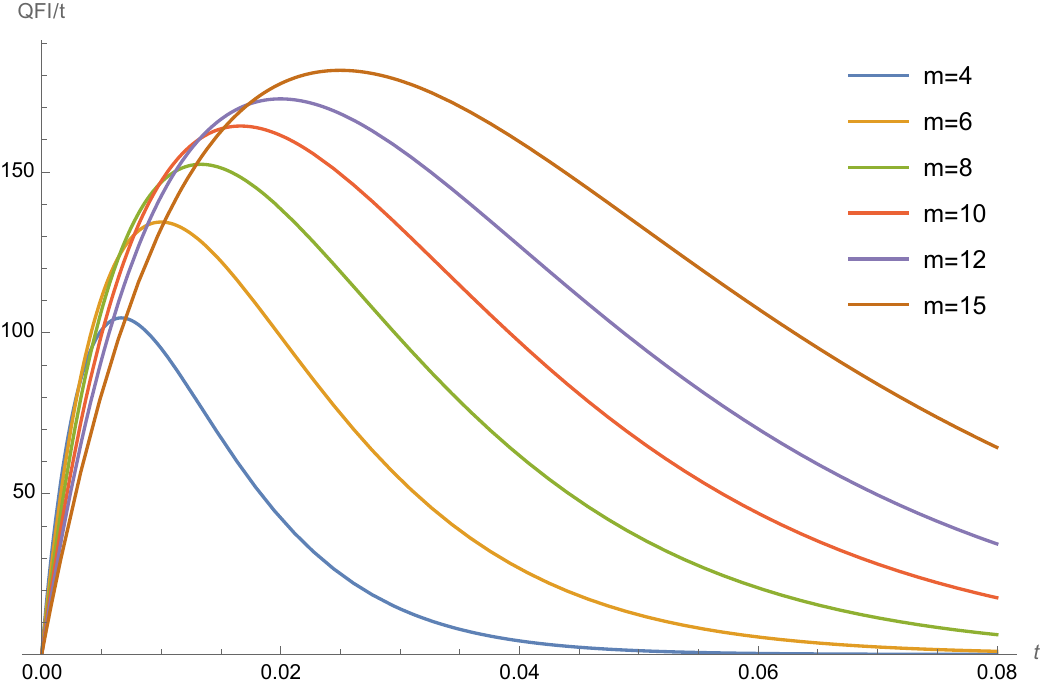}
    \caption{(Rescaled) QFI per unit evolution time as a function of evolution time $t$ (multiplied by $(2\gamma+\eta)$), under sensor loss and dephasing with loss detectability. Evolution time $t$ is in unit of $1/(2\gamma+\eta)$. We fix $F=1$, $k=99.5\%$ and $n=600$ for all the curves. The number of sub-ensembles is varied for $m=4,6,8,10,12,15$.}
    \label{fig:dephasing_QFI_per_t}
\end{figure}

In practice, there are also other cases where we do not just run the sensing cycle for a single shot, but instead the sensing cycle is repeated sequentially within a given total amount of time $T$ as resource, which is known as the sequential scheme of quantum sensing~\cite{huelga1997improvement}. Under such a setup, our objective then becomes to maximize the gained information within a fixed amount of time to minimize the parameter estimation variance. There is thus a natural trade-off between the duration of a single sensing cycle, and the amount of information obtained per sensing cycle. In the following, we ignore the time needed for initial state preparation and the final measurement of each sensing cycle, to highlight the effect of noisy parameter encoding evolution. With such an assumption, the accumulated information is the sum of QFI from each sensing cycle, from the additivity of QFI as each sensing cycle can be regarded as independent
\begin{align}
    \mathcal{I} = \sum_i \mathcal{F}(t_i),~ \sum_i t_i = T.
\end{align}
We may rewrite the sum as
\begin{align}
    \mathcal{I} = \sum_i \mathcal{F}(t_i) = \sum_i \frac{\mathcal{F}(t_i)}{t_i}t_i \leq T\max_t\frac{\mathcal{F}(t)}{t}.
\end{align}
Therefore, in the practical sequential scheme we are interested in the quantity $\mathcal{F}(t)/t$, \textit{QFI per unit (evolution) time} which characterizes the average information accumulation speed during an evolution with duration $t$. We consider $\mathcal{F}_\mathrm{loss,dp}(t)/t$ and its behavior as a function of evolution time $t$ is visualized in Fig.~\ref{fig:dephasing_QFI_per_t}. The shape has changed from the QFI dynamics itself, but some features remain. For instance, the peak value increases as $m$ increases, and at short times $t\rightarrow 0$ the partition number $m$ will affect how QFI per unit time changes as the evolution time per cycle varies. 

We can take the $t$ derivative of $\mathcal{F}_\mathrm{loss,dp}(t)/t$ as
\begin{align}
    &\frac{\partial}{\partial t}\left[\frac{\mathcal{F}_\mathrm{loss,dp}(t)}{t}\right]\\
    =& \frac{n^2\left[F(2k)^\frac{n}{m} - k\right]^2e^{-\frac{n}{m}(2\gamma+\eta) t}}{m^2k \left(2^\frac{n}{m} - 1\right)\left[F\left(2^\frac{n}{m} - 2\right) k^\frac{n}{m} + k\right]}[m - n(2\gamma+\eta) t]\nonumber.
\end{align}
The peak value of QFI per unit time is then achieved at the \textit{optimal time} $\tilde{t}^*_\mathrm{loss,dp} = m/[n(2\gamma+\eta)]$. Then the peak value is thus determined as
\begin{align}
    \frac{\mathcal{F}_\mathrm{loss,dp}(\tilde{t}^*_\mathrm{loss,dp})}{\tilde{t}^*_\mathrm{loss,dp}} \approx \frac{F}{ek(2\gamma+\eta)}k^\frac{n}{m}n.
\end{align}
It has a similar form to Eqn.~\ref{eqn:peak_qfi}, while the scaling behavior is different. For the sequential scheme, when it is free to choose the evolution time in every sensing cycle, the ultimate precision then has an apparent linear scaling with the total number of sensors $n$. On the other hand, higher state preparation infidelity comes together with larger probe state. Therefore, one should not blindly increase the sensing system size, but instead make the partition number $m$ and total number $n$ compatible by roughly satisfying $m=-n\ln k$. The increase of $m$ lead to advantage that stems from lower initial probe state preparation infidelity. Then we consider the regime of short evolution time per cycle, i.e. $t\rightarrow 0$, and the lowest order of the time derivative is
\begin{align}
    \frac{\partial}{\partial t}\left[\frac{\mathcal{F}_\mathrm{loss,dp}(t)}{t}\right] \approx \frac{F}{k}\frac{k^\frac{n}{m}n^2}{m} + O(t).
\end{align}
Similar to short time QFI, one can also verify that the zeroth-order term monotonically increases as $m$ increases before $m\geq -n\log k = m^*$. This means that partitioning offers advantage in the QFI per unit time if the evolution time in each sensing cycle is limited to be short.

Now we discuss the advantage and optimization of partitioning in the sequential scheme more comprehensively by considering the limit of evolution time in each sensing cycle, so that the evolution time is generally shorter than the typical coherence time of the sensor qubits. As we have shown previously, for any $m$, the QFI per unit time $\mathcal{F}_\mathrm{loss,dp}(t)/t$, which also depends on $m$, will monotonically increase as the evolution time per sensing cycle $t$ increases for $t\in[0,\tilde{t}_m^*]$ where $\tilde{t}_m^*\coloneqq \tilde{t}^*_\mathrm{loss,dp} = m/[n(2\gamma+\eta)]$. The advantage of partitioning and the optimization of partition number $m$ in the sequential scheme can then be investigated with the following strategy.
\begin{enumerate}
    \item The evolution time $t$ is restricted by $t\in[0,t_\mathrm{th}]$ with $t_\mathrm{th} < \tilde{t}_1^*$. We have that $\mathcal{F}_\mathrm{loss,dp}(t)/t$ monotonically increases throughout this interval for all $m$. Therefore, we only need to perform optimization $\arg\max_{m\geq1,m\in\mathbb{Z}}[\mathcal{F}_\mathrm{loss,dp}(t)/t]$ to find the optimized $m$. 
    \item The evolution time $t$ is restricted by $t\in[0,t_\mathrm{th}]$ with $t_\mathrm{th}\in[\tilde{t}_{l-1}^*,\tilde{t}_l^*)$, for $l>1,l\in\mathbb{Z}$. Then we know that $\mathcal{F}_\mathrm{loss,dp}(t)/t$ can achieve its maximum by choosing an evolution time within the interval for $m=1,2,\dots, l-1$, while taking $m=l-1$ achieves the highest highest maximum in comparison with other $m<k-1$. Then we should perform the optimization $m'=\arg\max_{m\geq1,m\in\mathbb{Z}}[\mathcal{F}_\mathrm{loss,dp}(t)/t]$, and then compare $\mathcal{F}_\mathrm{loss,dp}(t)/t$ for partition number $m=m'$ and the maximum for partition number $m=l-1$ to find the higher value.
\end{enumerate}

Then we may take the $m$ derivative of $\mathcal{F}_\mathrm{loss,dp}(t)/t$ to obtain that the optimal partition number $\tilde{m}^*$ for QFI per unit time with a fix evolution time $t$ is approximately
\begin{align}
    \tilde{m}^*(t) \approx n(2\gamma + \eta)t - n\ln k.
\end{align}
Notably, the first term corresponds to the form of optimal time $\tilde{t}^*_\mathrm{loss,dp}$, while the second term is exactly the optimal partition number $m^*$ when only considering state preparation errors as shown in Eqn.~\ref{eqn:opt_part_num_st_prep}. This demonstrates the intricate interplay among different perspectives of the GHZ partitioning strategy. Moreover, in general we have $\tilde{m}^* > l$ when the evolution time limitation $t_\mathrm{th}$ satisfies $t_\mathrm{th}\in[\tilde{t}_{l-1}^*,\tilde{t}_l^*)$ for $l\geq 1,k\in\mathbb{Z}$ with $\tilde{t}_{0}^*\coloneqq 0$. Then according to the above optimization strategy, the actual optimal partition number in the sequential scheme with limitation of evolution time per sensing cycle $t_\mathrm{th}$ should be an integer that is close to $\tilde{m}^*(t_\mathrm{th})$. It is also notable that as we are able to evolve for longer time in a single sensing cycle, the optimal partition number also increases. Assuming we can partition the sensors into $\approx\tilde{m}^*(t_\mathrm{th})$ sub-ensembles, the sensing performance will converge to
\begin{align}
    \mathcal{I} \rightarrow \frac{F}{e(2\gamma+\eta)}nT.
\end{align}
In comparison, if we cannot perform partitioning, i.e. fixing $m=1$, by optimizing the total number of sensors $n$ the sensing performance will converge to
\begin{align}
    \mathcal{I}_{m=1} \rightarrow \frac{F}{e(2\gamma+\eta)}\frac{-1}{ek\ln k}T,
\end{align}
which is only determined by system error parameters and independent of system size.

\subsection{Comparison with spin squeezed states}
GHZ states are the focus of this work, while it is also a fact that they are fragile against decoherence. Therefore, here we compare the performance of noisy GHZ states with \textit{noiseless} spin squeezed states (SSSs)~\cite{wineland1992spin,kitagawa1993squeezed,ma2011quantum} to demonstrate the regime in which the GHZ states can outperform robust SSSs even under noises. We focus on realistic experimental conditions, by restricting ourselves to the standard Ramsey spectroscopy~\cite{wineland1992spin,huelga1997improvement}, and one-axis twisting (OAT) SSS~\cite{wineland1992spin,kitagawa1993squeezed,ma2011quantum}. Wineland et al have derived the parameter estimation performance of SSS via Ramsey spectroscopy by proposing the spin squeezing parameter $\xi_R^2$ of the $n$-qubit SSS as the initial probe state, so that the phase ($\theta = \omega t$) estimation variance is given by~\cite{wineland1992spin,ma2011quantum}
\begin{align}\label{eqn:sss_est_var}
    \mathrm{Var}(\theta)= \frac{\xi_R^2}{n}.
\end{align}
On the other hand, Kitagawa and Ueda proposes another spin squeezing parameter $\xi_S^2$ in a different context than parameter estimation~\cite{kitagawa1993squeezed}, while these two spin squeezing parameters satisfies $\xi_S^2 \leq \xi_R^2$~\cite{ma2011quantum}. Therefore, we can use $\xi_S^2$ in Eqn.~\ref{eqn:sss_est_var} to give an lower bound of the estimation variance. Moreover, in the best case scenario of noiseless preparation of the OAT SSS, i.e. Hamiltonian evolution for time $t_H$ under the OAT Hamiltonian $H = \chi S_x^2$ with $S_x$ being the $x$-component of the collective spin, the closed-form expression of $\xi_S^2$ is known~\cite{kitagawa1993squeezed,ma2011quantum} as
\begin{align}
    \xi_S^2 = 1 - (n-1)C/4,
\end{align}
where
\begin{align}
    C =& \sqrt{\left(1 - \cos^{n-2}\phi\right)^2 + 16\sin^2\frac{\phi}{2}\cos^{2n-4}\frac{\phi}{2}}\nonumber\\
    &- (1 - \cos^{n-2}\phi),
\end{align}
with $\phi = 2\chi t_H$. Then we will use $\xi_S^2t^2/n$ as the lower bound on the variance estimation of the frequency $\omega$ when the Ramsey spectroscopy is run for $t$ time without decoherence. In the sequential scheme with $T=\mu t$, the total sensing performance will be bounded as
\begin{align}
    \mathrm{Var}(\omega)_\mathrm{OAT} \leq \frac{\xi_S^2}{n \mu t^2} = \frac{\xi_S^2}{n t T}.
\end{align}

On the other hand, for (partitioned) GHZ states, we consider the noisy Ramsey spectroscopy, including errors from state preparation, losses and dephasing, and assuming loss detectability for analytical convenience. Under partitioning, effectively we have $m$ smaller $n/m$-qubit GHZ states performing Ramsey spectroscopy in parallel. We first focus on the result conditioned on no loss, and include loss probability later given the loss detectability assumption. Recall that when we include individual sensor dephasing during the encoding dynamics for noisy $n$-qubit GHZ state, the effective initial probe state for unitary encoding is given by Eqn.~\ref{eqn:eff_probe_dephasing}. Then the final state before measurement is 
\begin{align}
    \rho_f(t) =& [1-\tilde{q}(t)]V(n)\mathrm{Proj}_{\mathrm{GHZ}^-(\omega t)}\nonumber\\
    &+ \tilde{q}(t)V(n)\mathrm{Proj}_{\mathrm{GHZ}^+(\omega t)} + [1 - V(n)] \frac{I_n}{2^n},
\end{align}
where $\mathrm{Proj}_{\mathrm{GHZ}^+(\omega t)}$ is the projector for $(|0\dots 0\rangle + e^{-in\omega t}|1\dots 1\rangle)/\sqrt{2}$, and $\mathrm{Proj}_{\mathrm{GHZ}^-(\omega t)}$ is the projector for $(|0\dots 0\rangle - e^{-in\omega t}|1\dots 1\rangle)/\sqrt{2}$, with $\tilde{q}(t) = (1 + e^{-n\gamma t})/2$ and $V(n) = (2^nFk^{n-1} - 1)/(2^n - 1)$. The final measurement can be implemented as first applying the CNOT network as the inverse of GHZ state preparation, and measuring the control qubit in $X$-basis~\cite{huelga1997improvement}. Note that as we have considered the GHZ preparation error, the final measurement circuit should also induce same errors to the state with fidelity $Fk^{n-1}$. The final observable is the probability for measuring 0, which should be 
\begin{align}
    P_0 =& Fk^{n-1}\tilde{q}(t)V(n)\frac{1 + \cos(n\omega t)}{2} \nonumber\\
    &+ Fk^{n-1}[1-\tilde{q}(t)]V(n)\frac{1 - \cos(n\omega t)}{2} \nonumber\\
    &+ \left[1 - Fk^{n-1}V(n)\right]\frac{1}{2} \nonumber\\
    =& \frac{1}{2} + \frac{Fk^{n-1}V(n)[2\tilde{q}(t) - 1]}{2}\cos(n\omega t).
\end{align}
Then according to error propagation~\cite{toth2014quantum} $\mathrm{Var}(\omega) = \mathrm{Var}(P_0)/|\partial P_0/\partial\omega|^2$ we can obtain the exact expression for the estimation variance of $\omega$ while also including sensor losses, as provided in the appendix. Furthermore, in the sequential scheme one can show that the optimal conditions are almost exactly the same as in~\cite{huelga1997improvement}, i.e. $n\omega t = (2j+1)\pi/2$ for $j=0,1,\dots$ and $n(2\gamma+\eta)t=1$. Then if we allow partitioning into $m$ sub-ensembles, the optimal sequential sensing performance will simply be $m$-fold reduced from monolithic performance with $n/m$ sensors, while the optimal evolution time should satisfy $n\omega t/m = (2j+1)\pi/2$. The optimal sensing performance for noisy GHZ Ramsey spectroscopy with $m$ sub-ensembles is then
\begin{align}
    \mathrm{Var}(\omega)_\mathrm{GHZ} = \frac{e\left(2^\frac{n}{m} - 1\right)^2(2\gamma+\eta)}{F^2k^{2\frac{n}{m}-4}\left[F(2k)^\frac{n}{m} - k\right]^2nT}.
\end{align}
One can see that partitioning demonstrate advantage from lower state preparation error.

\begin{figure}[t]
    \centering
    \includegraphics[width=\linewidth]{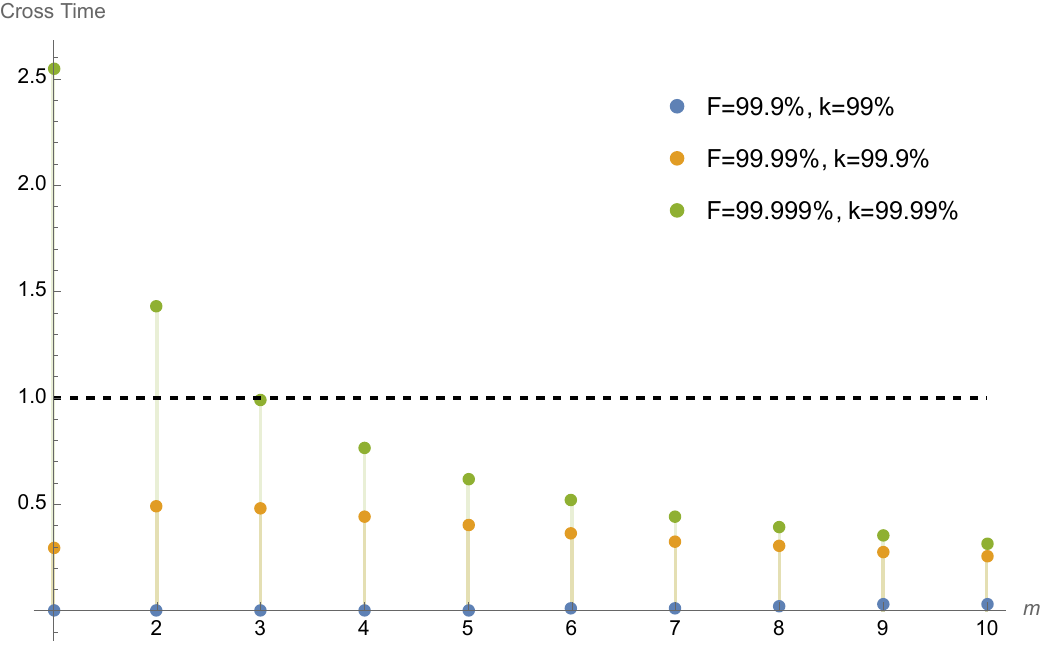}
    \caption{Comparison between scaled cross time $n(2\gamma+\eta)t_\mathrm{cross}/m$ and unit, for different partition numbers. Total number of sensors is fixed $n=600$. Different GHZ preparation error parameters are considered, including $F=99.9\%$ and $k=99\%$ (blue dots), $F=99.99\%$ and $k=99.9\%$ (yellow dots), and $F=99.999\%$ and $k=99.99\%$ (green dots). The black dashed line represents the unit baseline.}
    \label{fig:cross_time}
\end{figure}

Finally, we compare the sequential Ramsey spectroscopy performance between OAT SSS and partitioned GHZ state, when the evolution time per Ramsey cycle has varying limitation $t_\mathrm{th}$. Partitioned noisy GHZ state with $m$ sub-ensembles will have advantage over perfect OAT SSS if $\mathrm{Var}(\omega)_\mathrm{GHZ} \leq \mathrm{Var}(\omega)_\mathrm{OAT}$ when $n(2\gamma+\eta)t_\mathrm{th}/m \geq 1$, so that we can achieve the optimal Ramsey time for the GHZ state. To see if such advantage can be achieved, we find the \textit{crossing time} $t_\mathrm{cross}$ s.t. $\mathrm{Var}(\omega)_\mathrm{GHZ} = \mathrm{Var}(\omega)_\mathrm{OAT}$ and compare $n(2\gamma+\eta)t_\mathrm{cross}/m$ with unit. If $n(2\gamma+\eta)t_\mathrm{cross}/m \geq 1$ then there is advantage and vice versa. Now we numerically demonstrate such potential advantage. Similar to previous discussions on time dynamics, we consider a total number of sensors $n=600$. One can then numerically minimize $\xi_S^2$ to obtain the first non-trivial minimum as $\xi_S^2\approx 0.01464$ at $\phi\approx 0.0335$. We demonstrate the comparison between $n(2\gamma+\eta)t_\mathrm{cross}/m$ and unit in Fig.~\ref{fig:cross_time}. We consider 3 sets of different GHZ preparation error parameter values, from modest to a bit futuristic. We observe that advantage of GHZ state is achieved for monolithic and $m=2$ partition with futuristic parameter values, i.e. $F=99.99\%$ and $k=99.9\%$, while other more modest parameter values do not lead to advantage. Nevertheless, we emphasize that this does not diminish the meaningfulness of GHZ partitioning, because the comparison is made between noisy GHZ Ramsey spectroscopy estimation variance and the \textit{lower bound} of \textit{noiseless} OAT SSS Ramsey spectroscopy estimation variance. In contrary, it is noteworthy that with improved system hardware, noisy GHZ can outperform noiseless robust OAT SSS.

\section{Conclusion}\label{sec:conclusion}
In this paper, we propose an optimal partitioning strategy for GHZ state phase estimation that provides a clear and simple theoretical guidance for realizing quantum sensing advantage. 

We perform extensive analytical modeling of the sensing performance under various practically important sources of errors, namely state preparation errors which are ubiquitous for all quantum systems, qubit losses during encoding process which are common for atomic systems, and qubit dephasing which is typical in solid-state systems. We derive closed-form expressions of the QFI and obtain the following results.

\begin{enumerate}
    \item We first consider the scenario with state preparation errors only as the minimal example. We derive a simple, closed-form approximation for the optimal total number of sensor $n^*$ under a fixed partition number $m$ and the optimal partition number $m^*$ for a fixed total number of sensors $n$. We also demonstrate the impact of entangling gate fidelities on the highest achievable sensing performance, which can serve as additional motivation for pursuing higher gate fidelities. The advantage of partitioning is quantitatively analyzed under this scenario.
    \item We combine state preparation errors and qubit losses during the encoding process. Therein we consider two cases, one with the ability to detect losses in a nondemolition manner, and the other without. We derive approximations for the optimal total number of sensors $n^*$ and the optimal partition number $m^*$ when assuming loss detection capability. When loss detection is not possible, we explain why the sensing system should have fewer sensors. We also demonstrate the advantage of loss detection which increases monotonically as the loss probability increases.
    \item We then study a scenario that includes both state preparation errors and qubit dephasing during the encoding process. Again, we derive approximations for the optimal total number of sensors $n^*$ and the optimal partition number $m^*$. We derive that $n^*$ and $m^*$ are 4 times more sensitive to dephasing than the initial state preparation errors, which emphasizes the difference between noise sources, and the detrimental effect of dephasing.
    \item Last but not least, we perform comprehensive investigation into the explicit quantum sensing dynamics under losses and dephasing. We derive closed-form expressions for QFI dynamics, and demonstrate that partitioning continues to offer benefits to the maximal achievable QFI in the noisy evolution, the speed of QFI increase at short evolution times, and the parameter estimation performance in the practical sequential sensing scheme. Moreover, we compare GHZ state with the robust one-axis twisting spin squeezed state in the canonical Ramsey setup, and show that given further hardware fidelity improvement, noisy partitioned GHZ Ramsey can outperform noiseless OAT SSS Ramsey, when evolution time is limited.
\end{enumerate}

\begin{acknowledgments}
We thank Sisi Zhou for helpful discussions.
Part of this research was performed while the authors were visiting the Institute for Mathematical and Statistical Innovation (IMSI), which is supported by the National Science Foundation (Grant No. DMS-1929348).

The investigation of partitioned GHZ states' sensing performance under noise is based upon work supported by Q-NEXT (Grant No. DOE 1F-60579), one of the U.S. Department of Energy Office of Science National Quantum Information Science Research Centers.
A.Z. and T.Z. acknowledge support from the NSF Quantum Leap Challenge Institute for Hybrid Quantum Architectures and Networks (NSF Award 2016136).
T.-X.Z. and P.C.M. acknowledge support from Q-NEXT (Grant No. DOE 1F-60579).
\end{acknowledgments}

\bibliography{references}

\onecolumngrid
\appendix

\section{Derivations for the scenario with only state preparation errors}\label{app:derivation_state_prep}
\subsection{Optimal number of sensors}
To determine the optimal number of sensors, we would like to evaluate the partial derivative of the QFI with respect to $n$. Note that $k$ and $F$ should both be close to unity in practically meaningful regime, and the optimal number of sensors is generally large. We can thus make use of these additional aspects to obtain simpler approximation of the partial derivative.
\begin{align}
    &\frac{\left(2^{\frac{n}{m}}\cancel{-1}\right)^2k\left[\cancel{k + }F\left(2^{\frac{n}{m}}\cancel{-2}\right)k^{\frac{n}{m}}\right]^2m^2}{\left[F(2k)^{\frac{n}{m}}\cancel{ - k}\right]n} \left[\frac{\partial}{\partial n}\mathcal{F}(F,k,n,m)\right]\nonumber\\
    &= k^2\left[\cancel{2m + }2^{\frac{n}{m}}(n\ln 2-2m)\right] + F^2(2k^2)^{\frac{n}{m}}\left[2^{1+\frac{2n}{m}}m \cancel{+ 4(m+n\ln 2)} - 3\times 2^{\frac{n}{m}}(2m+n\ln 2)\right]\nonumber\\
    &~~~~+ Fk^{1+\frac{n}{m}}\left[2^{\frac{n}{m}}(4m-5n\ln 2) + 3\times 4^{\frac{n}{m}}n\ln 2\cancel{ - 4m}\right] + n\left(2^{\frac{n}{m}}\cancel{-1}\right)Fk^{\frac{n}{m}}\left[\left(3\times 2^{\frac{n}{m}}\cancel{-2}\right)k + \left(2^{\frac{n}{m}}\cancel{-2}\right)F(2k)^{\frac{n}{m}}\right]\ln k\nonumber\\
    \Rightarrow& \frac{Fkm^2}{n}(4k)^\frac{n}{m}\left[\frac{\partial}{\partial n}\mathcal{F}(F,k,n,m)\right] \approx \cancel{k^2(n\ln 2-2m) + }F^2\left(2k^2\right)^\frac{n}{m}\left[2^\frac{n}{m}(2m + n\ln k)\cancel{ - 6m - n\ln 8}\right]\nonumber\\
    &~~~~~~~~~~~~~~~~~~~~~~~~~~~~~~~~~~~~~~~~~~~~~~~~~~~~~~~~~~~~~ + Fk^{1+\frac{n}{m}}\left[\cancel{4m - 5n\ln 2 + }2^\frac{n}{m}\ln(8k^3)\right]\\
    \Rightarrow& \frac{\partial}{\partial n}\mathcal{F}(F,k,n,m)\approx \frac{n}{m^2}\left[Fk^{\frac{n}{m}-1}(2m + n\ln k)\cancel{ + 2^{-\frac{n}{m}}n\ln(8k^3)} \right]\\
    \Rightarrow& \frac{\partial}{\partial n}\mathcal{F}(F,k,n,m)\approx \frac{Fnk^{\frac{n}{m}-1}}{m^2}(2m + n\ln k).
\end{align}
Then from the above the equation $\frac{\partial}{\partial n}\mathcal{F}(F,k,n,m) = 0$ is approximately equivalent to $2m + n\ln k = 0$, whose solution is
\begin{align}
    n^* = -\frac{2m}{\ln k} \approx \frac{2m}{1-k},
\end{align}
as mentioned in the main text.

\subsection{Optimal number of sub-ensembles}
Now we take the partial derivative of $\mathcal{F}(F,k,n,m)$ with respect to $m$ and perform analytical approximations based on similar assumptions to the above.
\begin{align}
    &\frac{\left(2^{\frac{n}{m}}\cancel{-1}\right)^2k\left[\cancel{k + }F\left(2^{\frac{n}{m}}\cancel{-2}\right)k^{\frac{n}{m}}\right]^2m^3}{\left[\cancel{k} - F(2k)^{\frac{n}{m}}\right]n^2} \left[\frac{\partial}{\partial m}\mathcal{F}(F,k,n,m)\right]\nonumber\\
    &= Fk^{1+\frac{n}{m}}\left[2^{\frac{n}{m}}(2m-5n\ln 2) + 3\times 4^{\frac{n}{m}}n\ln 2\cancel{ - 2m}\right] + n\left(2^{\frac{n}{m}}\cancel{-1}\right)Fk^{\frac{n}{m}}\left[\left(3\times 2^{\frac{n}{m}}\cancel{-2}\right)k + \left(2^{\frac{n}{m}}\cancel{-2}\right)F(2k)^{\frac{n}{m}}\right]\ln k\nonumber\\
    &~~~~+ k^2\left[\cancel{m + }2^{\frac{n}{m}}(n\ln 2-m)\right] + F^2(2k^2)^{\frac{n}{m}}\left[4^\frac{n}{m}m \cancel{+ 2(m+2n\ln 2)} - 3\times 2^{\frac{n}{m}}(m+n\ln 2)\right]\\
    \Rightarrow& \frac{Fkm^3}{n^2}(4k)^\frac{n}{m}\left[\frac{\partial}{\partial m}\mathcal{F}(F,k,n,m)\right] \approx \cancel{k^2(n\ln 2-m)} - F^2\left(2k^2\right)^\frac{n}{m}\left[2^\frac{n}{m}(m + n\ln k)\cancel{ - 3m - 3n\ln 2}\right]\nonumber\\
    &~~~~~~~~~~~~~~~~~~~~~~~~~~~~~~~~~~~~~~~~~~~~~~~~~~~~~~~~~~~~~ - Fk^{1+\frac{n}{m}}\left[\cancel{2m - 5n\ln 2 + }2^\frac{n}{m}\ln(8k^3)\right]\\
    \Rightarrow& \frac{\partial}{\partial m}\mathcal{F}(F,k,n,m)\approx - \frac{n^2}{m^3}\left[Fk^{\frac{n}{m}-1}(m + n\ln k)\cancel{ + 2^{-\frac{n}{m}}n\ln(8k^3)} \right]\\
    \Rightarrow& \frac{\partial}{\partial m}\mathcal{F}(F,k,n,m)\approx -\frac{Fn^2k^{\frac{n}{m}-1}}{m^3}(m + n\ln k).
\end{align}
Then the equation $\frac{\partial}{\partial m}\mathcal{F}(F,k,n,m) = 0$ is approximately equivalent to $m + n\ln k = 0$, whose solution is
\begin{align}
    m^* = -n\ln k \approx n(1-k),
\end{align}
as mentioned in the main text.

\subsection{Advantage of partitioning}
We present the explicit expression of the ratio between optimally partitioned QFI and the monolithic QFI, using the closed-form approximation of $m^*$.
\begin{align}
    \frac{\mathcal{F}(F,k,n,m^*)}{\mathcal{F}(F,k,n)} \approx& \frac{(2^n\cancel{-1})\left(F\cancel{ - 2^\frac{1}{\ln k}ek}\right)^2[Fk^n(2^n\cancel{-2})\cancel{+k}]}{e\left(\cancel{2^\frac{1}{\ln k}}-1\right)[\cancel{k}-F(2k)^n]^2\left[F\cancel{ + (ek-2F)2^\frac{1}{\ln k}}\right]n\ln k} \nonumber\\
    \approx& -\frac{1}{e}\frac{k^{-n}}{n\ln k},
\end{align}
which is the one shown in the main text. We have used the fact that in practice $k\lesssim 1$ so that $\ln k\lesssim 0$, which makes $2^\frac{1}{\ln k}\ll 1$.

We also present the explicit expression of the ratio between optimally partitioned QFI and the monolithic QFI using the optimal total number of sensors, where we use the closed-form approximation of $n^*$.
\begin{align}
    \frac{\mathcal{F}(F,k,n,m^*)}{\mathcal{F}(F,k,n^*)} \approx& \frac{e\left(2^\frac{1}{\ln k} \cancel{+ 1}\right)\left(F\cancel{ - 2^\frac{1}{\ln k}ek}\right)^2\left[\cancel{(2F-e^2k)4^\frac{1}{\ln k}} - F\right]n\ln k}{4\left(F\cancel{ - 4^\frac{1}{\ln k}e^2k}\right)^2\left[F\cancel{ + (ek-2F)2^\frac{1}{\ln k}}\right]} \nonumber\\
    \approx& -\frac{e\ln k}{4} n,
\end{align}
which is the one shown in the main text. We have used the same assumption as above.

\subsection{Equal partition}
To justify the equal partition from concavity of $\mathcal{F}(F,k,n)$, we evaluate its second-order partial derivative with respect to $n$. After a tedious derivation, one is able to show
\begin{align}
    &-k(1-2^n)^3[(2^n-2)Fk^n + k]^3\left[\frac{\partial^2}{\partial n^2}\mathcal{F}(F,k,n)\right]\nonumber\\
    &= \left\{(2^n+1)(2^nk^nF-k)^2\ln^22 + 2(2^n-1)k^nF\ln 2k \left[2^nk\ln\frac{2}{k} + 2^{n+1}k^nF(2^n\ln k + \ln 2k) + k\ln 2k\right]\right\}\nonumber\\
    & ~~~~~~~~\times n^22^n[(2^n-2)k^nF + k]^2\nonumber\\
    & ~~~~+ \left\{[(2^n-2)Fk^n + k]\ln 2 + 2(2^n-1)k^nF\ln k \right\}\left\{2k - k^nF[4 - 2n\ln k + 2^n(n\ln 2k-2)]\right\}\nonumber\\
    & ~~~~~~~~\times n 2^{n+1}(2^n-1)(2^nk^nF-k)[(2^n-2)Fk^n + k]\nonumber\\
    & ~~~~+ (2^n-1)^2(2^nk^nF-k)^2\left\{2[(2^n-2)Fk^n + k]^2 - 4nk^nF[(2^n-2)Fk^n + k](2^n\ln 2k - 2\ln k) \right.\nonumber\\
    & ~~~~~~~~~~~~~~~~~~~~~~~~~~~~~~~~~~~~+ \left. n^2k^nF[2k^nF(2^n\ln 2k - 2\ln k)^2 - [(2^n-2)k^nF + k](2^n\ln^2 2k-2\ln^2 k)] \right\}
\end{align}
It is obvious that the factor on the left hand side of the above equation is positive in the regime of our interest. Then to determine the sign of the second-order partial derivative, we only need to evaluate the sign of the right hand side of the above equation. Explicitly we have the following analytical approximation
\begin{align}
    &\left\{(2^n\cancel{+1})(2^nk^nF\cancel{-k})^2\ln^22 + 2(2^n\cancel{-1})k^nF\ln 2k \left[2^nk\ln\frac{2}{k} + 2^{n+1}k^nF(2^n\ln k\cancel{ + \ln 2k})\cancel{ + k\ln 2k}\right]\right\}\nonumber\\
    & ~~~~~~~~\times n^22^n[(2^n\cancel{-2})k^nF\cancel{ + k}]^2\nonumber\\
    & ~~~~+ \left\{[(2^n\cancel{-2})Fk^n\cancel{ + k}]\ln 2 + 2(2^n\cancel{-1})k^nF\ln k \right\}\left\{\cancel{2k} - k^nF[\cancel{4 - 2n\ln k + }2^n(n\ln 2k-2)]\right\}\nonumber\\
    & ~~~~~~~~\times n 2^{n+1}(2^n\cancel{-1})(2^nk^nF\cancel{-k})[(2^n\cancel{-2})Fk^n\cancel{ + k}]\nonumber\\
    & ~~~~+ (2^n\cancel{-1})^2(2^nk^nF\cancel{-k})^2\left\{2[(2^n\cancel{-2})Fk^n\cancel{ + k}]^2 - 4nk^nF[(2^n\cancel{-2})Fk^n\cancel{ + k}](2^n\ln 2k\cancel{ - 2\ln k}) \right.\nonumber\\
    & ~~~~~~~~~~~~~~~~~~~~~~~~~~~~~~~~~~~~+ \left. n^2k^nF[2k^nF(2^n\ln 2k\cancel{ - 2\ln k})^2 - [(2^n\cancel{-2})k^nF\cancel{ + k}](2^n\ln^2 2k\cancel{-2\ln^2 k})] \right\} \nonumber\\
    \approx & 32^nk^{3n}F^3\left[\cancel{2kn^2(\ln^2 2 - \ln^2 k) + }2^nk^nF(2 + 4n\ln k + n^2\ln^2 k)\right]\\
    \approx & 2^{6n}k^{4n}F^4 (2 + 4n\ln k + n^2\ln^2 k),
\end{align}
where we consider in practice $k\lesssim 1$ and $n$ is not small, and we only omit exponentially suppressed terms. Then the root of the original right hand side expression is approximately determined by $2 + 4n\ln k + n^2\ln^2 k = 0$, which can be solved as
\begin{align}
    & n_\mathrm{low} \approx -\frac{2-\sqrt{2}}{\ln k},\\
    & n_\mathrm{up} \approx -\frac{2+\sqrt{2}}{\ln k},
\end{align}
as mentioned in the main text. It can be straightforwardly verified that $\mathcal{F}(F,k,n)$ is concave in $n$ for $n\in[n_\mathrm{low}, n_\mathrm{up}]$ to a good approximation.

\section{Derivations for the scenario with state preparation errors and particle losses}
\subsection{Effect of particle losses}
We first demonstrate that the result of any particle loss pattern is a GHZ-diagonal state with $\lambda_a=\lambda_b, \forall (a,b)\in\mathcal{S}$. Consider a GHZ basis state $(|i\rangle+|\overline{i}\rangle)/\sqrt{2}$, where $|i\rangle$ is an arbitrary computational basis state, and $|\overline{i}\rangle$ is the computational basis state corresponding to flipping every digit of $|i\rangle$. As mentioned in Sec.~\ref{sec:model}, if any qubits are lost from such a GHZ basis state under the $z$-direction phase accumulation, the resulting state for the remaining qubits is $(|i'\rangle\langle i'|+|\overline{i'}\rangle\langle \overline{i'}|)/2$, where $|i'\rangle(|\overline{i'}\rangle)$ is the computational basis state after removing the lost qubits from $|i\rangle(|\overline{i}\rangle)$. As we do not have any access to the lost qubits, we may represent their states by a maximally mixed state $(|0\rangle\langle 0| + |1\rangle\langle 1|)/2 = I/2$. Furthermore, we can also append the lost qubits back to the state of the remaining qubits through tensor product. For instance, consider the simplest 3-qubit GHZ state $(|000\rangle+|111\rangle)/\sqrt{2}$. Suppose the first qubit is lost, then the resulting state is 
\begin{align}
    \frac{1}{\sqrt{2}}(|000\rangle+|111\rangle) \xrightarrow[]{\text{1st qubit lost}}& \frac{1}{2}(|0\rangle\langle 0| + |1\rangle\langle 1|) \otimes \frac{1}{2}(|00\rangle\langle 00| + |11\rangle\langle 11|)\nonumber\\
    =& \frac{1}{4}(|000\rangle\langle 000| + |100\rangle\langle 100| + |011\rangle\langle 011| + |111\rangle\langle 111|)\nonumber\\
    =& \frac{1}{4}\left[\frac{1}{2}(|000\rangle+|111\rangle)(\langle 000|+\langle 111|) + \frac{1}{2}(|000\rangle-|111\rangle)(\langle 000|-\langle 111|)\right.\nonumber\\
    &~~~~\left.+ \frac{1}{2}(|011\rangle+|100\rangle)(\langle 011|+\langle 100|) + \frac{1}{2}(|011\rangle-|100\rangle)(\langle 011|-\langle 100|)\right]\nonumber\\
    =& \frac{1}{4}(|\mathrm{GHZ}^+_{000}\rangle\langle\mathrm{GHZ}^+_{000}| + |\mathrm{GHZ}^-_{000}\rangle\langle\mathrm{GHZ}^-_{000}| + |\mathrm{GHZ}^+_{011}\rangle\langle\mathrm{GHZ}^+_{011}| + |\mathrm{GHZ}^-_{011}\rangle\langle\mathrm{GHZ}^-_{011}|),
\end{align}
where the notation on the last line should be self-explanatory, while we may also use $|\psi_{a}\rangle$ to denote general GHZ basis states labeled by $a$ under arbitrary indexing method. One can see that the resulting state is indeed a GHZ-diagonal state which satisfies $\lambda_a=\lambda_b, \forall (a,b)\in\mathcal{S}$. Similarly, it can be straightforwardly verified that the total $n$-qubit state after any qubit losses will be an equal mixture of computational basis states, while $|i\rangle$ and $|\overline{i}\rangle$ always appear in pairs, thus always satisfying $\lambda_a=\lambda_b, \forall (a,b)\in\mathcal{S}$.

As we have assumed that the initial probe state is in GHZ-diagonal form $\rho_0 = \sum_a\lambda_{a}|\psi_{a}\rangle\langle\psi_{a}|$ (more specifically a depolarized GHZ state), then according to linearity the resulting state after particle losses should be
\begin{align}
    \rho' =& \mathcal{E}_\mathrm{loss}(\rho_0)\nonumber\\
    =& \sum_{l\in\mathcal{L}}p_l\mathcal{E}_l\left(\sum_a\lambda_{a}|\psi_{a}\rangle\langle\psi_{a}|\right)\nonumber\\
    =& p_\mathrm{no-loss}\rho_0 + \sum_a\sum_{l\in\overline{\mathcal{L}}}\lambda_{a}p_l\mathcal{E}_l\left(|\psi_{a}\rangle\langle\psi_{a}|\right),
\end{align}
where $\mathcal{L}$ denotes the set of all possible loss patterns which also includes the trivial no-loss pattern, and $\overline{\mathcal{L}}$ thus denotes the set of all possible loss patterns excluding the trivial no-loss pattern. Also $l\in\mathcal{L}$ denotes a specific loss pattern, while $p_l$ represents the probability for the loss pattern $l$ to occur. Under our assumption that each sensor qubit has equal probability $p$ to survive the encoding process, then the probability for an $n$-qubit GHZ state to survive as a whole is simply $p_\mathrm{no-loss} = p^n$.

Then we examine the impact of particle losses on the QFI. Using the QFI expression for GHZ-diagonal initial probe state we have
\begin{align}
    \mathcal{F}_\mathrm{loss} =& n^2\sum_{(a,b)\in\mathcal{S}}\frac{(\lambda_{a}'-\lambda_{b}')^2}{\lambda_{a}'+\lambda_{b}'}\nonumber\\
    =& n^2\sum_{(a,b)\in\mathcal{S}}\frac{(p_\mathrm{no-loss}\lambda_{a} + \delta\lambda_{a} - p_\mathrm{no-loss}\lambda_{b} - \delta\lambda_{b})^2}{p_\mathrm{no-loss}\lambda_{a} + \delta\lambda_{a} + p_\mathrm{no-loss}\lambda_{b} + \delta\lambda_{b}}\nonumber\\
    =& n^2\sum_{(a,b)\in\mathcal{S}}\frac{p_\mathrm{no-loss}^2(\lambda_{a} - \lambda_{b})^2}{p_\mathrm{no-loss}(\lambda_{a} + \lambda_{b}) + 2\delta\lambda_{a}}
\end{align}
where
\begin{align}
    \delta\lambda_{a/b} = \langle\psi_{a/b}|\left[\sum_a\sum_{l\in\overline{\mathcal{L}}}p_l\lambda_{a'}\mathcal{E}_l(|\psi_{a'}\rangle\langle\psi_{a'}|)\right]|\psi_{a/b}\rangle,
\end{align}
and we have also used the fact that $\delta\lambda_a=\delta\lambda_b$, given the assumption that the initial state is GHZ-diagonal and the above discussion on the effect of particle losses on GHZ basis states.

Getting back to our specific example of depolarized GHZ states, in the sum for QFI we still only have one term which corresponds to $(|00\dots 0\rangle\pm|11\dots 1\rangle)/\sqrt{2}$, and this is why we only need to care about the eigenvalue of $(|00\dots 0\rangle+|11\dots 1\rangle)/\sqrt{2} = |\mathrm{GHZ}_n\rangle$. Also notice that the maximally mixed state of $n$ qubits is a fixed point of the particle loss channel $\mathcal{E}_\mathrm{loss}$, and also $\sum_{l\in\mathcal{L}}p_l\mathcal{E}_l$. Therefore, we can write
\begin{align}
    &\sum_{l\in\overline{\mathcal{L}}}p_l\mathcal{E}_l\left(\sum_a\lambda_{a}|\psi_{a}\rangle\langle\psi_{a}|\right) = (\mathcal{E}_\mathrm{loss} - p_\mathrm{no-loss}\mathrm{Id})\left(\sum_a\lambda_{a}|\psi_{a}\rangle\langle\psi_{a}|\right)\nonumber\\
    =& \mathcal{E}_\mathrm{loss}\left[F(n)|\mathrm{GHZ}_n\rangle\langle\mathrm{GHZ}_n| + (1-F(n))I/2^n\right] - p_\mathrm{no-loss}[F(n)|\mathrm{GHZ}_n\rangle\langle\mathrm{GHZ}_n| + (1-F(n))I/2^n]\nonumber\\
    =& F(n)\left[\mathcal{E}_\mathrm{loss}(|\mathrm{GHZ}_n\rangle\langle\mathrm{GHZ}_n|) - p_\mathrm{no-loss}|\mathrm{GHZ}_n\rangle\langle\mathrm{GHZ}_n|\right] + (1-F(n))(1-p_\mathrm{no-loss})I/2^n,
\end{align}
where $F(n) = k^{n-1}F$ is the fidelity of the initially prepared $n$-qubit noisy GHZ state in the depolarized form, and $\mathrm{Id}$ denotes the identity channel that does not change the quantum system. Therefore, we only need to work out $\sum_{l\in\overline{\mathcal{L}}}p_l\mathcal{E}_l(|\mathrm{GHZ}_n\rangle\langle\mathrm{GHZ}_n|)$.

Suppose that $m$ qubits in $|\mathrm{GHZ}_n\rangle$ are lost, it is straightforward to verify that the resulting states will be an equal mixture of $2^{m+1}$ computational basis states, while $|i\rangle$ and $|\overline{i}\rangle$ exist in pairs. After normalization, the eigenvalue of $|\mathrm{GHZ}_n\rangle$ in such a mixed state is $\lambda_{m-\mathrm{lost}} = 2^{-(m+1)}$. Therefore, we have
\begin{align}
    \langle\mathrm{GHZ}_n|\left[\sum_{l\in\overline{\mathcal{L}}}p_l\mathcal{E}_l(|\mathrm{GHZ}_n\rangle\langle\mathrm{GHZ}_n|)\right]|\mathrm{GHZ}_n\rangle =& \sum_{m=1}^{n}p^{n-m}(1-p)^m\binom{n}{m}2^{-(m+1)}\nonumber\\
    =& \frac{(1+p)^n - (2p)^n}{2^{n+1}} = \delta\lambda.
\end{align}

Combining all the above, we can derive the QFI of an $n$-qubit GHZ state under independent particle losses where each qubit has probability $p$ to survive as
\begin{align}
    \mathcal{F}_\mathrm{loss}\coloneqq& \mathcal{F}_{\mathrm{eff},1}(F,k,p,n)\nonumber\\
    =& n^2\frac{p^{2n}\left[F(n) - \frac{1-F(n)}{2^n-1}\right]^2}{p^n\left[F(n) + \frac{1-F(n)}{2^n-1}\right] + \frac{(1+p)^n - (2p)^n}{2^n}}\nonumber\\
    =& \frac{[F(2k)^n-k]^2n^2(2p^2)^n}{(2^n-1)k\left[(2^n-1)(1+p)^nk + (2p)^n(2^n-2)(k-Fk^n)\right]},
\end{align}
which reduces to the QFI expression without particle loss if we take $p=1$, thus validating the expression.

\subsection{Optimal number of sensors}
We take the partial derivative with respect to $n$ of the QFI. We first examine the second case which assumes loss detection capability.
\begin{align}
    &\frac{km^2\left(2^\frac{n}{m}\cancel{-1}\right)^2\left[\cancel{k + }F\left(2^\frac{n}{m}\cancel{-2}\right)k^\frac{n}{m}\right]}{n\left[F(2k)^\frac{n}{m}\cancel{ - k}\right]p^\frac{n}{m}}\left[\frac{\partial}{\partial n}\mathcal{F}_{\mathrm{loss},2}(F,k,p,n,m)\right]\nonumber\\
    &= k^2\left[2^\frac{n}{m}\left(n\ln\frac{2}{p}-2m\right)\cancel{ + n\ln p + 2m}\right] + Fk^\frac{m+n}{m}\left[4^\frac{n}{m}n\ln(8k^3) + 2^\frac{n}{m}(4m -5n\ln(2k)+2n\ln p)\cancel{ + 2n\ln\frac{k}{p} - 4m}\right] \nonumber\\
    &~~~~ + F^2(2k^2)^\frac{n}{m}\left[4^\frac{n}{m}(n\ln(kp) + 2m) - 3\times 2^\frac{n}{m}(n\ln(2kp) + 2m)\cancel{ + 2n\ln(kp) + n\ln 16 + 4m}\right] \\
    \Rightarrow& \frac{Fkm^2(4k)^\frac{n}{m}}{np^\frac{n}{m}}\left[\frac{\partial}{\partial n}\mathcal{F}_{\mathrm{loss},2}(F,k,p,n,m)\right] \approx \cancel{k^2\left(2m + n\ln\frac{p}{2}\right)} - Fk^\frac{m+n}{m}\left[\cancel{4m - 5n\ln(2k) + }2^\frac{n}{m}n\ln(8k^3)\cancel{ + 2n\ln p}\right]\nonumber\\
    &~~~~ - F^2(k^2)^\frac{n}{m}\left[2^\frac{n}{m}(n\ln(kp) + 2n)\cancel{ - 3(n\ln(2kp) + 2m)}\right]\\
    \Rightarrow& \frac{Fkm^2(4k)^\frac{n}{m}}{np^\frac{n}{m}}\left[\frac{\partial}{\partial n}\mathcal{F}_{\mathrm{loss},2}(F,k,p,n,m)\right] \approx - F(2k)^\frac{n}{m}\left[\cancel{kn\ln(8k^3) + }F(2k)^\frac{n}{m}(2m + n\ln k + n\ln p) \right]\\
    \Rightarrow& \frac{\partial}{\partial n}\mathcal{F}_{\mathrm{loss},2}(F,k,p,n,m) \approx -\frac{Fn}{km^2}(kp)^\frac{n}{m}(2m + n\ln k + n\ln p),
\end{align}
where in the above for the approximations we only omit the exponentially suppressed terms. Then according to the above the equation $\frac{\partial}{\partial n}\mathcal{F}_{\mathrm{loss},2}=0$ is approximately equivalent to $2m + n\ln k + n\ln p = 0$, whose solution is 
\begin{align}
    n_{\mathrm{loss},2}^*\approx -\frac{2m}{\ln k + \ln p} \approx \frac{2m}{2 - k - p},
\end{align}
as mentioned in the main text.

Then we consider the first case without loss detection capability. 
\begin{align}
    & \frac{k\left(2^\frac{n}{m} \cancel{- 1}\right)^3\left(\cancel{1} - 2^\frac{n}{m}\right)\left[km\left(\frac{1+p}{2}\right)^\frac{n}{m} + \frac{m}{2}p^\frac{n}{m}\left(k - Fk^\frac{n}{m}\right)\left(\cancelto{1}{\coth\frac{n\ln 2}{2m}} - 3\right) \right]}{np^\frac{2n}{m}\left[\cancel{k} - F(2k)^\frac{n}{m}\right]} \left[\frac{\partial}{\partial n}\mathcal{F}_{\mathrm{loss},2}(F,k,p,n,m)\right]\nonumber\\
    &= 2^m\left(2^\frac{n}{m} \cancel{- 1}\right)\left[\cancel{k} - F(2k)^\frac{n}{m}\right]\left[p^\frac{n}{m}\left(2^\frac{n}{m} \cancel{- 2}\right)\left(k - Fk^\frac{n}{m}\right) - k\left(2^\frac{n}{m} \cancel{- 1}\right)\left(\frac{1+p}{2}\right)^\frac{n}{m} \right]\nonumber\\
    &~~~~ + 2n\left[k\left(2^\frac{n}{m} \cancel{- 1}\right)(1+p)^\frac{n}{m} - (2p)^\frac{n}{m}\left(2^\frac{n}{m} \cancel{- 2}\right)\left(k - Fk^\frac{n}{m}\right)\right]\left[k\ln 2 + Fk^\frac{n}{m}\left(2^\frac{n}{m}\ln k \cancel{- \ln 2k}\right) \right]\nonumber\\
    &~~~~ + 2n\ln p\left(2^\frac{n}{m} \cancel{- 1}\right)\left[\cancel{k} - F(2k)^\frac{n}{m}\right]\left[p^\frac{n}{m}\left(2^\frac{n}{m} \cancel{- 2}\right)\left(k - Fk^\frac{n}{m}\right) - k\left(2^\frac{n}{m} \cancel{- 1}\right)\left(\frac{1+p}{2}\right)^\frac{n}{m} \right]\nonumber\\
    &~~~~ + n\left[\cancel{k} - F(2k)^\frac{n}{m}\right]\left[F(2kp)^\frac{n}{m}\ln 2 + F(kp)^\frac{n}{m}\left(4^\frac{n}{m} - 3\times 2^\frac{n}{m} \cancel{+ 2}\right)\ln kp - k(2p)^\frac{n}{m}\ln\frac{3}{p^3} - kp^\frac{n}{m}\left(4^\frac{n}{m} \cancel{+ 2}\right)\ln p \right]\nonumber\\
    &~~~~ + kn\left(2^\frac{n}{m} \cancel{- 1}\right)^2\left[\cancel{k} - F(2k)^\frac{n}{m}\right]\left(\frac{1+p}{2}\right)^\frac{n}{m}\ln\frac{1+p}{2}\\
    \Rightarrow& \frac{km^2\left[k(1+p)^\frac{n}{m} - k(2p)^\frac{n}{m} + F(2kp)^\frac{n}{m}\right]^2}{F^2n(2k^2p^2)^\frac{n}{m}} \left[\frac{\partial}{\partial n}\mathcal{F}_{\mathrm{loss},2}(F,k,p,n,m)\right] \approx - p^\frac{n}{m}\left(k - Fk^\frac{n}{m}\right)\left(2^\frac{m+n}{m}m \cancel{- n\ln 2}\right)\nonumber\\
    &~~~~ + 2km(1+p)^\frac{n}{m} + n\left[F\left(2^\frac{n}{m} \cancel{- 3}\right)(kp)^\frac{n}{m}\ln kp - kp^\frac{n}{m}\left(2^\frac{n}{m}\ln k^2p \cancel{+ 3\ln p}\right) + k(1+p)^\frac{n}{m}\ln\frac{2k^2p^2}{1+p} \right]\\
    \Rightarrow& \frac{km^2\left[k(1+p)^\frac{n}{m} - k(2p)^\frac{n}{m} + F(2kp)^\frac{n}{m}\right]^2}{F^2n(2k^2p^2)^\frac{n}{m}} \left[\frac{\partial}{\partial n}\mathcal{F}_{\mathrm{loss},2}(F,k,p,n,m)\right]\nonumber\\
    &\approx F(2kp)^\frac{n}{m}\left(2m + n\ln kp\right) - k(2p)^\frac{n}{m}\left(2m + n\ln k^2p\right) + k(1+p)^\frac{n}{m}\left(2m + n\ln \frac{2k^2p^2}{1+p}\right),
\end{align}
where we consider that $\coth x\approx 1$ for $x\gg 1$. Also note that when we focus on the optimal total number of sensors, $Fk^\frac{n}{m}\lesssim k$ should not be simply omitted. Then according to the above the equation $\frac{\partial}{\partial n}\mathcal{F}_{\mathrm{loss},1}=0$ is approximately equivalent
\begin{align}
    0 = F(2kp)^\frac{n}{m}\left(2m + n\ln kp\right) - k(2p)^\frac{n}{m}\left(2m + n\ln k^2p\right) + k(1+p)^\frac{n}{m}\left(2m + n\ln \frac{2k^2p^2}{1+p}\right),
\end{align}
as mentioned in the main text.

\subsection{Optimal number of sub-ensembles}
We first examine the second case which assumes loss detection capability.
\begin{align}
    &\frac{km^3\left(2^\frac{n}{m} \cancel{- 1}\right)^2\left[\cancel{k} + Fk^\frac{n}{m}\left(2^\frac{n}{m} \cancel{- 2}\right)\right]^2}{kn^2p^\frac{n}{m}\left[\cancel{1} - (2k)^\frac{n}{m}\right]}\left[\frac{\partial}{\partial m}\mathcal{F}_{\mathrm{loss},2}(F,k,p,n,m)\right]\nonumber\\
    &= k^2\left[\cancel{m + n\ln p} + 2^\frac{n}{m}\left(n\ln\frac{2}{p} - m\right)\right] + fk^\frac{m+n}{m}\left[4^\frac{n}{m}n\ln(8k^3) + 2^\frac{n}{m}(2m + 2n\ln p - 5n\ln 2k) + \cancel{2n\ln\frac{k}{p} - 2m}\right]\nonumber\\
    &~~~~ + F^2(2k^2)^\frac{n}{m}\left[4^\frac{n}{m}(m + n\ln kp) - 3\times 2^\frac{n}{m}(m + n\ln 2kp) + \cancel{2(m + n\ln 4kp)}\right]\\
    \Rightarrow& \frac{-kFm^3(4k)^\frac{n}{m}}{n^2p^\frac{n}{m}}\left[\frac{\partial}{\partial m}\mathcal{F}_{\mathrm{loss},2}(F,k,p,n,m)\right] \approx \cancel{k^2(n\ln\frac{2}{p} - m)} + Fk^\frac{m+n}{m}\left[2^\frac{n}{m} n\ln(8k^3) \cancel{ - 5n\ln 2k + 2(m + n\ln p)} \right]\nonumber\\
    &~~~~ + F^2k^\frac{2n}{m}\left[4^\frac{n}{m}(m+n\ln kp) - 3\times 2^\frac{n}{m}(m + n\ln 2kp) \right]\\
    \Rightarrow& \frac{-kFm^3(4k)^\frac{n}{m}}{n^2p^\frac{n}{m}}\left[\frac{\partial}{\partial m}\mathcal{F}_{\mathrm{loss},2}(F,k,p,n,m)\right] \approx F(2k)^\frac{n}{m}\left[\cancel{kn\ln(8k^3)} + F(2k)^\frac{n}{m}(m + n\ln kp) \cancel{- 3Fk^\frac{n}{m}(m + n\ln 2kp)}\right]\\
    \Rightarrow& \frac{-kFm^3(4k)^\frac{n}{m}}{n^2p^\frac{n}{m}}\left[\frac{\partial}{\partial m}\mathcal{F}_{\mathrm{loss},2}(F,k,p,n,m)\right] \approx F^2F(2k)^\frac{2n}{m}[m + n(\ln k +\ln p)],
\end{align}
where in the above for the approximations we only omit the exponentially suppressed terms. Then according to the above the equation $\frac{\partial}{\partial m}\mathcal{F}_{\mathrm{loss},2}=0$ is approximately equivalent to $m + n\ln k + n\ln p = 0$, whose solution is 
\begin{align}
    m_{\mathrm{loss},2}^*\approx -n(\ln k + \ln p) \approx n(2 - k - p),
\end{align}
as mentioned in the main text.

Then we consider the first case without loss detection capability. 
\begin{align}
    &km^3\left(\cancel{1} - 2^\frac{n}{m}\right)\left[\left(\frac{1+p}{2}\right)^\frac{n}{m} + \frac{p^\frac{n}{m}\left(k - Fk^\frac{n}{m}\right)\left(\cancelto{1}{\coth\frac{n\ln 2}{2m}}-3\right)}{2k}\right]^2\left[\frac{\partial}{\partial m}\mathcal{F}_{\mathrm{loss},1}(F,k,p,n,m)\right]\nonumber\\
    &= n^2\left[\cancel{k} - F(2k)^\frac{n}{m}\right]p^\frac{2n}{m}\left\{\frac{m\left[F(2k)^\frac{n}{m} - k\right]\left[\left(\frac{1+p}{2}\right)^\frac{n}{m} + \frac{p^\frac{n}{m}\left(k - Fk^\frac{n}{m}\right)\left(\cancelto{1}{\coth\frac{n\ln 2}{2m}}-3\right)}{2k}\right]}{k\left(\cancel{1} - 2^\frac{n}{m}\right)}  \right.\nonumber\\
    &~~~~~~~~ + \frac{2n\left[2^\frac{n}{m}\left(2^\frac{n}{m} \cancel{- 2}\right)\left(k - Fk^\frac{n}{m}\right)p^\frac{n}{m} - k\left(2^\frac{n}{m} \cancel{- 1}\right)(1+p)^\frac{n}{m} \right]\left[k\ln 2 + Fk^\frac{n}{m}\left(2^\frac{n}{m}\ln k - \ln 2k\right) \right]}{k^2\left(2^\frac{n}{m} \cancel{- 1}\right)^3}\nonumber\\
    &~~~~~~~~ + \frac{2n\ln p\left[\cancel{k} - F(2k)^\frac{n}{m}\right]\left[k\left(2^\frac{n}{m} \cancel{- 1}\right)\left(\frac{1+p}{2}\right)^\frac{n}{m} - \left(2^\frac{n}{m} \cancel{- 2}\right)\left(k - Fk^\frac{n}{m}\right)p^\frac{n}{m} \right]}{k^2\left(2^\frac{n}{m} \cancel{- 1}\right)^2}\nonumber\\
    &~~~~~~~~ + \frac{n\left[F(2k)^\frac{n}{m} \cancel{- k}\right]}{k^2\left(2^\frac{n}{m} \cancel{- 1}\right)^3}\left[F(2kp)^\frac{n}{m}\ln 2 + F(kp)^\frac{n}{m}(4^\frac{n}{m} - 3\times 2^\frac{n}{m} \cancel{+ 2})\ln kp - kp^\frac{n}{m}\left(2^\frac{n}{m}\ln\frac{2}{p^3} + (4^\frac{n}{m} \cancel{+ 2})\ln p\right) \right.\nonumber\\
    &~~~~~~~~~~~~~~~~~~~~~~~~~~~~~~~~  \left. \left. - k\left(2^\frac{n}{m} \cancel{- 1}\right)^2\left(\frac{1+p}{2}\right)^\frac{n}{m}\ln\frac{2}{1+p} \right] \right\}\\
    \Rightarrow& \frac{km^3\left[k(1+p)^\frac{n}{m} - k(2p)^\frac{n}{m} + F(2kp)^\frac{n}{m}\right]^2}{F^2n^2(2k^2p^2)^\frac{n}{m}}\left[\frac{\partial}{\partial m}\mathcal{F}_{\mathrm{loss},1}(F,k,p,n,m)\right] \approx p^\frac{n}{m}\left(k - Fk^\frac{n}{m}\right)\left(2^\frac{n}{m}m \cancel{- n\ln 2}\right)\nonumber\\
    &~~~~ + knp^\frac{n}{m}\left(2^\frac{n}{m}\ln k^2p \cancel{+ 3\ln p}\right) - km(1+p)^\frac{n}{m} - Fn\left(2^\frac{n}{m} \cancel{+ 3}\right)(kp)^\frac{n}{m}\ln kp - kn(1+p)^\frac{n}{m}\ln \frac{2k^2p^2}{1+p}\\
    \Rightarrow& \frac{km^3\left[k(1+p)^\frac{n}{m} - k(2p)^\frac{n}{m} + F(2kp)^\frac{n}{m}\right]^2}{F^2n^2(2k^2p^2)^\frac{n}{m}}\left[\frac{\partial}{\partial m}\mathcal{F}_{\mathrm{loss},1}(F,k,p,n,m)\right]\nonumber\\
    &\approx - F(2kp)^\frac{n}{m}\left(m + n\ln kp\right) + k(2p)^\frac{n}{m}\left(m + n\ln k^2p\right) - k(1+p)^\frac{n}{m}\left(m + n\ln \frac{2k^2p^2}{1+p}\right),
\end{align}
where we consider that $\coth x\approx 1$ for $x\gg 1$. Also note that when we focus on the optimal total number of sensors, $Fk^\frac{n}{m}\lesssim k$ should not be simply omitted. Then according to the above the equation $\frac{\partial}{\partial m}\mathcal{F}_{\mathrm{loss},1}=0$ is approximately equivalent
\begin{align}
    0 = F(2kp)^\frac{n}{m}\left(m + n\ln kp\right) - k(2p)^\frac{n}{m}\left(m + n\ln k^2p\right) + k(1+p)^\frac{n}{m}\left(m + n\ln \frac{2k^2p^2}{1+p}\right),
\end{align}
as mentioned in the main text.

\subsection{Advantage of loss detection}
In this section we compare the sensing performance in the two considered cases, that is, $\mathcal{F}_{\mathrm{eff},1}(F,k,p,n/m)$ and $p^{n/m}\mathcal{F}(F,k,n/m)$. 

According to previous derivation details, we have 
\begin{align}
    \mathcal{F}_{\mathrm{eff},1}(F,k,p,n) =& n^2\frac{p^{2n}\left[F(n) - \frac{1-F(n)}{2^n-1}\right]^2}{p^n\left[F(n) + \frac{1-F(n)}{2^n-1}\right] + \frac{(1+p)^n - (2p)^n}{2^n}}\nonumber\\
    \leq& n^2\frac{p^{2n}\left[F(n) - \frac{1-F(n)}{2^n-1}\right]^2}{p^n\left[F(n) + \frac{1-F(n)}{2^n-1}\right]}\nonumber\\
    =& p^n \mathcal{F}(F,k,n),
\end{align}
where the equality only holds when $p=1$, i.e. there is no particle loss. Therefore, we naturally have $\mathcal{F}_{\mathrm{eff},1}(F,k,p,n/m) \leq p^{n/m}\mathcal{F}(F,k,n/m)$, which means
\begin{align}
    \mathcal{F}_1(F,k,p,n,m) \leq \mathcal{F}_2(F,k,p,n,m).
\end{align}

Then we quantitatively compare the QFI's for the two cases. Specifically, we may take the ratio between the two QFI's
\begin{align}
    \frac{\mathcal{F}_2(F,k,p,n,m)}{\mathcal{F}_1(F,k,p,n,m)} =& \frac{p^{n/m}\mathcal{F}(F,k,n/m)}{\mathcal{F}_{\mathrm{eff},1}(F,k,p,n/m)}\nonumber\\
    =& \frac{F\left(2^\frac{n}{m}\cancel{ - 2}\right)k^\frac{n}{m} + k\left[\left(\frac{1+p}{p}\right)^\frac{n}{m} - \left(\frac{1+p}{2p}\right)^\frac{n}{m} - 2^\frac{n}{m}\cancel{ + 2} \right]}{\cancel{k + }F\left(2^\frac{n}{m}\cancel{-2}\right)k^\frac{n}{m}}\nonumber\\
    \approx& 1 + \frac{k}{F}\left(\frac{1+p}{4kp}\right)^\frac{n}{m}\left[2^\frac{n}{m} - \left(\frac{4p}{1+p}\right)^\frac{n}{m}\cancel{ - 1}\right]\nonumber\\
    \approx& 1 + \frac{k}{F}\left(\frac{1+p}{4kp}\right)^\frac{n}{m}\left[2^\frac{n}{m} - \left(\frac{4p}{1+p}\right)^\frac{n}{m}\right],
\end{align}
where we have mainly omitted exponentially suppressed constant terms. Furthermore, for relatively large particle loss probability $1-p$, the above expression can be further approximated as
\begin{align}
    \frac{\mathcal{F}_2(F,k,p,n,m)}{\mathcal{F}_1(F,k,p,n,m)} \approx& 1 + \frac{k}{F}\left(\frac{1+p}{4kp}\right)^\frac{n}{m}\left[2^\frac{n}{m} \cancel{- \left(\frac{4p}{1+p}\right)^\frac{n}{m}}\right]\nonumber\\
    \approx& 1 + \frac{k}{F}\left(\frac{1+p}{2kp}\right)^\frac{n}{m}.
\end{align}

\subsection{Advantage of partitioning}
The explicit expression of the ratio between optimally partitioned QFI and the monolithic QFI, and corresponding approximation are as follows
\begin{align}
    \frac{\mathcal{F}_{\mathrm{loss},2}(F,k,p,n,m_{\mathrm{loss},2}^*)}{\mathcal{F}_{\mathrm{loss},2}(F,k,p,n)} \approx& \frac{\left(2^n \cancel{- 1}\right)\left[k + F\left(2^n \cancel{- 2}\right)k^n\right]\left[F \cancel{- k(2k)^\frac{1}{\ln k + \ln p}}\right]^2}{en(\ln k + \ln p)p^n\left(\cancel{2^\frac{1}{\ln k + \ln p}} - 1\right)\left[\cancel{k} - F(2k)^n\right]^2\left[F \cancel{- (2F)2^\frac{1}{\ln k + \ln p} + k(2k)^\frac{1}{\ln k + \ln p}}\right]}\nonumber\\
    \approx& -\frac{1}{e}\frac{(kp)^{-n}}{n(\ln k + \ln p)},
\end{align}
which is the one shown in the main text. We have used the fact that in practice $k,p\lesssim 1$ so that $\ln k,\ln p\lesssim 0$, which makes $x^\frac{1}{\ln k + \ln p}\ll 1$ for $x>1$.

The explicit expression of the ratio between optimally partitioned QFI and the monolithic QFI using the optimal total number of sensors is as follows
\begin{align}
    \frac{\mathcal{F}_{\mathrm{loss},2}(F,k,p,n,m_{\mathrm{loss},2}^*)}{\mathcal{F}_{\mathrm{loss},2}(F,k,p,n_{\mathrm{loss},2}^*)} \approx& \frac{-en(\ln k + \ln p)\left(\cancel{2^\frac{1}{\ln k + \ln p}} + 1\right)\left[F \cancel{- k(2k)^\frac{1}{\ln k + \ln p}}\right]^2\left[F \cancel{- (2F)4^\frac{1}{\ln k + \ln p} + k(4k^2)^\frac{1}{\ln k + \ln p}}\right]}{4\left[F \cancel{- (2F)2^\frac{1}{\ln k + \ln p} + k(2k)^\frac{1}{\ln k + \ln p}}\right]\left[F \cancel{- k(4k^2)^\frac{1}{\ln k + \ln p}}\right]^2}\nonumber\\
    \approx& -\frac{e(\ln k + \ln p)}{4} n,
\end{align}
which is the one shown in the main text. We have used the same assumptions as above.

\section{Derivations for the scenario with state preparation errors and dephasing}
\subsection{Spin-off: elementary representation of a hypergeometric function}
Recall that when deriving the effect of dephasing on sensing performance we encountered a series
\begin{align}
    \tilde{q} =& \sum_{\mathrm{even}~\#Z}\binom{n}{\#Z}p_{\#Z}\nonumber\\
    =& \sum_{i=0}^{\lfloor n/2\rfloor}\binom{n}{2i}q^{n-2i}(1-q)^{2i}\nonumber\\
    =& \frac{1 + (2q-1)^n}{2}.
\end{align}
The derivation of the series is straightforward. For the summation with only even terms in the binomial expansion, we use the following fact
\begin{align}
    (a+b)^n + (a-b)^n =& \sum_{i=0}^n\binom{n}{k}a^{n-k}b^k + \sum_{i=0}^n\binom{n}{k}a^{n-k}b^k(-1)^k\nonumber\\
    =& 2\sum_{\mathrm{even}\ k}\binom{n}{k}a^{n-k}b^k.
\end{align}
For the series of our interest, we have $a = q$ and $b = 1 - q$. Therefore, we have
\begin{align}
    \sum_{\mathrm{even}\ j}\binom{n}{j}q^{n-j}(1-q)^{j} =& \frac{1}{2}\left[(q + 1 - q)^n + (q - 1 + q)^n\right]\nonumber\\
    =& \frac{1 + (2q - 1)^n}{2}.
\end{align}

From the above series we accidentally (re)discovered an elementary representation of hypergeometric function (HGF) ${}_2F_1(a,b,c;z)$~\cite{egorychev1984integral} with a special set of parameters similar to the spin-off in~\cite{zang2025no}
\begin{align}
    {}_2F_1\left(\frac{1}{2}-\frac{n}{2},-\frac{n}{2};\frac{1}{2};\left(\frac{1-q}{q}\right)^2\right) = \frac{1 + (2q-1)^n}{2q^n},
\end{align}
which is a less standard result. Here we derive this result.

The (Gaussian/ordinary) HGF is defined as follows:
\begin{equation}
    {}_2F_1(a,b;c;z) = \sum_{n=0}^\infty\frac{(a)_n(b)_n}{(c)_n}\frac{z^n}{n!},
\end{equation}
where $(x)_n$ is the so-called rising factorial, i.e. $(x)_n=\prod_{k=0}^{n-1}(x+k)$ with $(x)_0=1$. Then we have
\begin{align}
     {}_2F_1\left(\frac{1}{2}-\frac{n}{2},-\frac{n}{2};\frac{1}{2};\left(\frac{1-q}{q}\right)^2\right) =& \sum_{i=0}^\infty\frac{\left(\frac{1-n}{2}\right)_i\left(-\frac{n}{2}\right)_i}{\left(\frac{1}{2}\right)_i}\frac{\left(\frac{1-q}{q}\right)^{2i}}{i!}\nonumber\\
     =& \sum_{i=0}^\infty\left[\prod_{j=0}^{i-1}\frac{\left(\frac{1-n}{2} + j\right)\left(-\frac{n}{2} + j\right)}{\left(\frac{1}{2} + j\right)}\right]\frac{\left(\frac{1-q}{q}\right)^{2i}}{i!}\nonumber.
\end{align}
Note that for both even and odd $n$, the continued multiplication will become zero for $i$ above a certain threshold. For even $n$, the multiplication will become zero for $i\geq n/2+1$ because $-n/2 + j=0$ for $j = n/2$, while for odd n, the the multiplication will become zero for $i\geq (n+1)/2$ because $(1-n)/2 + j=0$ for $j = (n-1)/2$. Then we have
\begin{align}
     {}_2F_1\left(\frac{1}{2}-\frac{n}{2},-\frac{n}{2};\frac{1}{2};\left(\frac{1-q}{q}\right)^2\right) =& \sum_{i=0}^{\lfloor n/2\rfloor}\left[\prod_{j=0}^{i-1}\frac{\left(\frac{1-n}{2} + j\right)\left(-\frac{n}{2} + j\right)}{\left(\frac{1}{2} + j\right)}\right]\frac{1}{i!}\left(\frac{1-q}{q}\right)^{2i}.
\end{align}
The term before $[(1-q)/q]^{2i}$ can be written as
\begin{align}
    \frac{1}{i!}\prod_{j=0}^{i-1}\frac{\left(\frac{1-n}{2} + j\right)\left(-\frac{n}{2} + j\right)}{\left(\frac{1}{2} + j\right)} =& \frac{1}{i!}\prod_{j=0}^{i-1}\frac{(2j + 1 - n)(2j - n)}{2(2j + 1)} \nonumber\\
    =& \prod_{j=0}^{i-1}\frac{(n - 2j - 1)(n - 2j)}{2(j+1)(2j + 1)}\nonumber\\
    =& \frac{(n-1)!!n!!}{(n-2i-1)!!(n-2i)!!(2i)!!(2i-1)!!}\nonumber\\
    =& \frac{n!}{(n-2i)!(2i)!} = \binom{n}{2i},
\end{align}
where $x!!=\sum_{k=0}^{\lceil n/2\rceil - 1}(n-2k)$ is the double factorial, and we have use the fact that $x!!(x-1)!!=x!$, for integer $x$. Then it is clear that
\begin{align}
    {}_2F_1\left(\frac{1}{2}-\frac{n}{2},-\frac{n}{2};\frac{1}{2};\left(\frac{1-q}{q}\right)^2\right) =& \sum_{i=0}^{n/2}\binom{n}{2i}q^{-2i}(1-q)^{2i}\nonumber\\
    =& q^{-n}\sum_{i=0}^{n/2}\binom{n}{2i}q^{n-2i}(1-q)^{2i}\nonumber\\
    =& \frac{1 + (2q-1)^n}{2q^n}.
\end{align}
It is noteworthy that even though the above derivation is for integer $n$, one can verify numerically the elementary representation of the HGF is also highly accurate for non-integer $n$. We thus generalize the above derivation to a conjecture:
\begin{conjecture}
    ${}_2F_1\left(\frac{1}{2}-\frac{n}{2},-\frac{n}{2};\frac{1}{2};\left(\frac{b}{a}\right)^2\right) = \frac{(a+b)^n + (a-b)^n}{2a^n}$, $\forall a,b,n\in\mathbb{R}$.
\end{conjecture}

\subsection{Optimal number of sensors}
Now we would like to evaluate the partial derivative of the QFI with respect to $n$ to determine the optimal total number of sensors.
\begin{align}
    &\frac{km^2\left(2^\frac{n}{m} \cancel{- 1}\right)^2\left[\cancel{k +} Fk^\frac{n}{m}\left(2^\frac{n}{m} \cancel{- 2}\right)\right]^2}{n\left[F(2k)^\frac{n}{m} \cancel{- k}\right](2q-1)^\frac{2n}{m}}\left[\frac{\partial}{\partial n}\mathcal{F}_\mathrm{dp}(F,k,q,n,m)\right]\nonumber\\
    &= F^2(2k^2)^\frac{n}{m}\left[\cancel{4m +} m2^{\frac{2n}{m}+1} - 3\times 2^\frac{n}{m}(2m+n\ln 2) \cancel{+ n\ln 16} + n\left(4^\frac{n}{m} - 3\times 2^\frac{n}{m} \cancel{+ 2}\right) (2\ln(2q-1) + \ln k ) \right]\nonumber\\
    &~~~~ + Fk^\frac{m+n}{m}\left[2n\ln k + n2^\frac{n}{m}\left(3\times 2^\frac{n}{m} \cancel{- 5}\right)\ln 2k + 4\left(2^\frac{n}{m} \cancel{- 1}\right)(n\ln(2q-1) + m)\right]\nonumber\\
    &~~~~ + k^2\left[\cancel{2n\ln(2q-1) + 2m} - 2^\frac{n}{m}(2n\ln(2q-1) + 2m - n\ln 2)\right]\\
    \Rightarrow& \frac{Fkm^2(8k)^\frac{n}{m}}{n(2q-1)^\frac{2n}{m}}\left[\frac{\partial}{\partial n}\mathcal{F}_\mathrm{dp}(F,k,q,n,m)\right] \approx F^2(4k^2)^\frac{n}{m}\left[2\left(2^\frac{n}{m} \cancel{- 3}\right)m \cancel{- n\ln 8} + n\left(2^\frac{n}{m} \cancel{- 3}\right)(2\ln(2q-1) + \ln k) \right]\nonumber\\
    &~~~~ + k^22^\frac{n}{m}[n\ln 2 - 2n\ln(2q-1) - 2m] + Fk^\frac{m+n}{m}\left[\cancel{2n\ln k +} 3n4^\frac{n}{m}\ln 2k + 2^{2+\frac{n}{m}}(n\ln(2q-1) + m)\right]\\
    \Rightarrow& \frac{Fkm^2(4k)^\frac{n}{m}}{n(2q-1)^\frac{2n}{m}}\left[\frac{\partial}{\partial n}\mathcal{F}_\mathrm{dp}(F,k,q,n,m)\right] \approx F^2(4k^2)^\frac{n}{m}[n\ln k + 2n\ln(2q-1) + 2m]\nonumber\\
    &~~~~ + Fk^\frac{m+n}{m}\left[3n4^\frac{n}{m}\ln 2k \cancel{+ 4n\ln(2q-1) + 4m}\right] \cancel{+ k^2[n\ln 2 - 2n\ln(2q-1) - 2m]} \\
    \Rightarrow& \frac{km^2 2^\frac{n}{m}}{n(2q-1)^\frac{2n}{m}}\left[\frac{\partial}{\partial n}\mathcal{F}_\mathrm{dp}(F,k,q,n,m)\right] \approx \cancel{3kn\ln 2k +} F(2k)^\frac{n}{m}[2m + n\ln k + 2n\ln(2q-1)]\\
    \Rightarrow& \frac{\partial}{\partial n}\mathcal{F}_\mathrm{dp}(F,k,q,n,m) \approx \frac{Fnk^\frac{n}{m}(2q-1)^\frac{2n}{m}}{km^2} [2m + n\ln k + 2n\ln(2q-1)].
\end{align}
Then the equation $\frac{\partial}{\partial n}\mathcal{F}_\mathrm{dp}(F,k,q,n,m) = 0$ is approximately equivalent to $2m + n\ln k + 2n\ln(2q-1) = 0$, whose solution is 
\begin{align}
    n_\mathrm{dp}^*\approx -\frac{2m}{\ln k + 2\ln(2q-1)} \approx \frac{2m}{5-k-4q},
\end{align}
as mentioned in the main text.

\subsection{Optimal number of sub-ensembles}
Here we evaluate the partial derivative of the QFI with respect to $m$ to determine the optimal partition number. 
\begin{align}
    &\frac{km^3\left(2^\frac{n}{m} \cancel{- 1}\right)^2\left[\cancel{k +} Fk^\frac{n}{m}\left(2^\frac{n}{m} \cancel{- 2}\right)\right]^2}{n^2\left[\cancel{k} - F(2k)^\frac{n}{m}\right](2q-1)^\frac{2n}{m}}\left[\frac{\partial}{\partial m}\mathcal{F}_\mathrm{dp}(F,k,q,n,m)\right] = k^2\left[\cancel{m + 2n\ln(2q-1)} - 2^\frac{n}{m}(2n\ln(2q-1) - n\ln 2 + m)\right]\nonumber\\
    &~~~~ + F^2(2k^2)^\frac{n}{m}\left[4^\frac{n}{m}(2n\ln(2q-1) + n\ln k + m) - 3\times 2^\frac{n}{m}(2n\ln(2q-1) + n\ln 2k + m) \cancel{+ 2(2n\ln(4q-2) + n\ln k + m)}\right]\nonumber\\
    &~~~~ + Fk^\frac{m+n}{m}\left[\cancel{2n\ln k +} 4^\frac{n}{m}n\ln 8k^3 \cancel{- 2(2n\ln(2q-1)+m)} + 2^\frac{n}{m}(4n\ln(2q-1) - 5n\ln 2k + 2m)\right]\\
    \Rightarrow& -\frac{Fkm^3(4k)^\frac{n}{m}}{n^2(2q-1)^\frac{2n}{m}}\left[\frac{\partial}{\partial m}\mathcal{F}_\mathrm{dp}(F,k,q,n,m)\right] \approx Fk^\frac{m+n}{m}\left[\cancel{2m - 5n\ln 2k + 4n\ln(2q-1) +} 2^\frac{n}{m}n\ln 8k^3\right]\nonumber\\
    &~~~~ + F^2(2k^2)^\frac{n}{m}\left[2^\frac{n}{m}n\ln k + \left(2^\frac{n}{m} \cancel{- 3}\right)(2n\ln(2q-1) + m) \cancel{- 3n\ln 2k}\right] \cancel{- k^2[m + 2n\ln(2q-1) - n\ln 2]}\\
    \Rightarrow& -\frac{km^3 2^\frac{n}{m}}{n^2(2q-1)^\frac{2n}{m}}\left[\frac{\partial}{\partial m}\mathcal{F}_\mathrm{dp}(F,k,q,n,m)\right] \approx \cancel{kn\ln 8k^3 +} F(2k)^\frac{n}{m}[m + n\ln k + 2n\ln(2q-1)]\\
    \Rightarrow& \frac{\partial}{\partial m}\mathcal{F}_\mathrm{dp}(F,k,q,n,m) \approx -\frac{Fn^2k^\frac{n}{m}(2q-1)^\frac{2n}{m}}{km^3} [m + n\ln k + 2n\ln(2q-1)].
\end{align}
Then the equation $\frac{\partial}{\partial m}\mathcal{F}_\mathrm{dp}(F,k,q,n,m) = 0$ is approximately equivalent to $m + n\ln k + 2n\ln(2q-1) = 0$, whose solution is 
\begin{align}
    m^*_\mathrm{dp} \approx -n\left[\ln k + 2\ln(2q-1)\right] \approx n(5-k-4q),
\end{align}
as mentioned in the main text.

\subsection{Advantage of partitioning}
The explicit expression of the ratio between optimally partitioned QFI and the monolithic QFI, and corresponding approximation are as follows
\begin{align}
    \frac{\mathcal{F}_\mathrm{dp}(F,k,q,n,m_\mathrm{dp}^*)}{\mathcal{F}_\mathrm{dp}(F,k,q,n)} \approx& - \frac{\left(2^n \cancel{- 1}\right)\left[\cancel{k +} F\left(2^n \cancel{- 2}\right)k^n\right]\left[\cancel{k} - F(2k)^\frac{-1}{\ln k + 2\ln(2q-1)}\right]^2(2q-1)^{-2n-\frac{2}{\ln k + 2\ln(2q-1)}}}{n[\ln k + \ln(2q-1)]\left(2^\frac{-1}{\ln k + 2\ln(2q-1)} \cancel{- 1}\right)\left[\cancel{k} - F(2k)^n\right]^2\left[\cancel{k +} F\left(2^\frac{-1}{\ln k + 2\ln(2q-1)} \cancel{- 2}\right)k^\frac{-1}{\ln k + 2\ln(2q-1)}\right]}\nonumber\\
    \approx& -\frac{1}{e}\frac{\left[k(2q-1)^2\right]^{-n}}{n\left[\ln k + 2\ln(2q-1)\right]},
\end{align}
which is the one shown in the main text. We have used the fact that in practice $k,q\lesssim 1$ so that $\ln k,\ln(2q-1)\lesssim 0$, which makes $x^\frac{-1}{\ln k + 2\ln(2q-1)}\gg 1$ for $x>1$.

The explicit expression of the ratio between optimally partitioned QFI and the monolithic QFI using the optimal total number of sensors is as follows
\begin{align}
    &\frac{4}{n\left[\ln k + 2\ln(2q-1)\right]}\frac{\mathcal{F}_\mathrm{dp}(F,k,q,n,m_\mathrm{dp}^*)}{\mathcal{F}_\mathrm{dp}(F,k,q,n_\mathrm{dp}^*)}\nonumber\\
    &\approx \frac{\left(4^\frac{-1}{\ln k + 2\ln(2q-1)} \cancel{- 1}\right)\left[\cancel{k +} F\left(4^\frac{-1}{\ln k + 2\ln(2q-1)} \cancel{- 2}\right)k^\frac{-1}{\ln k + 2\ln(2q-1)}\right]\left[\cancel{k} - F(2k)^\frac{-1}{\ln k + 2\ln(2q-1)}\right]^2 (2q-1)^\frac{2}{\ln k + 2\ln(2q-1)}}{\left(2^\frac{-1}{\ln k + 2\ln(2q-1)} \cancel{- 1}\right)\left[\cancel{k} - F(4k)^\frac{-1}{\ln k + 2\ln(2q-1)}\right]^2\left[\cancel{k +} F\left(2^\frac{-1}{\ln k + 2\ln(2q-1)} \cancel{- 2}\right)k^\frac{-1}{\ln k + 2\ln(2q-1)}\right]}\nonumber\\
    \Rightarrow& \frac{\mathcal{F}_\mathrm{dp}(F,k,q,n,m_\mathrm{dp}^*)}{\mathcal{F}_\mathrm{dp}(F,k,q,n_\mathrm{dp}^*)} \approx -\frac{e\left[\ln k + 2\ln(2q-1)\right]}{4} n,
\end{align}
which is the one shown in the main text. We have used the same assumptions as above.

\section{Derivations for QFI dynamics}

\subsection{Sensor loss - without loss detection}
\subsubsection{Monotonicity of QFI with initial state preparation quality}
We prove that higher $k$ leads to higher $\mathcal{F}_{\mathrm{loss},1}(t)$ by taking the partial derivative with respect to $k$ as
\begin{align}
    \frac{\partial}{\partial k}\mathcal{F}_{\mathrm{loss},1}(t) =& \frac{\frac{F}{k^2}(n-m)\left(4ke^{-2\eta t}\right)^\frac{n}{m}\left[F(2k)^\frac{n}{m} - k\right] n^2 t^2}{\left(2^\frac{n}{m} - 1\right)\left[mk\left[\left(2^\frac{n}{m} - 1\right) \left(1 + e^{-\eta t}\right)^\frac{n}{m} - \left(2^\frac{n}{m} - 2\right) \left(2e^{-\eta t}\right)^\frac{n}{m}\right] + mF\left(2^\frac{n}{m} - 2\right) \left(2ke^{-\eta t}\right)^\frac{n}{m}\right]^2}\nonumber\\
    &\times \left[F\left(2^\frac{n}{m} - 2\right) \left(2ke^{-\eta t}\right)^\frac{n}{m} + 2k\left(2^\frac{n}{m} - 1\right) \left(1 + e^{-\eta t}\right)^\frac{n}{m} - k\left(2\times 4^\frac{n}{m} - 5\times 2^\frac{n}{m} + 2\right)e^{-\frac{n}{m}\eta t} \right],
\end{align}
where fraction on the first line is obviously non-negative for the practical parameter regime of interest. Therefore, we only need to evaluate the sign of the bracket term on the second line.
\begin{align}
    & F\left(2^\frac{n}{m} - 2\right) \left(2ke^{-\eta t}\right)^\frac{n}{m} + 2k\left(2^\frac{n}{m} - 1\right) \left(1 + e^{-\eta t}\right)^\frac{n}{m} - k\left(2\times 4^\frac{n}{m} - 5\times 2^\frac{n}{m} + 2\right)e^{-\frac{n}{m}\eta t}\nonumber\\
    \geq & F\left(2^\frac{n}{m} - 2\right) \left(2ke^{-\eta t}\right)^\frac{n}{m} + 2k\left(2^\frac{n}{m} - 1\right) \left(2e^{-\eta t}\right)^\frac{n}{m} - k\left(2\times 4^\frac{n}{m} - 5\times 2^\frac{n}{m} + 2\right)e^{-\frac{n}{m}\eta t}\nonumber\\
    =& e^{-\frac{n}{m}\eta t} \left[k\left(3\times 2^\frac{n}{m} - 2\right) + F\left(2^\frac{n}{m} - 2\right) (2k)^\frac{n}{m} \right] \geq 0,
\end{align}
which completes the proof.

\subsubsection{Peak time}
We consider the peak time by taking the partial derivative  with respect to $t$ as
\begin{align}
    &\frac{\left(2^\frac{n}{m} - 1\right) \left(1 + e^{\eta t}\right) \left[F\left(2^\frac{n}{m} - 2\right) \left(2ke^{-\eta t}\right)^\frac{n}{m} + k \left(2^\frac{n}{m} - 1\right) \left(1 + e^{-\eta t}\right)^\frac{n}{m} - k \left(2^\frac{n}{m} - 2\right) \left(2e^{-\eta t}\right)^\frac{n}{m}\right]^2}{\left(2e^{-2\eta t}\right)^\frac{n}{m} \left[F(2k)^\frac{n}{m} - k\right]^2 n^2 t / (m^2k)} \left[\frac{\partial}{\partial t}\mathcal{F}_{\mathrm{loss},1}(t)\right]\nonumber\\
    =& \left[k \left(2^\frac{n}{m} \cancel{- 1}\right) \left(1 + e^{-\eta t}\right)^\frac{n}{m} + F \left(1 + e^{\eta t}\right) \left(2^\frac{n}{m} \cancel{- 2}\right) \left(2ke^{-\eta t}\right)^\frac{n}{m} - k \left(1 + e^{\eta t}\right) \left(2^\frac{n}{m} \cancel{- 2}\right) \left(2e^{-\eta t}\right)^\frac{n}{m}\right] (2m - n\eta t)\nonumber\\
    &+ 2k e^{\eta t} \left(2^\frac{n}{m} \cancel{- 1}\right) \left(1 + e^{-\eta t}\right)^\frac{n}{m} (m - n\eta t)\\
    \approx& 2k e^{\eta t} 2^\frac{n}{m} \left(1 + e^{-\eta t}\right)^\frac{n}{m} \left(\left[\frac{1}{2e^{\eta t}} + \frac{Fk^\frac{n}{m} - k}{k}\frac{1 + e^{\eta t}}{2e^{\eta t}} \left(\frac{2e^{-\eta t}}{1 + e^{-\eta t}}\right)^\frac{n}{m}\right](2m - n\eta t) + (m - n\eta t)\right).
\end{align}
Therefore, the equation $\partial\mathcal{F}_{\mathrm{loss},1}(t)/\partial k=0$ is approximated by
\begin{align}
    \left[\frac{1}{2e^{\eta t}} + \frac{Fk^\frac{n}{m} - k}{k}\frac{1 + e^{\eta t}}{2e^{\eta t}} \left(\frac{2e^{-\eta t}}{1 + e^{-\eta t}}\right)^\frac{n}{m}\right](2m - n\eta t) + (m - n\eta t) = 0,
\end{align}
which is the same as the equation mentioned in the main text, after substituting $\eta t$ with $t$. Also as mentioned in the main text, the above approximate equation can be solved analytically by Taylor expanding the terms that depend on $t$ non-linearly. Here we consider the first-order Taylor expansions:
\begin{align}
    & \frac{1}{2e^{\eta t}} \approx \frac{1 - t}{2} + O(t^2),\\
    & \frac{1 + e^{\eta t}}{2e^{\eta t}} \left(\frac{2e^{-\eta t}}{1 + e^{-\eta t}}\right)^\frac{n}{m} \approx 1 - \frac{m+n}{2m}t + O(t^2).
\end{align}
Then one is able to obtain a closed form expression of approximate peak time
\begin{align}
    t_{\mathrm{loss},1}^* \approx \frac{2m}{n}\frac{kn - 2F(2n+m)k^\frac{n}{m} + \sqrt{4F^2m^2k^\frac{2n}{m} + 4F(2n-m)n k^\frac{m+n}{m} + k^2 n^2}}{4\left[kn - k^\frac{n}{m}F(n+m)\right]}.
\end{align}
Through comparison with $t_{\mathrm{loss},2}^*$, we can see that $t_{\mathrm{loss},1}^*$ is $t_{\mathrm{loss},2}^*$ modified by a complicated factor as shown by the second fraction above.

\subsection{Sensor loss - with loss detection}
We prove that higher $k$ leads to higher $\mathcal{F}_{\mathrm{loss},2}(t)$ by taking the partial derivative with respect to $k$ as
\begin{align}
    \frac{\partial}{\partial k}\mathcal{F}_{\mathrm{loss},2}(t) =& \frac{F n^2 (n-m) \left[F(2k)^\frac{n}{m} - k\right] e^{-\frac{n}{m}\eta t} t^2}{m^2 k^2 \left(2^\frac{n}{m} - 1\right) \left[F \left(2^\frac{n}{m} - 2\right) k^\frac{n}{m} + k\right]^2} \left[F(2k)^\frac{n}{m} \left(2^\frac{n}{m} - 2\right) + k\left(3\times 2^\frac{n}{m} - 2\right) \right].
\end{align}
Obviously the fraction on the right hand side of the above equation is non-negative. Meanwhile, the bracket term is also positive in the parameter regime of our interest, specifically $n/m>1$. Therefore, the QFI at a fixed evolution time $t$ monotonically increases as $k$ increases.

\subsection{Highest achievable QFI}
Here we present the exact expression of the highest achievable QFI when combining sensor loss and dephasing, and also assuming loss detectability.
\begin{align}
    \mathcal{F}_\mathrm{loss,dp}(t^*_\mathrm{loss,dp}) =& \frac{4m\left[F(2k)^\frac{n}{m} \cancel{- k}\right]^2}{e^2k \left(2^\frac{n}{m} \cancel{- 1}\right)\left[F\left(2^\frac{n}{m} \cancel{- 2}\right) k^\frac{n}{m} \cancel{+ k}\right](2\gamma + \eta)^2}\nonumber\\
    \approx& \frac{4F}{ek(2\gamma + \eta)^2}k^\frac{n}{m}m,
\end{align}
which is the shown in the main text, where we have omitted exponentially smaller terms.

\subsection{Short time QFI}
Here we examine the $t$ derivative of the QFI at early evolution times $t\rightarrow 0$.
\begin{align}
    \frac{\partial}{\partial t}\mathcal{F}_\mathrm{loss,dp}(t) =& \frac{2n^2\left[F(2k)^\frac{n}{m} - k\right]^2}{mk \left(2^\frac{n}{m} - 1\right)\left[F\left(2^\frac{n}{m} - 2\right) k^\frac{n}{m} + k\right]}t + O(t^2).
\end{align}
We are interested in how the factor of the linear term change when $m$ changes, so we take its $m$ derivative
\begin{align}
    &\frac{m^3k\left(2^\frac{n}{m} \cancel{- 1}\right)\left[F\left(2^\frac{n}{m} \cancel{- 2}\right) k^\frac{n}{m} \cancel{+ k}\right]}{2n^2\left[\cancel{k} - F(2k)^\frac{n}{m}\right]} \left[\frac{\partial}{\partial m}\frac{2n^2\left[F(2k)^\frac{n}{m} - k\right]^2}{mk \left(2^\frac{n}{m} - 1\right)\left[F\left(2^\frac{n}{m} - 2\right) k^\frac{n}{m} + k\right]}\right]\nonumber\\
    &= nFk^\frac{n}{m}\ln k\left(2^\frac{n}{m} \cancel{- 1}\right)\left[\left(3\times 2^\frac{n}{m} \cancel{- 2}\right)k + F(2k)^\frac{n}{m}\left(2^\frac{n}{m} \cancel{- 2}\right)\right] + F^\frac{m+n}{m}\left[2^\frac{n}{m}(2m-5n\ln2) + 4^\frac{n}{m}n\ln 8 \cancel{- 2m} \right]\nonumber\\
    &~~~~+ k^2\left[\cancel{m +} 2^\frac{n}{m}(n\ln 2 - m)\right] + F^2(2k^2)^\frac{n}{m}\left[4^\frac{n}{m}m - 3\times 2^\frac{n}{m}(m+n\ln 2) \cancel{+ 2(m+n\ln 4)}\right]\\
    \Rightarrow& \frac{m^3Fk(4k)^\frac{n}{m}}{2n^2} \left[\frac{\partial}{\partial m}\frac{2n^2\left[F(2k)^\frac{n}{m} - k\right]^2}{mk \left(2^\frac{n}{m} - 1\right)\left[F\left(2^\frac{n}{m} - 2\right) k^\frac{n}{m} + k\right]}\right]\nonumber\\
    &\approx \cancel{k^2(m - n\ln 2)} - F^2(2k^2)^\frac{n}{m}\left[2^\frac{n}{m}(m+n\ln k) \cancel{- (3m + n\ln 8)} \right] - Fk^\frac{m+n}{m}\left[3\times 2^\frac{n}{m}n \ln(2k) \cancel{+ (2m - 5n\ln 2)}\right]\\
    \Rightarrow& \frac{\partial}{\partial m}\frac{2n^2\left[F(2k)^\frac{n}{m} - k\right]^2}{mk \left(2^\frac{n}{m} - 1\right)\left[F\left(2^\frac{n}{m} - 2\right) k^\frac{n}{m} + k\right]} \approx - \frac{2n^2}{m^3k2^\frac{n}{m}}\left[F(2k)^\frac{n}{m}(m+n\ln k) \cancel{+ 3kn\ln(2k)}\right]\\
    \Rightarrow& \frac{\partial}{\partial m}\frac{2n^2\left[F(2k)^\frac{n}{m} - k\right]^2}{mk \left(2^\frac{n}{m} - 1\right)\left[F\left(2^\frac{n}{m} - 2\right) k^\frac{n}{m} + k\right]} \approx - \frac{2Fk^\frac{n}{m}n^2}{m^3k}(m+n\ln k).
\end{align}
It is obvious that when $m \leq -n\ln k$ the $m$ derivative is positive, which means that increasing the partition number will result in faster information accumulation at short times. This is exactly the claim in the main text.

\subsection{Sequential scheme}
\subsubsection{Highest QFI per unit time}
Here we present the exact expression of the highest achievable QFI when combining sensor loss and dephasing, and also assuming loss detectability.
\begin{align}
    \frac{\mathcal{F}_\mathrm{loss,dp}(\tilde{t}^*_\mathrm{loss,dp})}{\tilde{t}^*_\mathrm{loss,dp}} =& \frac{n\left[F(2k)^\frac{n}{m} \cancel{- k}\right]^2}{ek \left(2^\frac{n}{m} \cancel{- 1}\right)\left[F\left(2^\frac{n}{m} \cancel{- 2}\right) k^\frac{n}{m} \cancel{+ k}\right](2\gamma + \eta)} \nonumber\\
    \approx& \frac{F}{ek(2\gamma+\eta)}k^\frac{n}{m}n.
\end{align}
which is the shown in the main text, where we have omitted exponentially smaller terms.

\subsubsection{Time derivative of QFI per unit time}
The $t$ derivative of $\mathcal{F}_\mathrm{loss,dp}(t)/t$ is
\begin{align}
    \frac{\partial}{\partial t}\left[\frac{\mathcal{F}_\mathrm{loss,dp}(t)}{t}\right] =& \frac{n^2\left[F(2k)^\frac{n}{m} - k\right]^2e^{-\frac{n}{m}(2\gamma+\eta) t}}{m^2k \left(2^\frac{n}{m} - 1\right)\left[F\left(2^\frac{n}{m} - 2\right) k^\frac{n}{m} + k\right]}[m - n(2\gamma+\eta) t]\nonumber\\
    =& \frac{\left[F(2k)^\frac{n}{m} - k\right]^2n^2}{mk \left(2^\frac{n}{m} - 1\right)\left[F\left(2^\frac{n}{m} - 2\right) k^\frac{n}{m} + k\right]} + O(t).
\end{align}
We then take the $m$ derivative of the zero-th order
\begin{align}
    &\frac{m^3k \left(2^\frac{n}{m} \cancel{- 1}\right)^2\left[F\left(2^\frac{n}{m} \cancel{- 2}\right) k^\frac{n}{m} \cancel{+ k}\right]^2}{n^2 \left[F(2k)^\frac{n}{m} \cancel{- k}\right]}\left[\frac{\partial}{\partial m}\frac{\left[F(2k)^\frac{n}{m} - k\right]^2n^2}{mk \left(2^\frac{n}{m} - 1\right)\left[F\left(2^\frac{n}{m} - 2\right) k^\frac{n}{m} + k\right]}\right]\nonumber\\
    &= k^2\left[\cancel{m +} 2^\frac{n}{m}(n\ln 2-m)\right] + F^2(2k^2)^\frac{n}{m}\left[4^\frac{n}{m}m - 3\times 2^\frac{n}{m}(m+n\ln 2) \cancel{+ 2(m+n\ln 4)}\right]\nonumber\\
    &~~~~+ Fk^\frac{n}{m}\ln k\left(2^\frac{n}{m} \cancel{- 1}\right)n\left[k\left(3\times 2^\frac{n}{m} \cancel{- 2}\right) + F(2k)^\frac{n}{m}\left(2^\frac{n}{m} \cancel{- 2}\right)\right]\nonumber\\
    &~~~~+ Fk^\frac{m+n}{m}\left[2^\frac{n}{m}(2m - 5n\ln 2) + 4^\frac{n}{m}n\ln 8 \cancel{- 2m}\right]\\
    \Rightarrow& \frac{m^3Fk(4k)^\frac{n}{m}}{n^2}\left[\frac{\partial}{\partial m}\frac{\left[F(2k)^\frac{n}{m} - k\right]^2n^2}{mk \left(2^\frac{n}{m} - 1\right)\left[F\left(2^\frac{n}{m} - 2\right) k^\frac{n}{m} + k\right]}\right]\nonumber\\
    &\approx Fk^\frac{m+n}{m}\left[\cancel{n\ln 32 - 2m} - 3\times 2^\frac{n}{m}n\ln(2k)\right] - F^2(2k^2)^\frac{n}{m}\left[2^\frac{n}{m}(m+n\ln k) \cancel{- 3m - n\ln 8}\right] \cancel{+ k^2(m-n\ln 2)}\\
    \Rightarrow& \frac{\partial}{\partial m}\frac{\left[F(2k)^\frac{n}{m} - k\right]^2n^2}{mk \left(2^\frac{n}{m} - 1\right)\left[F\left(2^\frac{n}{m} - 2\right) k^\frac{n}{m} + k\right]} \approx -\frac{n^2}{2^\frac{n}{m}m^3}\left[\frac{F}{k}(2k)^\frac{n}{m}(m + n\ln k) \cancel{+ 3n\ln(2k)}\right]\\
    \Rightarrow& \frac{\partial}{\partial m}\frac{\left[F(2k)^\frac{n}{m} - k\right]^2n^2}{mk \left(2^\frac{n}{m} - 1\right)\left[F\left(2^\frac{n}{m} - 2\right) k^\frac{n}{m} + k\right]} \approx -\frac{n^2F(2k)^\frac{n}{m}}{2^\frac{n}{m}m^3k}(m + n\ln k).
\end{align}
It is obvious that when $m \leq -n\ln k$ the $m$ derivative is positive, which means that increasing the partition number will result in faster information accumulation at short times. This is exactly the claim in the main text.

\subsubsection{Optimal partition number}
Here we explicitly take the $m$ derivative of $\mathcal{F}_\mathrm{loss,dp}(t)/t$
\begin{align}
    &\frac{km^3\left(2^\frac{n}{m} \cancel{- 1}\right)^2\left[F\left(2^\frac{n}{m} \cancel{- 2}\right) k^\frac{n}{m} \cancel{+ k}\right]^2}{n^2 \left[\cancel{k} - F(2k)^\frac{n}{m}\right]e^{-\frac{n}{m}(2\gamma+\eta)t}t}\frac{\partial}{\partial m}\left[\frac{\mathcal{F}_\mathrm{loss,dp}(t)}{t}\right]\nonumber\\
    &= k^2\left[\cancel{m - 2n\gamma t - n\eta t} + 2^\frac{n}{m}(2n\gamma t + n\eta t + n\ln 2 - m)\right] + Fk^\frac{n}{m}\ln k\left(2^\frac{n}{m} \cancel{- 1}\right)n\left[k\left(3\times 2^\frac{n}{m} \cancel{- 2}\right) + F(2k)^\frac{n}{m}\left(2^\frac{n}{m} \cancel{- 2}\right)\right]\nonumber\\
    &~~~~+ F^2(2k^2)^\frac{n}{m}\left[\cancel{2m - 4n\gamma t - 2n\eta t} + 4^\frac{n}{m}(m - 2n\gamma t - n\eta t) + 2^\frac{n}{m}(6n\gamma t + 3n\eta t - 3m - n\ln 8) \cancel{+ n\ln 16}\right]\nonumber\\
    &~~~~+ Fk^\frac{m+n}{m}\left[\cancel{4n\gamma t + 2n\eta t - 2m }+ 4^\frac{n}{m}n\ln 8 + 2^\frac{n}{m}(2m - n\ln 32 - 4n\gamma t - 2n\eta t)\right]\\
    \Rightarrow& -\frac{m^3F(4k)^\frac{n}{m}e^{\frac{n}{m}(2\gamma+\eta)t}}{kn^2 t}\frac{\partial}{\partial m}\left[\frac{\mathcal{F}_\mathrm{loss,dp}(t)}{t}\right]\nonumber\\
    &\approx \cancel{k^2(2n\gamma t + n\eta t + n\ln 2 - m)} + F^2(2k^2)\left[\left(2^\frac{n}{m} \cancel{- 3}\right)m - \left(2^\frac{n}{m} \cancel{- 3}\right)n(2\gamma + \eta)t \cancel{- n\ln 8}\right]\nonumber\\
    &~~~~+ Fk^\frac{m+n}{m}\left[2^\frac{n}{m}n\ln 8 \cancel{+ 2m - 2n(2\gamma + \eta)t - n\ln 32}\right] + F(2k)^\frac{n}{m}\left[F(2k)^\frac{n}{m} \cancel{+ 3k}\right]n\ln k\\
    \Rightarrow& -\frac{m^32^\frac{n}{m}e^{\frac{n}{m}(2\gamma+\eta)t}}{kn^2 t}\frac{\partial}{\partial m}\left[\frac{\mathcal{F}_\mathrm{loss,dp}(t)}{t}\right]\approx F(2k)^\frac{n}{m}[m - n(2\gamma + \eta)t + n\ln k] \cancel{+ kn\ln 8}\\
    \Rightarrow& \frac{\partial}{\partial m}\left[\frac{\mathcal{F}_\mathrm{loss,dp}(t)}{t}\right]\approx - \frac{Fk^\frac{m+n}{m}n^2 t}{m^3e^{\frac{n}{m}(2\gamma+\eta)t}}[m - n(2\gamma + \eta)t + n\ln k].
\end{align}
Then the equation $\frac{\partial}{\partial m}[\mathcal{F}_\mathrm{loss,dp}(t)/t]=0$ is approximately equivalent to $m - n(2\gamma + \eta)t + n\ln k = 0$, whose solution is 
\begin{align}
    \tilde{m}^*(t) \approx n(2\gamma + \eta)t - n\ln k,
\end{align}
as mentioned in the main text.

\subsection{Comparison with spin squeezed states}
Here we provide the derivations for GHZ Ramsey spectroscopy performance. Recall that the probability to measure 0 for $n$-qubit noisy GHZ Ramsey is
\begin{align}
    P_0 = \frac{1}{2} + \frac{Fk^{n-1}V(n)[2\tilde{q}(t) - 1]}{2}\cos(n\omega t),
\end{align}
with $\tilde{q}(t) = (1 + e^{-n\gamma t})/2$ and $V(n) = (2^nFk^{n-1} - 1)/(2^n - 1)$. Then according to error propagation we have the explicit expression of frequency estimation variance
\begin{align}
    \mathrm{Var}(\omega) =& \frac{P_0(1-P_0)}{\left\vert\frac{\partial P_0}{\partial \omega}\right\vert^2}\nonumber\\
    =& \frac{k^{4-2n}\left(2^n - 1\right)^2e^{2n\gamma t} - F^2\left[F(2k)^n - k\right]^2\cos^2(n\omega t)}{n^2F^2\left[F(2k)^n - k\right]^2t^2\sin^2(n\omega t)}.
\end{align}
Note that the measurement is binary so its variance is just $\mathrm{Var}(P_0) = P_0(1-P_0)$. Then in the sequential scheme where the total duration is $T=\mu t$, i.e. repeating Ramsey with $t$ duration for $\mu$ times, the final estimation variance is
\begin{align}
    \mathrm{Var}(\omega) \rightarrow& \mathrm{Var}(\omega)\frac{1}{\mu}\nonumber\\
    =& \frac{k^{4-2n}\left(2^n - 1\right)^2e^{2n\gamma t} - F^2\left[F(2k)^n - k\right]^2\cos^2(n\omega t)}{n^2F^2\left[F(2k)^n - k\right]^2\sin^2(n\omega t)tT}.
\end{align}
The above derivation is conditioned on no loss with only dephasing. Then under the assumption of loss detectability, we can include loss by
\begin{align}
    \mathrm{Var}(\omega) \rightarrow& e^{n\eta t}\mathrm{Var}(\omega)\nonumber\\
    =& \frac{k^{4-2n}\left(2^n - 1\right)^2e^{n(2\gamma+\eta)t} - F^2\left[F(2k)^n - k\right]^2\cos^2(n\omega t)e^{n\eta t}}{n^2F^2\left[F(2k)^n - k\right]^2\sin^2(n\omega t)tT}.
\end{align}
Then to obtain the optimal conditions, we take the $t$ derivative of the above sequential scheme estimation variance
\begin{align}
    &n^2F^2k^{2n}\left[F(2k)^n - k\right]^2Te^{-n\eta t}t\left[\frac{\partial}{\partial t}\mathrm{Var}(\omega)\right]\nonumber\\
    =& 2F^2k^{2n}\left[F(2k)^n - k\right]^2n\omega t\cot^3(n\omega t) - F^2k^{2n}(n\eta t-1)\cot^2(n\omega t) + 2F^2k^{2n}\left[F(2k)^n - k\right]^2n\omega t\cot(n\omega t)\nonumber\\
    &- 2k^4 \left(2^n - 1\right)^2 e^{2n\gamma t}n\omega t\csc^2(n\omega t)\cot(n\omega t) + k^4 \left(2^n - 1\right)^2 e^{2n\gamma t}\csc^2(n\omega t)[n(2\gamma+\eta)t - 1].
\end{align}
Then it is obvious that the optimal conditions are
\begin{align}
    & n\omega t = (2j+1)\pi/2,~ 0\leq j\in\mathbb{Z},\\
    & n(2\gamma+\eta)t=1,
\end{align}
as mentioned in the main text. Then by substituting the optimal conditions into the original expression, we obtain the optimal estimation variance for sequential noisy Ramsey with $n$-qubit GHZ state
\begin{align}
    \mathrm{Var}(\omega)_\mathrm{GHZ} = \frac{e\left(2^n - 1\right)^2(2\gamma+\eta)}{F^2k^{2n-4}\left[F(2k)^n - k\right]^2nT}.
\end{align}
When we consider partitioning, the final estimation variance is simply
\begin{align}
    m\left.\mathrm{Var}(\omega)_\mathrm{GHZ}\right\vert_{n\rightarrow \frac{n}{m}} = \frac{e\left(2^\frac{n}{m} - 1\right)^2(2\gamma+\eta)}{F^2k^{2\frac{n}{m}-4}\left[F(2k)^\frac{n}{m} - k\right]^2nT},
\end{align}
as mentioned in the main text. 

Finally, for the comparison with OAT SSS, the crossing time can be found by solving the equation
\begin{align}
    \frac{\xi_S^2}{n t_\mathrm{cross} T} = \frac{e\left(2^\frac{n}{m} - 1\right)^2(2\gamma+\eta)}{F^2k^{2\frac{n}{m}-4}\left[F(2k)^\frac{n}{m} - k\right]^2nT},
\end{align}
which leads to the scaled cross time
\begin{align}
    \frac{n}{m}(2\gamma+\eta)t_\mathrm{cross} = \frac{nF^2k^{2\frac{n}{m}-4}\left[F(2k)^\frac{n}{m} - k\right]^2}{em\left(2^\frac{n}{m} - 1\right)^2}\xi_S^2,
\end{align}
which we compare with unit.

\end{document}